%% file: draft.tex
\documentclass[a4,12pt]{book}
\usepackage{epsfig,amssymb,graphicx,subfigure,fancyheadings}

\newcommand{\be}{\begin{equation}}
\newcommand{\ee}{\end{equation}}
\newcommand{\bea}{\begin{eqnarray}}
\newcommand{\eea}{\end{eqnarray}}

\newcommand{\eqref}[1]{(\ref{#1})}
\newcommand{\cL}{{\cal L}}

\renewcommand{\d}{\partial}

\newcommand{\x}{{\bf x}}
\newcommand{\y}{{\bf y}}
\newcommand{\ph}{\varphi}

\newcommand{\bP}{\bar\Phi}
\newcommand{\ep}{\varepsilon}

\newcommand{\q}{{\bf q}}
\renewcommand{\k}{{\bf k}}
\newcommand{\p}{{\bf p}}

\renewcommand{\Im}{\,\textrm{Im}\,}
\renewcommand{\Re}{\,\textrm{Re}\,}

\newcommand{\pint}[2]{{\int\!\frac{d^{#1}#2}{(2\pi)^#1}\,}}

\newcommand{\bd}{\begin{displaymath}}
\newcommand{\ed}{\end{displaymath}}
\newcommand{\bb}{\left[}
\newcommand{\eb}{\right]}
\newcommand{\kabs}{|\textbf{k}|}
\newcommand{\xvon}{\textbf{x}}
\newcommand{\arctg}{\textrm{arctg}}

\def\tort#1#2{{#1\over #2}}

\def\avr#1{\langle{#1}\rangle}
\def\textmini#1{\textrm{\footnotesize{#1}}}
\def\textbfit#1{\textrm{\bfseries\itshape{#1}}}
\def\nn{\\ \nonumber}
\newcommand{\nonu}{\nonumber\\}

\newcommand{\eq}[1]{equation~(\ref{#1})}

\def\bbox{\partial^2}

\newcommand{\lp}{\left(}
\newcommand{\rp}{\right)}

\def\la{\langle}
\def\ra{\rangle}

\def\eq#1{(\ref{#1})}

\def\vavr#1{\overline{#1}^V}
\def\addtext#1{{#1}}
\def\removetext#1{{}}

\def\at#1{\addtext{#1}}
\def\rt#1{\removetext{#1}}

\newcommand{\clearemptydoublepage}{\newpage{\pagestyle{empty}\cleardoublepage}}

\begin{document}

\thispagestyle{empty}
\vspace*{4cm}
{\Large\bf\centerline{Real Time Dynamics of Symmetry Breaking}}
\vskip2.5truecm
\centerline{\large Zsolt Sz\'ep}
\vskip0.5truecm
\centerline{\large Ph.D. Thesis}
\vskip1.5truecm
\centerline{\large Supervisor: Andr\'as Patk{\'o}s}
\vspace*{1.5cm}
\centerline{\large physics program}
\centerline{\large program leader: Horv\'ath Zal\'an}
\centerline{\large subprogram leader: George P\'ocsik}

\vskip2truecm
\centerline{\large Department of Atomic Physics}
\centerline{\large E\"otv{\"o}s Lor\'and University, Budapest}
\vskip0.5truecm
\centerline{\large September 2001}

\clearemptydoublepage

\pagestyle{fancyplain}
\pagenumbering{roman}

\tableofcontents
\clearemptydoublepage
\pagestyle{plain}
\input{intro.tex}

\clearemptydoublepage
\pagestyle{fancyplain}
\input{formalism.tex}

\clearemptydoublepage
\input{scalar1l.tex}
\renewcommand{\sectionmark}[1]%
{\markright{\MakeUppercase{%
\thesection. Effective OP dynamics and the decay of a metastable
vacuum \dots }}}
\input{scalar1nl.tex}

\clearemptydoublepage
\renewcommand{\sectionmark}[1]%
{\markright{\MakeUppercase{%
\thesection. #1}}}
\input{scalar2l.tex}

\renewcommand{\sectionmark}[1]%
{\markright{\MakeUppercase{%
\thesection. Nonlinear relaxation of classical O(2) symmetric fields}}}
\clearemptydoublepage
\input{scalar2nl.tex}

\clearemptydoublepage
\renewcommand{\sectionmark}[1]%
{\markright{\MakeUppercase{%
\thesection. #1}}}
\input{higgs.tex}
\clearemptydoublepage
\input{acknowledgments.tex}
\clearemptydoublepage
\appendix
\input{appendix.tex}

\clearemptydoublepage

\clearemptydoublepage
\input{results.tex}

\clearemptydoublepage
\input{eredmenyek.tex}
\end{document}

%% file: intro.tex
\chapter{Introduction and motivation}
\pagenumbering{arabic}
\pagestyle{plain}

In the last few years remarkable progress has been made in the investigation
of the non-equilibrium behaviour of finite temperature systems, both for
scalar and gauge fields. Actually, after understanding equilibrium features
of these systems many of the workers in the field have eagerly blazed away
at investigating non-equilibrium aspects of finite temperature Quantum Field
Theories, with a special emphasis on the real time simulation of its high
temperature limit. There are a few good reasons for doing this.

First of all, time dependent phenomena in hot relativistic quantum field
theory play an important role in applications to cosmology, in the context
of baryogenesis and inflation, and in the heavy ion collisions. Not all of
these phenomena are non-perturbative (for example relaxation phenomena to
equilibrium from states slightly out of equilibrium can be treated
perturbatively) but as the departure from equilibrium increases
non-perturbative tools are needed. Non-perturbative calculations under
general circumstances are very difficult. Real time numerical simulations in
the QFT would face the problem of complex weights appearing in the quantum
expectation value of time dependent quantities. Fortunately a classical
description can be accurate for physical quantities which are determined by
low momentum modes of the theory, because these modes are highly occupied at
high enough temperature.  For high momentum modes, the classical treatment
is incorrect, but in general the influence of these modes can be taken into
account perturbatively.  Initially proposed in \cite{GriRu88} for the
computation of the rate of sphaleron transitions, the numerical solution of
the classical approximation proved to be eventually a powerful tool in
evaluating quantities for systems developing in time.

\vspace*{0.5cm}
\noindent
{\bf Preheating and the possibility of Electroweak baryogenesis}

An intensively investigated field is the transition from an inflationary to
the radiation dominated epoch in the early Universe. In the context of
chaotic inflation \cite{KofLiSta94} it was realised that the initially
homogeneous inflaton field can decay very rapidly into narrow bands of its 
own low momentum modes or into modes of other scalar fields through parametric 
resonance. This phenomenon goes by the name of preheating.

The large amplitudes of the parametrically amplified modes make this problem
non-pertur\-ba\-tive, but at the same time the large occupation number of these
modes opens the possibility to study this problem in the classical field
approximation on lattice (for example \cite{Khlebnikov} and references
therein). It was then realised that in addition to the standard
high-temperature phase transitions that is assumed to occur in the state of
thermal equilibrium \cite{Lindebook}, there exists a new class of phase
transitions which may occur at the intermediate stage between the end of
inflation and the establishing of thermal equilibrium
\cite{KofLiSta96,Khelbnikov98}.  These new types of phase transitions are
referred as to cosmological non-thermal phase transitions after preheating,
and can restore symmetry on scales up to $10^{16}$ GeV, due to large
amplitude non-thermal fluctuations. An important feature of the non-thermal
phase transition based on preheating is that the classical non-linear
dynamics of inflaton field coupled with other boson fields can induce
symmetry breaking for the latter.

Symmetry breaking is a fundamental component of theoretical particle physics
because through the Higgs effect it accounts for the masses of the
elementary particles. These particles have got their masses during the
electroweak phase transition. The importance of this phase transition is
revealed by its impact on the observed baryon asymmetry in our Universe. A
successful theory of baryogenesis requires beyond the existence of baryon
number violating processes C and CP violating processes and departure from
thermal equilibrium. These three requirements are known as Sakharov
conditions. In the standard model (SM) baryon number is not conserved
because of the non-perturbative processes in the gauge sector that involve
the quantum anomaly. The usual scenario assumes that the universe was in
thermal equilibrium before and after the electroweak phase transition, and
far from it during the phase transition. In order to significantly drive the
primordial plasma out of equilibrium a strongly first order phase transition
is needed. This is also a condition for the sphaleron processes to stop
quickly after the transition in order to prevent the washing out of the
created baryon asymmetry. The order of the EW phase transition was
thoroughly analysed, both perturbatively \cite{Fodor95} and with MC
simulations \cite{Kajantie97,Fodor99} and the conclusion was, that there is
an end point on the line of the first order phase transition separating the
symmetric and broken phases at a Higgs mass of about $\approx$ 72 GeV. For
greater Higgs mass values there is no first order PT that makes the things
hopeless in view of the LEP experiments that exclude a Higgs mass below 113
GeV. So, the possibility of baryogenesis in the usual scenario is ruled out
in the SM. This also means that if the EWPT occurred in thermal equilibrium
then the onset of the Higgs effect was a smooth process.

In the usual scenario one can go to a minimal supersymmetric extension of
the SM. The phase structure of MSSM is much more difficult to explore
because of its large parameter space. According to perturbation theory and
3-dimensional lattice simulations \cite{Laine} as well as 4-dimensional
lattice simulation \cite{Katz}, there could be an EWPT in MSSM that is strong 
enough for baryogenesis up to a value of $\approx$ 105 GeV for the lightest 
Higgs mass.

Recently it was shown that the baryogenesis can take place below the EW
scale in a very economical extension of the SM in which only one more scalar
$SU(2)\times U(1)$-singlet field is required, to be taken the inflaton
\cite{Garcia99,Trodden}.  The novel scenario make use of the possibilities
that reside in the phenomena of preheating. The preheating precedes the
establishment of thermal equilibrium. With an appropriate choice of the
parameters of the inflation one can achieve that the reheating temperature
remains below the EW scale, therefore no intensive sphaleron processes are
allowed. There is an important baryon asymmetry production during
rescattering after preheating due to the fact that the low modes that carry
a large amount of the energy quickly ``thermalize'' at a high effective
temperature (a few times the value of the EW scale) where the rate of
sphaleron transitions is very effective.

\vspace*{0.5cm}
\noindent
{\bf Thermalization and relaxation}

The approach to equilibrium of initially out-of-equilibrium states is a
highly important issue in many branches of physics ranging from inflationary
cosmology (the spectrum of density fluctuation in the early universe)
through particle physics (the problem of baryogenesis, formation of DCC in
heavy ion collisions) to statistical physics (dynamics of phase transitions,
realization of Boltzmann's conjecture that is an ensemble of isolated
interacting systems eventually approaches thermal equilibrium at large times.)

In the recent field theoretical studies of thermalization and relaxation
classical fields are intensively investigated \cite{aartsPRD63}. The
interest stems from the fact that in order to explore the reliability of
truncation and expansion schemes they should be benchmarked against this
exact solution. Recently, evolution of equal-time 1PI correlation functions
derived from the effective time dependent action \cite{Wetterich99} were
confronted with the results of the exact time evolution. By solving
equations non-local in time, obtained from 2PI effective action
thermalization of quantum fields was demonstrated \cite{cox}. These
approximate methods can be formulated both for classical and quantum cases.
In the quantum case the analogue of the classical ensemble averaging over
the initial conditions is the quantum expectation value. With this
correspondence the formal derivations, and hence the results, are quite
similar \cite{aarts98}.

\rt{
The many apparently disjoint parts of this Thesis reflect the interest of
the research group at a particular moment of time. Still, there was a topic
of constant interest, that of developing a systematic formalism in which
eventually one can keep track both analytically and numerically of the
setting in of the Higgs effect and in which the induced gauge current can be
evaluated in the presence of the Higgs condensate.}

\vspace*{0.5cm}
\noindent
{\bf Heavy Ion collision}

Two phenomena of the QCD chiral phase transition have received much
attention in the past decade due to the fact that both may be within reach
of discovery by the current or forthcoming relativistic heavy-ion collision
experiments: the formation of disoriented chiral condensate (DCC) and the end 
point (the critical point $E$) of the first order phase transition line
for decreasing chemical potential $\mu$ in the $\mu-T$ QCD phase diagram.

When both areas are studied in the context of heavy-ion collision
experiments a knowledge of the evolution of far from equilibrium dynamical
system of strongly interacting quantum fields of bosonic and fermionic
degrees of freedom is needed. During the evolution the system rapidly
expands and cools creating and emitting baryons and mesons.

As investigating such systems and in particular phase transitions directly in
QCD is out of question at the moment, a model that encodes relevant aspects
of QCD in a faithful manner is needed.

It was argued in \cite{WilRa} that the O(4)-symmetric linear $\sigma$-model
is representative of the physics relevant for the formation of DCC.
This model is in the same static universality class as the QCD with two
massless quarks, that is a good approximation to the world at temperatures
and energies below
$\lambda_{QCD}$. The basic idea of DCC formation is that regions of
misaligned vacuum might occur. These are regions where the field
$\Phi^a\equiv(\sigma,\vec\pi)$ instead of taking the true ground state value
$(v,0)$ is partially aligned with the $\pi$ direction. It is expected that
the relaxation of the misaligned vacuum region to the true vacuum would
proceed through coherent pion emission with a 
charge distribution violating the isospin symmetry.
For energetical reasons (see Ref. \cite{WilRa} ) the emission of a large number
of pions that produce an easily detectable signal could only be envisaged in
a highly non-thermal chiral phase transition, that is when the system is far
out of equilibrium. This situation is approximated with a sudden quench from
high to low temperature during which the long wavelength modes of the pion
fields become unstable and grow relative to the short wavelength ones.

At the critical point $E$ the phase transition is second order and belongs
to the Ising universality class. The pions remain massive but the correlation 
length of the $\sigma$ field, that plays the role of the order parameter, 
diverges due to growing long wavelength fluctuations.
An increase in the correlation length compared with the one corresponding to
the equilibrium energy density of the QCD plasma can be observed by
analysing the low transversal momentum pions in which the $\sigma$ fields
decay.
 
At present it is not possible to construct the exact mapping between the axes 
of QCD phase diagrams and the axes of the Ising phase diagrams.
With the new possibility of simulating the QCD phase diagram at non-zero
chemical potential proposed in \cite{katz_mu} this question may be solved
in the near future.

Critical slowing down that occurs near the critical point $E$ requires long
equilibration times meaning that the plasma will inevitably slow out of
equilibrium if the plasma is cooled. Since the rate of
cooling is faster than the rate at which the system can adapt to this
change, a non-equilibrium evolution is guaranteed.

The  phenomenon we are interested in both cases cited above is related to
the growth of the correlation length basically produced by the long
wavelength, classical modes. This is an argument favouring classical real
time numerical simulations. The inclusion of quantum correction is an
important question and has been implemented in mean-field approximation.

Recently the effect of dynamical behaviour of the $\sigma$ correlation length 
around the critical point $E$ was studied in Ref. \cite{BeRaj}. An intuitive
mapping between QCD and the Ising model was proposed identifying the
magnetic field and the reduced temperature of the Ising model with the
temperature of the QCD and the chemical potential, respectively. The result
of the investigation, a factor of 2-3 increase in the correlation length
proved to be insensitive to moderate tilt in this mapping. In Ref.
\cite{Sexty} basically the same investigation was performed with a second
order dynamics in the context of $\Phi^4$ theory.

\vspace*{0.5cm}
\noindent
{\bf The classical approximation and its problems.} 

It is well known that at high temperature the thermodynamics of a QFT in
($3+1$) dimensions is successfully described through dimensional reduction
(DR) by an effective classical 3 dimensional theory derived for the static
modes of the original theory. Its couplings are determined by the
integration over the non-static modes \cite{matching,PPI}.  Corrections to
the dimensional reduction are small when the thermal mass and
external momenta are small compared to the temperature, a requirement that
is satisfied when the theory is weakly coupled. Unlike the case of QFT at
finite temperature where the Bose-Einstein distribution effectively
introduces an UV-cutoff, the formulation of a classical field theory is
meaningful only if a $\Lambda$ UV-cutoff is present, otherwise one
encounters the Rayleigh-Jeans divergences. The introduction of
$\Lambda$ means that the parameters of the effective theory must
depend on it in such a way as to cancel the $\Lambda$-dependence of
the regularized loop integrals in the effective theory.
    
Encouraged by the success of DR one could naively expect that the dynamics
of soft modes is described by classical Hamiltonian dynamics i.e.
calculation of time dependent correlation functions implies solving 
classical Hamiltonian equations of motion for arbitrary initial conditions,
over which a Boltzmann weighted averaging with the 3 dimensional DR action is
performed. This was the essence of the proposal of Grigoriev and Rubakov
\cite{GriRu88} for calculating time dependent correlation functions. 
The expectation above works only for scalar $\Phi^4$ theory and for 
Abelian-gauge theory in ($1+1$) dimension. In the former case it was shown 
that the ultraviolet divergences of the time dependent  correlation functions 
can be absorbed by the same mass counterterms as needed in the static 
theory  \cite{Aarts97}.

For non-Abelian gauge theory in higher spatial dimensions the dynamics of
infrared fields is classical, but they are not described by classical
Hamiltonian dynamics. Here the problem of UV divergences is more involved.
Actually, the existence of a classical theory poses two questions: 1. what
is the divergence structure of the hot real time classical Yang-Mills theory
2. if there are well-separated scales in the theory (for example in a gauge
theory) what is the interplay between them.  The answer to the first
question is, that at one loop the linear divergences of the classical theory
are related to the quantum hard thermal loops (HTL) discovered by Braaten
and Pisarski. For example, the divergent part of the classical self-energy
can be obtained as the classical limit of HTL self energy
\cite{bodeker95,ASY95,AaNauta99}. It was realised in \cite{bodeker95} that
there are UV divergences in the classical thermal Yang-Mills theory which are
non-local in space and time (on the lattice they are sensitive to the
lattice geometry), they do not occur in equal time correlation functions and
so they cannot be absorbed by local counterterms. The second question led 
in the end to the serious challenge of constructing an effective theory for
the low-momentum modes by incorporating as accurately as possible the effect
of the quantum but perturbative high-momentum modes. This step has to be
done if the classical theory exhibits strong cut-off dependence, or
equivalently lattice-spacing dependence,
because if this occurs it shows that the classical dynamics is
sensitive to quantum modes with momenta of the order of the cutoff.

In non-Abelian gauge theory at temperatures where the weak coupling limit is
valid, there is a hierarchy of three momentum scales that played an important
role already in the static case \cite{LeBellac,Braaten_Nieto}. There is the
``hard'' scale $T$ of the typical momentum of a particle, there is the
``soft'' scale $gT$ associated with colour-electric screening and finally
there is the ``ultrasoft'' scale $g^2T$ associated with colour-magnetic
screening and where non-perturbative effects appear.

The first numerical simulations were done in a purely Yang-Mills classical
theory with a cutoff $\Lambda$ of order $T$, but it raised the suspicion of
a strong cutoff dependence of the measured quantity, the sphaleron rate.
This indeed was the case as shown for example in \cite{RummMore99}.

One can incorporate the effect of the ``hard'' modes of order $\sim T$ on
the dynamics of softer modes using a classical theory based on the HTL
effective action obtained after integrating out these modes. There
are difficulties with this approach due to the fact that the HTL equations
of motions are non-local in space and time and a local Hamiltonian form
exists only in the continuum.

The classical effective HTL theory leads to UV divergences and has to be
defined with an UV-cutoff $gT\ll\Lambda\ll T$. In order to ensure the UV
insensitivity, the coefficients of the effective theory (effective HTL
Hamiltonian) must be calculated with an infrared cutoff $\Lambda$, so that
the cutoff dependence of the parameters in the effective theory cancels
against the cutoff dependence of the classical thermal loops. The problem
is, that on the lattice it is very difficult to derive the expression of the
$\Lambda$ (IR cutoff)-dependent piece in the HTL Hamiltonian.

In order to reduce the sensitivity to the UV scale $\Lambda$ one can go
further and integrating out the soft degrees of freedom down to scale
$g^2T\ll\Lambda\ll gT$ starting from the classical effective theory for the 
soft fields. This was first performed by B{\"o}deker, who showed that the 
resulting theory at the scale $g^2T$ takes the form of a Boltzmann-Langevin 
equation \cite{Bodeker_log}. It was demonstrated later that this equation is
insensitive to the ultraviolet fluctuations \cite{ASY98}.

\vspace*{0.5cm}
\noindent
{\bf Presentation of the thesis}

This thesis reflects the route followed by its author towards the
derivation of the dynamical equations allowing the study of the real time
onset of the Higgs-effect. Since this study would involve both analytical 
and numerical investigations we have developed and tested both techniques in
simpler models.

First, I have investigated the dynamics of the one-component scalar theory
both in the linear-response approximation and then for large deviations from
equilibrium.

Next, I proceeded to the real time characterisation of the Goldstone effect
when a condensate breaks the O(N) symmetry of the field theory. Here again
both quantum and classical systems were studied.

Finally, I have studied the corrections to the Hard Thermal Loop dynamics in
the Abelian Higgs model, which reflect the presence of the scalar
condensate.

I have five publications related to my thesis:
\begin{itemize}
\item
A. Patk{\'o}s, Zs. Sz{\'e}p, Phys. Lett. {\bf B446} (1999) 272-277
\item
A. Jakov\'ac, A. Patk{\'o}s, P. Petreczky, Zs. Sz{\'e}p, 
Phys. Rev. {\bf D61} (2000) 025006
\item
Sz. Bors\'anyi, A. Patk{\'o}s, Zs. Sz{\'e}p,
Phys. Lett. {\bf B469} (1999) 188-192
\item 
Sz. Bors\'anyi, A. Patk{\'o}s, A. Polonyi, Zs. Sz{\'e}p,
Phys. Rev. {\bf D62} (2000) 085013
\item
Sz. Bors\'anyi, Zs. Sz{\'e}p,
Phys. Lett. {\bf B508} (2001) 109-116
\end{itemize}

The structure of the thesis is as follows. 

In Chapter 2 I briefly review the formalism we use in the following
Chapters: the Green-function approach to transport theory of quantum scalar
fields developed by Danielewicz and Mr{\'o}wczy\'nski and the classical
linear response theory of Jakov\'ac and Buchm\"uller. I show the connection
between the former method with others existing in the literature.

In Chapter 3 I present the work done within the one-component scalar
$\Phi^4$ model. An effective theory of low frequency fluctuations of
selfinteracting scalar fields is constructed in the broken symmetry phase
coupled to the particles of a relativistic scalar gas via their locally
variable mass. The non-local dynamics of Landau damping is investigated in a
kinetic gas model approach.

Next, I present a numerical investigation of the thermalization in the
broken phase of classical $\Phi^4$ theory in $2+1$ dimensions. The $\Phi$
field is coupled with a homogeneous external ``magnetic'' field that induces
a transition from a metastable state to the stable ground state. The
dynamics of the system is described using an effective equation of the order
parameter. This description is consistent with the nucleation theory in a
first order phase transition.
 
Chapter 4 also consists of two Sections. In the first one an effective
theory of the soft modes in the broken phase of $O(N)$ symmetric
$\Phi^4$ model is presented to linear approximation in the background when
the effect of high-frequency fluctuations is taken into account at one-loop
level. The damping of Higgs and Goldstone modes is studied together with
large time asymptotic decay of an arbitrary initial configuration.

In the second Section the real time thermalization and relaxation phenomena
are numerically studied in a $2+1$ dimensional classical $O(2)$ symmetric
scalar theory. The near-equilibrium decay rate of on-shell waves and the
power law governing the large time asymptotics of the off-shell relaxation
is checked to agree with the analytic results based on linear response
theory. The realisation of the Mermin-Wagner theorem is also studied in the
final equilibrium ensemble.

Chapter 5  deals  with the real time dynamics of the Higgs effect. The
effective equations of motion for low-frequency mean gauge fields in the
Abelian Higgs model are investigated in the presence of a scalar condensate,
near the high temperature equilibrium.

More details on the presentation can be found at the beginning of each Section.

%% file: formalism.tex
\chapter{General formalism for non-equilibrium QFT}

We describe in this Chapter different techniques applied in the literature to
the investigation of the real time behaviour of quantum and classical fields. 
We begin with the derivation of the exact Schwinger-Dyson equation in a 
general framework. We introduce approximation schemes and transform
the SD equations into simpler kinetic equations.
Next, we demonstrate to leading order of the perturbative expansion 
with respect to the powers of the
coupling the equivalence of the Schwinger-Dyson equations
with the iterative solution of the Heisenberg equation of motion.
The method of mode-function expansion is also presented.

We finish by presenting the linear response theory of a classical system,
regarded as an approximation of a high-temperature quantum system.
  
\section{Schwinger-Dyson approach}

In order to understand the mechanism by which the system approaches
equilibrium one needs to keep track of real time processes. The most
general framework for dealing with field theory in a non-equilibrium real
time settings is the Schwinger-Keldysh, or close time path (CTP) formalism.
For a review se for example Ref. \cite{PR145} and references therein.


Generally we need to evaluate the expectation value of time ordered products 
of Heisenberg fields for given initial density matrix $\rho(t_0)$:
\be
\avr{T\Phi_H(t_1)\Phi_H(t_2)...\Phi_H(t_n)}:=
\textrm{Tr}\left\{T\rho(t_0) \Phi_H(t_1)\Phi_H(t_2)...\Phi_H(t_n)\right\}.
\ee
These n-point function can be obtained formally with the help of the formula
\be
\avr{T\Phi_H(t_1)\Phi_H(t_2)...\Phi_H(t_n)}=\frac{1}{Z[0]}
\frac{\delta^n Z[j]}{i \delta j(x_1)\cdots i\delta j(x_n)}\Bigg|_{j=0},
\ee
from the generator
\be
Z[j(x)]=\textrm{Tr}\left\{\rho(t_0)T\exp\left[
i\int_{-\infty}^{+\infty}d^4xj(x)\Phi_H(x)
\right]\right\}.
\ee
If we want to evaluate this quantity perturbatively then we keep to the
usual way of doing this by splitting the Hamiltonian into a free and an
interacting part and by switching to the interaction picture. The different
pictures coincide at time $t_0$, which means $\rho_H(t_0)=\rho_I(t_0)$. 
In the interaction picture the density matrix is evolved from its value in 
the remote past by the unitary time evolution operator 
$U(t_b,t_a)=\exp{-i[\int_{t_a}^{t_b}dt H_{int}^I(\Phi_I(t))]}$ 
according to the relation:
\be
\rho(t_0)=U(t_0,-\infty)\rho_I(t=-\infty)U^+(t_0,-\infty).
\label{Eq:rho}
\ee
We denote the density matrix in the remote past by $\rho_0$ that represents
the density matrix of a free system if we switch on the
interactions adiabatically.

Using the relation between the fields in the interaction and Heisenberg
picture, namely $\Phi_H=U^+(t,t_0)\Phi_I(t)U(t,t_0)$, Eq.(\ref{Eq:rho}) 
the cyclic invariance of the trace and the properties of the 
time evolution operator we find:
\bea
\avr{T\Phi_H(t_1)\Phi_H(t_2)...\Phi_H(t_n)}=
\textrm{Tr}\left\{\rho_0TU^+(t_f,-\infty)
\Phi_I(t_1)\Phi_I(t_2)...\Phi_I(t_n) U(t_f,-\infty)\right\}.
\eea
We arrived at this formula by inserting the unit operator 
${\bf 1}=U(t_n,t_f) U(t_f,t_n)$ after $\Phi_I(t_n)$. The time $t_f$ is 
arbitrary but its value must be above the largest time argument 
of the n-point function to be evaluated. Eventually its value will be set to 
$+\infty$.

In order to deal with the $U^+(t_f,-\infty)$ (for a nonequilibrium system no 
state of the past can be related with a state in the future) the 
Schwinger-Keldysh contour of Fig. 1 is introduced, the use of which is 
mandatory in an off-equilibrium situations. As one can see the contour
goes first from $-\infty$ to $t_f=+\infty$ and then back from $t_f=+\infty$
to $-\infty$. The free fields (the fields in the interaction picture) 
appearing in the interaction Hamiltonian are living on the contour.
For physical observables the time values are on the upper branch, yet 
in a self-consistent formalism both upper and lower branches will come into 
play at intermediate steps of the calculation. This fact explains the need 
for four Green functions. 

\begin{figure}
\begin{center}
\includegraphics{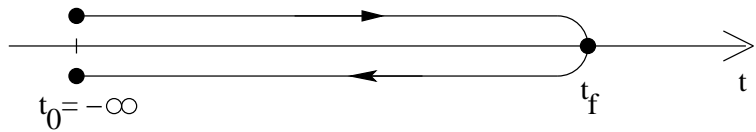}
\end{center}
\vspace*{-0.8cm}
\caption{The Schwinger-Keldysh contour.}
\label{contour}
\end{figure}
The time ordering on the contour is defined through the relation:
\be
\label{Eq:corder}
T_c\Phi_I(x)\Phi_I(y)=\Phi_I(x)\Phi_I(y)\Theta_c(x_0,y_0)+
\Phi_I(y)\Phi_I(x)\Theta_c(y_0,x_0),
\ee
where $\Theta_c(x_0,y_0)=1$, if $y_0$ precedes $x_0$ on the contour,
otherways it is zero. The relation above also means that on the upper
branch we have chronological ordering while on the lower branch we have
anti-chronological ordering.

Summarising we have for the time-ordered product of fields :
\be
\avr{T\Phi_H(t_1)\Phi_H(t_2)...\Phi_H(t_n)}=
\textrm{Tr}\left\{\rho_0T_c\exp{\left[-i\int_c dt H_{int}^I(\Phi_i(t))
\right]}\Phi_I(t_1)\Phi_I(t_2)...\Phi_I(t_n)\right\}, 
\ee
where by definition
\be
\label{Eq:integral}
\int_c dt=\int_{-\infty}^{\infty} dt_1-\int_{-\infty}^{\infty} dt_2.
\ee
Here $1$ stands for the upper and $2$ for the lower branch of the contour.

This once again can be obtained from the generating functional:
\be
Z[j(x)]=\exp\left[-i \int_c dt H_{int}^I\left(
\frac{\delta}{i\delta j(t)}
\right)\right]
T_c\left[\exp{i\int_c d^4xj(x) \Phi_I(x)}\right],
\ee
where $\int_c d^4x=\int_c dt\int d^3x$ and
$j(x)$ is defined also on the contour and has different values on the
upper ($1$) and lower ($2$) branch of it:
\be
\label{Eq:curent}
j(x)=\Theta_c(t_f,x_0) j_1(x)-\Theta_c(x_0,t_f)j_2(x).
\ee 

Using the Wick theorem (see \cite{Itzykson}) which is an operator identity
of the form:
\bea
\label{Eq:Wick}
T_c\left[\exp{i\int_c d^4xj(x) \Phi_I(x)}\right]&=&
\exp\left(
-\int_c d^4x \int_c d^4x'j(x) G_c(x-x')j(x')\right)\nn
&&\textrm{Tr}\left\{
\rho_0:\exp\left[i\int_c d^4x j(x)\Phi_I(x)\right]:\right\}
\eea
one obtains the following expression for the generating functional:
\be
Z[j(x)]=
\exp\left[-i \int_c dt H_{int}^I\left(
\frac{\delta}{i\delta j(t)}
\right)\right]
Z_0[j(x)] N_c[j(x)],
\ee
$Z_0[j(x)]$ being the first factor  and $N_c[j(x)]$ the second factor on the 
r.h.s. of Eq.(\ref{Eq:Wick}).

In the preceding equations, $G_c(x-x')$ is the vacuum or causal propagator
\be
G_c(x-x')=\avr{0|T_c\Phi_I(x)\Phi_I(x')|0}
\ee
and $N_c[j(x)]$ describes the ``initial correlations''.
It's interesting to see that due to the fact that $\Phi_I$ satisfies the
homogeneous equation $(p_0^2-{\bf p}^2-m^2)\Phi_I(p)=0$, the correlations
contribute to the propagator only on the mass-shell, and that each component
of the propagator gets the same additional term. 

At equilibrium and in the non-equilibrium case at least in the kinetic
approximation (see \cite{mabilat}) $N_c[j(x)]$ can be expressed in terms of the corresponding 
thermal propagator $G_T$
\be
N_c[j(x)]^{eq}=\exp\left(-\int_c d^4x \int_c d^4x'j(x) G_T(x-x')j(x')\right).
\ee       
So we can write:
\be
Z_0[j(x)]=\exp\left(-\int_c d^4x_1\int_c d^4x_2j(x_1)
\Delta_c^0(x_1-x_2)j(x_2)\right),
\ee
where $\Delta_c^0(x_1-x_2)$ is defined by 
\be
\label{Eq:Delta_c}  
\Delta_c^0(x_1-x_2)=
\Theta_c(x_0,y_0)\avr{\Phi_I(x)\Phi_I(y)}+\Theta_c(y_0,x_0)
\avr{\Phi_I(y)\Phi_I(x)}
\ee
and consists of two parts: a T-dependent an a T-independent part
$\Delta_c^0(x_1-x_2)=G_c(x_1-x_2)+G_T(x_1-x_2)$.

Using the expression Eq.(\ref{Eq:curent}) for the current and
Eq. (\ref{Eq:integral}) we obtain
\be
\int_c dt\int_c dt'j(x)\Delta_c^0(x_1-x_2)j(x')=
\sum_{r,s=1}^2\int_{-\infty}^{\infty}dt\int_{-\infty}^{\infty}dt'
 j_r(x)\Delta_{rs}^0(x_1-x_2)j_s(x')
\ee
with
\bea
\label{Eq:delta11}
\Delta_{11}^0(x,x')&=&\avr{T\Phi_I(x)\Phi_I(x')}=
\Theta(t-t')\Delta^>_0(x,x')+\Theta(t'-t)\Delta^<_0(x,x')\\
\label{Eq:delta12}
\Delta_{12}^0(x,x')&=&\Delta^<_0(x,x')\\
\label{Eq:delta21}
\Delta_{21}^0(x,x')&=&\Delta^>_0(x,x')\\
\label{Eq:delta22}
\Delta_{22}^0(x,x')&=&\avr{T^*\Phi_I(x)\Phi_I(x')}=
\Theta(t'-t)\Delta^>_0(x,x')+\Theta(t-t')\Delta^<_0(x,x')
\eea
where $T$ is the time ordering, $T^*$ is the anti time-ordering, and the
Wightman functions  $\Delta^<_0(x,x'),$ $\Delta^>_0(x,x')$ stands for two 
distinct ordering of  $\Phi(x)$ and  $\Phi(x')$
\be
\label{Eq:deltasb}
\Delta^<_0(x,x')=\avr{\Phi_I(x') \Phi_I(x)},\qquad
\Delta^>_0(x,x')=\avr{\Phi_I(x) \Phi_I(x')}.
\ee
The index of the propagator denotes the branch from which the first and the
second time argument is taken.

Extending the definition of the time ordering on the contour (\ref{Eq:corder})
to fields in Heisenberg picture we introduce the exact Green-function of 
fields that possess a non-vanishing expectation value  to be treated as
a classical or ``mean'' field, in the form: 
\be
i\Delta(x,y)=\avr{T_c\Phi(x)\Phi(y)}-\avr{\Phi(x)}\avr{\Phi(y)},
\ee
where the $H$ subscript was omitted.  Then, depending from which branch the 
arguments of the propagator are the same relation as in 
Eqs. (\ref{Eq:delta11}) $\dots$ (\ref{Eq:deltasb}) hold.

The two functions defined in (\ref{Eq:deltasb}) are related to each-other
through the commutator of the field:
\be
\Delta^>(x,y)-\Delta^<(x,y)=-i\avr{[\Phi(x),\Phi(y)]}=:\avr{D(x,y)},
\ee
where for the free case 
\be
iD(x,y)=\int\frac{d^3k}{(2\pi)^32\omega_k}
\left({e^{-ik(x-y)}-e^{ik(x-y)}
}\right),
\ee
with $\omega_k=(m^2+\textbfit{k}^2)^{1/2}$ and $k_0=\omega_k$.

The equation of motion for the 2-point function can be obtained in the form:
\bea
\label{Eq:eom1}
{[}\partial^2_x+m^2{]}\Delta(x,y)&=&-\delta^{(4)}(x,y)
+\int_Cd^4x'\Pi(x,x')\Delta(x',y),\\
\label{Eq:eom2}
{[}\partial^2_y+m^2{]}\Delta(x,y)&=&-\delta^{(4)}(x,y)
+\int_Cd^4x'\Delta(x,x')\Pi(x',y),
\eea
where for the case of scalar theory with a given interaction Lagrangian 
term ${\mathcal L}_{int}$
\bea
\label{Eq:Pidef1}
i\int_Cd^4x'\Pi(x,x')\Delta(x',y)=
\avr{T_c\frac{d{\mathcal L}_{int}}{d\Phi(x)}\Phi(y)}
-\avr{\frac{d{\mathcal L}_{int}}{d\Phi(x)}}\avr{\Phi(y)},\\
\label{Eq:Pidef2}
i\int_Cd^4x'\Delta(x,x')\Pi(x',y)=
\avr{T_c\frac{d{\mathcal L}_{int}}{d\Phi(y)}\Phi(x)}
-\avr{\frac{d{\mathcal L}_{int}}{d\Phi(y)}}\avr{\Phi(x)},
\eea
as can be seen doing perturbative expansion in the r.h.s. of 
(\ref{Eq:Pidef1}),(\ref{Eq:Pidef2}).

The effect of the mean field can be separated from the self-energy by
writing:
\be
\Pi(x,y)=\Pi_{MF}(x)\delta_c^{(4)}(x,y)+\Pi^>(x,y)\Theta_c(x_0,y_0)
+\Pi^<(x,y)\Theta_c(y_0,x_0)
\ee
where
\be
\delta_c^{(4)}(x,y)=
\cases{
\delta^{(4)}(x-y)& \textrm{for $x_0$, $y_0$ from the upper branch,}\cr
0& \textrm{ for $x_0$, $y_0$ from different branches,}\cr
-\delta^{(4)}(x-y)& \textrm{for $x_0$, $y_0$ from the lower branch},
}
\ee
and $\Pi_{MF}(x)$ is the mean field self-energy while $\Pi^{>(<)}(x,y)$ 
is the collisional self-energy which provides the collision terms in the
transport equations. In the case of a $\Phi^4$ theory we will see at the end
of Subsection \ref{ss:iteralas} what exactly is $\Pi_{MF}(x)$ and the leading 
expression of $\Pi^{>(<)}(x,y)$ in a perturbative expansion.

Due to the fact that $\Delta_0(x,y)$, the free Green function satisfies the
equation
\be
{[}\partial_x^2+m^2{]}\Delta_0(x,y)=-\delta_c^{(4)}(x,y),
\ee
we can rewrite Eqs. (\ref{Eq:eom1}) and (\ref{Eq:eom2})  in the form of a 
Dyson--Schwinger equation:
\be
\Delta=\Delta_0-\Delta_0\Pi\Delta.
\ee

Using an equation analogous to (\ref{Eq:Delta_c}) corresponding to Heisenberg 
fields:
\be
\Delta(x,y)=\Theta_c(x_0,y_0)\Delta^>(x,y)+\Theta_c(y_0,x_0)\Delta^<(x,y),
\ee
Eq. (\ref{Eq:integral}), 
and the fact that $x_0, y_0$ are on the upper branch of the contour but $x'_0$ 
could be as well on the upper as on the lower branch, we obtain:
\bea
\nonumber
\label{Eq:Delta<>1}
{[}\partial_x^2+m^2-\Pi_{MF}(x){]}\Delta^{>(<)}(x,y)&=&
\int_{-\infty}^{y_0}d^4x'\Pi^{>(<)}(x,x')
[\Delta^<(x',y)-\Delta^>(x',y)]+\\
&&\int_{-\infty}^{x_0}d^4x'
[\Pi^>(x,x')-\Pi^<(x,x')]\Delta^{>(<)}(x',y),\nn
{[}\partial_y^2+m^2-\Pi_{MF}(y){]}\Delta^{>(<)}(x,y)&=&
\int_{-\infty}^{y_0}d^4x'\Delta^{>(<)}(x,x')
[\Pi^<(x',y)-\Pi^>(x',y)]+\\
&&\int_{-\infty}^{x_0}d^4x'
[\Delta^>(x,x')-\Delta^<(x,x')]\Pi^{>(<)}(x',y).
\eea
Introducing the retarded and advanced quantities
\bea
\label{Eq:retar}
\Pi_R(x,y)&=&(\Pi^>(x,y)-\Pi^<(x,y))\Theta(x_0-y_0),\\
\Delta_R(x,y)&=&(\Delta^>(x,y)-\Delta^<(x,y))\Theta(x_0-y_0),\\
\Pi_A(x,y)&=&(\Pi^<(x,y)-\Pi^>(x,y))\Theta(y_0-x_0),\\
\Delta_A(x,y)&=&(\Delta^<(x,y)-\Delta^>(x,y))\Theta(y_0-x_0)
\label{Eq:avanj}
\eea
we obtain the following equations of motions:
\bea
\nonumber
\left[\partial_x^2+m^2-\Pi_{MF}(x)\right]\Delta^{>(<)}(x,y)&=&
\int d^4x'\left[\Pi^{>(<)}(x,x')\Delta_A(x',y)\right.\\
&&\qquad\qquad\qquad\qquad
\left.+\Pi_R(x,x')\Delta^{>(<)}(x',y)\right]\label{mrow4.12}
\label{Eq:Deltax<>}\\
\nonumber
\left[\partial_y^2+m^2-\Pi_{MF}(x)\right]\Delta^{>(<)}(x,y)&=&
\int d^4x'\left[\Delta^{>(<)}(x,x')\Pi_A(x',y)\right.\\
&&\qquad\qquad\qquad\qquad
\left.+\Delta_R(x,x')\Pi^{>(<)}(x',y)\right]\label{mrow4.14}
\label{Eq:Deltay<>}
\eea
where the time integration runs no more on the contour $C$ but from $-\infty$ 
to $+\infty$. 

Using the defining equation for the advanced and retarded two-point function 
and the commutation relation of the fields we can express the derivative of
$\Delta_{R(A)}$ function with $\Delta^{>(<)}$, 
for example
\be
\partial_x^2\Delta_R(x,y)=\left(\partial_x^2\Delta^>(x,y)-
\partial_x^2\Delta^<(x,y)\right)\Theta(x_0-y_0)-\delta^{(4)}(x_0-y_0).
\ee
Then using Eqs. (\ref{Eq:Deltax<>}) and (\ref{Eq:Deltay<>}) 
an equation of motion for the advanced and retarded two-point
function can be derived
\bea
\label{Eq:DeltaxRA}
\hspace*{-0.5cm}
\left[\partial_x^2+m^2-\Pi_{MF}(x)\right]\Delta_{R(A)}(x,y)&=& 
-\delta^{(4)}(x-y)+
\int d^4x' \Pi_{R(A)}(x,x')\Delta_{R(A)}(x',y),\\
\label{Eq:DeltayRA}
\hspace*{-0.5cm}
\left[\partial_y^2+m^2-\Pi_{MF}(y)\right]\Delta_{R(A)}(x,y)&=& 
-\delta^{(4)}(x-y)+
\int d^4x' \Delta_{R(A)}(x,x')\Pi_{R(A)}(x',y).
\eea

Equations (\ref{Eq:Deltax<>}), (\ref{Eq:Deltay<>}), (\ref{Eq:DeltaxRA}),
(\ref{Eq:DeltayRA}) are exact and they are equivalent to the field
equations of motion and they are known as the Kadanoff--Baym equations 
first derived in the framework of non-relativistic many-body theory
\cite{K-B}.

In order to solve these equations one has to implement an approximation
scheme. Usually this is done in a way to transform the general equations
into much simpler kinetic equations for the two-point function. The 
approximation scheme involves gradient expansion, quasi-particle approximation 
and the perturbative expansion of the self-energy. We present it
in Subsection \ref{ss:kinapprox}, following the treatment of 
Ref. \cite{BJ_PR}. We can consider 
the classical equation of motion for the two-point function
as an approximation of the quantum Kadanoff--Baym equations.
 Using linear
response theory this is presented in Section \ref{s:linresp}.

Another approximation is to perform a perturbative expansion of the self-energy. 
By doing this $\Pi^{<(>)}(x,y)$ can be expressed with the help of the
two-point function $\Delta^{>(<)}$ closing in this way the equation of
motion for $\Delta^{>(<)}$ (note that $\Delta^>(x,y)=\Delta^<(y,x)$). 

\subsection{\label{ss:kinapprox}Kinetic equations for the two-point
functions}

In thermal equilibrium the two-point functions depend only on the relative
coordinates $u_\mu=x_\mu-y_\mu$ and are strongly peaked around $u_\mu=0$ with a 
range of variation determined by the wavelength of a particle with a typical 
momentum $k$ in the plasma. At high temperature $k\sim T$ and the particle's 
thermal wavelength is $\lambda_T=1/k\sim T$.

Out of equilibrium the two-point function depends on both coordinates, and
if the deviations from equilibrium are slowly varying in space and time it
pays out to introduce two new variables: the relative coordinate 
$u_\mu=x_\mu-y_\mu$ and the center-of-mass coordinate 
$X_\mu=\frac{x_\mu+y_\mu}{2}$ 
in terms of which the kinetic equation can be derived. For slowly varying 
disturbances with $\lambda\gg\lambda_T$, one can 
expect that the $u$ dependence of the two-point function is close to that of 
equilibrium and $\partial_u\sim k\sim T$ and $\partial_X\sim 1/\lambda\ll T$.

For a function $f(X,u)$ that varies slowly with $X$ and is strongly peaked for 
$u\approx 0$ one can approximate $f(X+u,u)$ as
\be
f(X+u,u)\approx f(X,u)+u^\mu\frac{\partial f(X,u)}{\partial X^\mu},
\ee
i.e. only terms involving at most one derivative $\partial_X$ are kept. This
approximation is referred to as the gradient expansion.
 
Then one defines the Wigner transform
\be
\Delta^{>}(x,y)\stackrel{\textrm{W.tr}}{\longrightarrow}
\Delta^{>}(X,p):=\int du^4 e^{ip\cdot u}\Delta^{>}
(X+\frac{u}{2},X-\frac{u}{2})
\ee
with its inverse
\be
\Delta^{>}(x,y)=\int\frac{d^4 p}{(2\pi)^4}e^{-ip(x-y)}
\Delta^{>}(\frac{x+y}{2},p).
\ee
The properties of the Wigner transformation, useful for our calculation, are
the following:
\bea
&&\hspace*{-1.5cm}
{\partial}_x^2\Delta^{>}(x,y)
\stackrel{\textrm{W.tr.}}{\longrightarrow}
\left(\tort{1}{2}\partial_X-ip\right)^2\Delta^{>}(X,p),\\
&&\hspace*{-1.5cm}
{\partial}_y^2\Delta^{>}(x,y)
\stackrel{\textrm{W.tr}}{\longrightarrow}
\left(\tort{1}{2}\partial_X+ip\right)^2\Delta^{>}(X,p),\\
\label{Eq:exactW1}
&&\hspace*{-1.5cm}
a(x)f(x,y)
\stackrel{\textrm{W.tr.}}{\longrightarrow}
\int \tort{d^4 q}{(2\pi)^4}
a(q)f\left(X,p-\tort{q}{2}\right)e^{-iqX}\approx a(X)f(X,p)-i
\frac{1}{2}\frac{\partial a(X)}{\partial X^\mu}
\frac{\partial f(X,p)}{\partial p_\mu},\\
\label{Eq:exactW2}
&&\hspace*{-1.5cm}
a(y)f(x,y)
\stackrel{\textrm{W.tr.}}{\longrightarrow}
\int \tort{d^4 q}{(2\pi)^4}
a(q)f\left(X,p+\tort{q}{2}\right)e^{-iqX}\approx a(X)f(X,p)+i
\frac{1}{2}\frac{\partial a(X)}{\partial X^\mu}
\frac{\partial f(X,p)}{\partial p_\mu},\\
&&\hspace*{-1.5cm}
\int d^4z f(x,z)g(z,y)
\stackrel{\textrm{W.tr.}}{\longrightarrow}
f(X,p)g(X,p)+\frac{i}{2}\left\{f,g\right\}_{P.B.}+\dots\,.
\eea
In the last formula (for the derivation of it we refer to
Section 4.2. of Ref. \cite{Danielewicz})
$\left\{f,g\right\}_{P.B.}$ denotes a Poisson bracket
\be
\left\{f,g\right\}_{P.B.}:=\partial_p f\cdot\partial_X g-\partial_X f
\cdot\partial_p g,
\ee
and the dots mean that the gradient expansion was used. The first expressions 
on the r.h.s of Eqs. (\ref{Eq:exactW1}), (\ref{Eq:exactW2}) represent the 
exact result of the Wigner transform, while the second expressions the result 
from the gradient expansion. 

Of course, for other two-point functions and the self energies analogous
relations hold.

The kinetic equations for the two-point function $\Delta^{>(<)}$
is obtained by subtracting the Wigner transform of Eqs. 
(\ref{Eq:Deltax<>}) and (\ref{Eq:Deltay<>}) while for the two-point function 
$\Delta_{R(A)}$  by adding the Wigner transform of Eqs. (\ref{Eq:DeltaxRA}) 
and (\ref{Eq:DeltayRA}). 

Using the properties 
\bea
\Delta_R(X,p)-\Delta_A(X,p)=\Delta^>(X,p)-\Delta^<(X,p),\\
\Pi_R(X,p)-\Pi_A(X,p)=\Pi^>(X,p)-\Pi^<(X,p),
\eea
that follows from the definition (\ref{Eq:retar})$\dots$(\ref{Eq:avanj}) and 
\be
\Delta_R(X,p)+\Delta_A(X,p)=2\Re\Delta_R(X,p),\quad\qquad\quad 
\Pi_R(X,p)+\Pi_A(X,p)=2\Re\Pi_R(X,p),
\ee
one obtains the following kinetic equation
\be
2\left(p^\mu+\frac{\partial\Re\Pi}{\partial p_\mu}\right)
\frac{\Delta^{>(<)}}{\partial X^\mu}-
\frac{\partial\Pi_{MF}}{\partial X_\mu}
\frac{\partial\Delta^{>(<)}}{\partial p^\mu}
-\left\{\Re\Delta_R,\Pi^{>(<)}\right\}_{P.B.}
=i\left(\Pi^<\Delta^>-\Delta^<\Pi^>\right),
\ee
where $\Re\Pi(X,p):=\Pi_{MF}+\Re\Pi_R(X,p).$ 

This equation represents the quantum generalisation of the Boltzmann
equation. The terms of this equation have the following physical
interpretation (see Ref. \cite{BJ_PR}). The Wigner function $\Delta$ plays
the role of the phase-space distribution function. The drift term on the
l.h.s. generalises the kinetic drift term by including two types of self
energy corrections. One is due to the real part of the self-energy that acts
as an effective potential whose space-time derivative provides the ``force''
$(\partial_X^\mu \Re\Pi)(\partial_\mu^p\Delta^<).$  
The second correction is due to the momentum dependence of the self-energy
that modifies the velocity of the particles. The terms on the r.h.s.
describe collisions, while the term containing the Poisson bracket accounts
for the off-equilibrium shape of the spectral density as one can see in Eq.
(\ref{Eq:spectral}).

For the spectral density $\rho(X,p)=i\Delta^>(X,p)-i\Delta^<(X,p)$ we obtain
\be
2\left(p^\mu+\frac{\partial\Re\Pi}{\partial p_\mu}\right)
\frac{\partial\rho}{\partial X^\mu}-
\frac{\partial\Pi_{MF}}{\partial X_\mu}
\frac{\partial\rho}{\partial p^\mu}=
i\left\{\Re\Delta_R,\Pi^>-\Pi^<\right\}_{P.B.}
\label{Eq:spectral}
\ee

One arrives at the quasi-particle approximation when the term on the r.h.s. is
neglected. In this case the spectral density solution of Eq.
(\ref{Eq:spectral})  read as
\be
\rho(X,p)=2\pi\epsilon(p_0)\delta\left(p^2-m^2+\Re\Pi(X,p)
\right).
\label{Eq:rho_hivat}
\ee

In a recent work by Aarts and Berges \cite{Aarts_Berges} non-equilibrium time 
evolution of the spectral function was studied numerically. There was observed 
that the spectral function develops a non-vanishing width. This  
points towards the need of relaxing the assumption of zero-width 
of the spectral
density if a consistent quantum-Boltzmann equation is aimed.

In the mean field approximation, that is when the collision term is neglected
altogether we obtain:
\bea
&&
\left[p\cdot\partial_X+\frac{1}{2}(\partial_X^\mu\Pi_{MF})
\partial_\mu^p\right]\Delta^<(X,p)=0,\\
&&\rho(X,p)=2\pi\epsilon(p_0)\delta\left(p^2-m^2+\Re\Pi_{MF}(X,p)
\right).
\eea

Neglecting the collision term corresponds in the Dyson-Schwinger formalism
to the truncation of the hierarchy of the n-point function at the level of
the 2-point function, all higher n-point functions are neglected.

We will use this approximation in the thesis.

\subsection{\label{ss:iteralas}Iterative solution of the Heisenberg equations}
We can arrive at the approximated Kadanoff--Baym equations with a 
different method too.

We start with the equation of motion of the full quantum operator 
\be
(\partial^2+m^2)\hat\Phi ({\bf x},t)+{\lambda\over 6}\hat\Phi^3({\bf
x},t)=0,
\label{full_eq}
\ee
then we split the field into the sum of its average $\phi(x)$ and the 
quantum fluctuation around it $\varphi(x)$
\be
\hat\Phi ({\bf x},t)=\phi({\bf x},t)+\varphi ({\bf x},t),
\qquad \phi({\bf x},t)=\langle\hat\Phi ({\bf x},t)\rangle.
\label{qm_cl_szetvalasztas}
\ee
Upon averaging the Heisenberg equation one obtains the equation of the
classical field $\phi(x)$, and subtracting this from the original
Heisenberg equation we obtain the  equation from the fluctuation $\varphi(x)$.
These two equations reads as:
\bea
\hspace*{-1.5cm}
&&\biggl(\partial^2+m^2+\frac{\lambda}{6}\phi^2(x)+\frac{\lambda}{2}
\avr{\varphi^2(x)}\biggr)\phi(x)=-\frac{\lambda}{6}\avr{\varphi^3(x)},\\
\hspace*{-1.5cm}
&&\biggl(\partial^2+m^2+\frac{\lambda}{2}\phi^2(x)+ \frac{\lambda}{2}
\avr{\varphi^2(x)}\biggr)\varphi(x)=
-\frac{\lambda}{2}\phi(x)\left[\varphi^2(x)\right]-\frac{\lambda}{6}
\left[\varphi^3(x)\right]+\frac{\lambda}{2}\avr{\varphi^2(x)}\varphi(x),
\eea
where $[\cdot]=\cdot-\avr{\cdot}$. We have added to both sides of the equation 
of $\varphi(x)$ the term $\frac{\lambda}{2}\avr{\varphi^2(x)}\varphi(x)$.
This will mean a resummation and provides the mean-field part of the 
self-energy, the $\Pi_{MF}(x)$ when considering the equation of motion for
$\avr{\varphi(x)\varphi(y)}$.

Introducing the notations:
\be
j(x):=-\frac{1}{2}\phi(x)\left[\varphi^2(x)\right]-\frac{1}{6}
\left[\varphi^3(x)\right]+\frac{1}{2}\avr{\varphi^2(x)}\varphi(x),
\ee
we follow Ref. \cite{Boyanovsky98} and solve the equation 
\be
\label{unalom}
\biggl(\partial^2+m^2+\frac{\lambda}{2}\phi^2(x)+
\frac{\lambda}{2}\avr{\varphi^2(x)}\biggr)\varphi(x)=\lambda j(x),
\ee
in the form
\be
\label{Eq:kifejtes}
\varphi(x)=\varphi_0(x)-\lambda\int dx' G_R(x,x')j(x'),
\ee 
with $\varphi_0$ the solution of the homogeneous equation 
\be
\label{Eq:homogeneous}
\biggl(\partial^2+m^2+\frac{\lambda}{2}\phi^2(x)+ \frac{\lambda}{2}
\avr{\varphi^2(x)}\biggr)\varphi_0(x)=0,
\ee
and $G_R$ the retarded Green-function associated with this equation
$$\biggl(\partial^2+m^2+\frac{\lambda}{2}\phi^2(x)+
\frac{\lambda}{2}\avr{\varphi^2(x)}\biggr)G_R(x,y)=-\delta^{(4)}(x,y).$$

Using the homogeneous equation and the commutator relation of the free field
$\varphi_0(x)$ it's easy to see that:
\bea
\label{Eq:G_Rdef}
G_R(x,y)=(G^>(x,y)-G^<(x,y))\Theta(x_0-y_0), 
&\qquad iG^>(x,y)=\avr{\varphi_0(x)\varphi_0(y)},&\\
&\qquad iG^<(x,y)=\avr{\varphi_0(y)\varphi_0(x)}.&
\eea   

For the equation of $\phi$ we calculate to linear order in $\lambda$ 
the quantities $\avr{\varphi^2(x)}$ and  $\avr{\varphi^3(x)}$ using equation 
(\ref{Eq:kifejtes}) and obtain:
\bea
\avr{\varphi^2(x)}&=&\avr{\varphi_0^2(x)}+{\mathcal O}(\lambda^2),\\\nonumber
\avr{\varphi^3(x)}&=&\avr{\varphi_0^3(x)}+
\lambda\int d^4x' \phi(x') G_R(x,x')
\left[G^>(x',x)^2\right.\\
&&\qquad\qquad\qquad\left.+G^>(x,x')G^>(x',x)+G^>(x,x')^2 \right]+
{\mathcal O}(\lambda^2).
\label{Eq:Phikob}
\eea
  
The equation for $\avr{\varphi(x)\varphi(y)}$ is obtained after multiplying
Eq. (\ref{unalom}) with $\varphi(y)$ and taking the average. We need to
evaluate 
\be
\lambda\avr{j(x)\varphi(y)}=
-\frac{\lambda}{2}\phi(x)\avr{\varphi^2(x)\varphi(y)}-\frac{\lambda}{6}
\avr{\varphi^3(x)\varphi(y)}+\frac{\lambda}{2}\avr{\varphi^2(x)}
\avr{\varphi(x)\varphi(y)},
\ee
making use of Eq. (\ref{Eq:kifejtes}). The
${\mathcal O}(\lambda)$ contribution is zero. 

Taking into account that
$\avr{j[\varphi_0(x')]\varphi_0(x)}=0$ up to ${\mathcal O}(\lambda^2)$ we
obtain:
\bea
\nonumber
\lambda\avr{j(x)\varphi(y)}&=&-\lambda^2\int dy' G_R(y,y')
\avr{j[\varphi_0(x)]j[\varphi_0(y')]}\nn
&&+\frac{\lambda^2}{2}\phi(x)\int dx' G_R(x,x') \left(
\avr{j[\varphi_0(x')]\varphi_0(x)\varphi_0(y)}+
\avr{\varphi_0(x)j[\varphi_0(x')]\varphi_0(y)}
\right)\nn
&&+\frac{\lambda^2}{6}\phi(x)\int dx' G_R(x,x') \biggl(
\avr{j[\varphi_0(x')]\varphi^2_0(x)\varphi_0(y)}+
\avr{\varphi_0(x)j[\varphi_0(x')]\varphi_0(x)\varphi_0(y)}\\
&&\qquad\qquad\qquad
+\avr{\varphi_0^2(x)j[\varphi_0(x')]\varphi_0(y)} \biggl).
\eea
The averages on the r.h.s. can be evaluated with the help of the Wick theorem 
leading to:
\bea
\nonumber
&&\hspace*{-0,75cm}-\frac{\lambda^2}{2}\int dx'\phi(x)\phi(x')
\biggl[ G_R(y,x')G^>(x,x')^2+G_R(x,x')G^>(x',y)
\biggl(G^>(x',x)+G^>(x,x')\biggr)\biggr]\nn
&&\hspace*{-0,75cm}-\frac{\lambda^2}{6}\int dx'\left[G_R(y,x')G^>(x,x')^3+ 
G_R(x,x')G^>(x',y) \left(G^>(x',x)^2\right.\right.\\\nonumber
&&\left.\left.\qquad\qquad\qquad\qquad\qquad\qquad
+G^>(x,x')G^>(x',x)+G^>(x,x')^2\right)
\right].
\eea 
We observe the same structure in the last line of the above equation as the
one appearing in the Eq. (\ref{Eq:Phikob}). Using the definition 
(\ref{Eq:G_Rdef}), the property $G^>(x,y)=G^<(y,x)$ and 
some algebraic manipulation with $G^{>(<)}$ one arrives to express all terms
with some power of $G^{>(<)}$.
One obtain:
\bea
\nonumber
\lambda\avr{j(x)\varphi(y)}&=&
-\frac{\lambda^2}{2}\phi(x)\int_{-\infty}^{y_0}d^4x'
G^>(x,x')^2\biggl[G^<(x',y)-G^>(x',y)\biggr]\phi(x')\nn
&&-\frac{\lambda^2}{2}\phi(x)\int_{-\infty}^{x_0}d^4x'
\left[G^>(x,x')^2-G^<(x,x')^2\right]G^>(x',y)\phi(x')\nn
&&-\frac{\lambda^2}{6}\int_{-\infty}^{y_0}d^4x'
G^>(x,x')^3\biggl[G^<(x',y)-G^>(x',y)\biggr]\\
&&-\frac{\lambda^2}{6}\int_{-\infty}^{x_0}d^4x'
\left[G^>(x,x')^3-G^<(x,x')^3\right]G^>(x',y)+{\mathcal O}(\lambda^3).
\eea
Now we can replace the Green's function 
$iG^>(x,y)=\avr{\varphi_0(x)\varphi_0(y)}$ with the exact one since the
corrections will be ${\mathcal O}(\lambda^2)$ implying ${\mathcal
O}(\lambda^4)$ corrections in the equation above.

\begin{figure}
\begin{center}
\includegraphics[width=8cm]{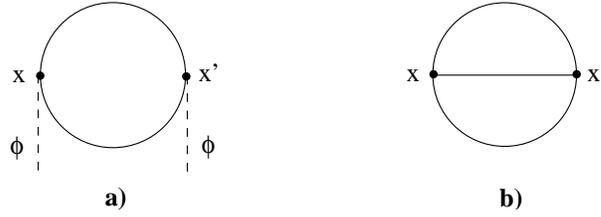}
\end{center}
\vspace*{-0.8cm}
\caption{One-loop a) and two-loop b) self energies.}
\label{Hurkak}
\end{figure}

Introducing for the one and two-loop self-energy components 
that emerges in a perturbation expansion in the broken phase the
notation
\be
\Pi_{2a}^{>(<)}(x,x'):=
-\frac{\lambda^2}{2}\phi(x) G^{>(<)}(x,x')^2\phi(x'),\qquad
\Pi_{2b}^{>(<)}(x,x'):=-\frac{\lambda^2}{6}G^{>(<)}(x,x')^3,
\ee
up to order $\lambda^2$ the self-energy takes the form:
\be
\Pi_{2}^{>(<)}(x,x')=\Pi_{2a}^{>(<)}(x,x')+\Pi_{2b}^{>(<)}(x,x').
\ee
The components of the self-energy labelled with $a$ and $b$ correspond to 
the graphs of Fig. \ref{Hurkak}\,. 

With the help of the above relations we can write down to order $\lambda^2$
the following set of equations:
\bea
&&\hspace*{-1.75cm}
\biggl(\partial^2+m^2+\frac{\lambda}{6}\phi^2+\frac{\lambda}{2}
\avr{\varphi^2}\biggr)\phi(x)+\frac{\lambda}{6}\phi^3(x)+
\frac{\lambda^2}{6}\int_{-\infty}^{x_0}dx'
\left(\Pi^>_{2b}(x,x')-\Pi^<_{2b}(x,x')\right)\phi(x')=0,\nn
\nonumber
&&\hspace*{-1.75cm}
\biggl(\partial^2+m^2+\frac{\lambda}{2}\phi^2(x)+\frac{\lambda}{2}
\avr{\varphi^2(x)} \biggr)G^>(x,y)=
\int_{-\infty}^{y_0}d^4x'\Pi^>_{2}(x,x')
\biggl[G^<(x',y)-G^>(x',y)\biggr]\\
&&\qquad\qquad\qquad\qquad\qquad\qquad\qquad+\int_{-\infty}^{x_0}d^4x'
\left[\Pi^>_2(x,x')-\Pi^<_2(x,x')\right]G^>(x',y).
\label{Eq:Glambda2}
\eea

We see that Eq. (\ref{Eq:Glambda2}) for the  $\avr{\varphi(x)\varphi(y)}$ has 
the same structure as Eq. (\ref{Eq:Delta<>1}).
Actually they coincide if $\Delta^{>(<)}$ is identify with $G^{>(<)}$
and $-\Pi_{MF}(x)$ with 
$\frac{\lambda}{2}\phi^2(x)+\frac{\lambda}{2} \avr{\varphi^2(x)}$ 
and if one expand the self-energy in Eq. (\ref{Eq:Delta<>1}) up to $\lambda^2$.

If one express the self-energy $\Pi^{>(<)}$ in term of $G^>$ alone the
equation for $G^>$ turns into an equation with the same structural form as
the one obtained from the three-loop $2PI$ 
effective action by Berges and Cox \cite{cox}, although the relation between
the two approximation schemes is not yet clear.
 With the help of this equation 
these authors has shown in a numerical simulation in $1+1$ dimensions 
the thermalization of a quantum system.

\rt{
What we do when calculating the effects of the quantum fluctuations on the
equation of motion for the classical field
corresponds to put $j(x)=0$ (if the ominous term is not
added), not considering at all the
non-linear interaction of the quantum fluctuation. That means taking 
$\varphi(x)=\varphi_0(x).$
And we studied the effect of the classical background up to linear order on
the equation of the quantum fluctuation.
\be
\left(\partial^2+m^2+\frac{\lambda}{2}\bar\phi\right)\varphi_0(x)=
-\lambda\bar\phi\phi(x)\varphi_0(x)
\ee
whose iterative solution is
\be
\varphi_0^{(1)}(x)=\varphi_0^{(0)}(x)+\lambda\bar\phi\int dz
G_\varphi^R(x-z)\phi(z)\varphi_0^{(0)}(z) \quad\textrm{with}\quad
G_\varphi^R(p)=\frac{1}{p^2-m^2-\lambda\bar\phi^2/2}
\ee
Then the leading effect of the quantum fluctuations in the evolution of the
background is given by $\avr{\varphi_0(x)\varphi_0(x)}^{(1)}$:
\bea
\left(\partial^2+\frac{\lambda}{2}\bar\phi^2+\frac{\lambda}{2}
\avr{\varphi_0^{(0)}(x)\varphi_0^{(0)}(x)}\right)\phi(x)&=&
-\frac{\lambda}{2}\bar\phi\avr{\varphi_0(x)\varphi_0(x)}^{(1)},\\
m^2\bar\phi+\frac{\lambda}{6}\bar\phi^3+\frac{\lambda}{2}\bar\phi
\avr{\varphi_0^{(0)}(x)\varphi_0^{(0)}(x)}&=&0 
\eea
This is evaluated in the following chapters.}

\rt{
\bea 
\nonumber
\avr{\varphi_0(x)\varphi_0(y)}^{(1)}\Big|_{x=y}&=&
\lambda\bar\phi\int dz \phi(z)\left[
G_\varphi^R(x-z)\avr{\varphi_0^{(0)}(z)\varphi_0^{(0)}(y)}+
G_\varphi^R(y-z)\avr{\varphi_0^{(0)}(x)\varphi_0^{(0)}(z)}\right]_{x=y}\\
&=&\frac{\lambda}{2}\bar\phi\int_Q e^{-iQ\cdot X}\phi(Q)\int_P
\frac{\Delta(P-Q/2)-\Delta(P+Q/2)}{Q\cdot P}\\
&=&
-\frac{\lambda}{2}\bar\phi\int_Q e^{-iQ\cdot X}\phi(Q)\int_P
\frac{Q_\epsilon}{Q\cdot P}\frac{\partial\Delta_0(P)}{\partial P_\epsilon}
\eea
}

\subsection{Mode-function expansion}

The method of mode-function expansion was used in the investigation of the
evolution towards equilibrium of non-equilibrium states produced during
preheating. This was done in the Hartee-Fock approximation in which the
dynamics is described in terms of a mean-field and
the fluctuations around it are characterised by the
two-point function only. The two-point function can be constructed in terms
of a complete set of mode functions that, if the mean field is homogeneous,
can be chosen as plane waves labeled by the wave vector $\k$. During
preheating, only mode functions in a narrow $\k$-band will be excited by the
time-dependent homogeneous mean field, via the phenomenon called parametric
resonance. In an early stage, the so called linear stage, when the occupation
number of different modes, and consequently the fluctuation is not too large
the dynamics of the system is well approximated by the equations of motion
linearised with respect to the fluctuation. In the second stage, called
back-reaction the fluctuation grows exponentially and one has to switch to a
fully non-linear dynamics (see for example \cite{Khlebnikov}).

The system eventually becomes stationary but due to the lack of scattering
the method of the mode function expansion in a homogeneous background cannot
lead to thermalization. This is reflected by the shape of the distribution
function that does not approach the Bose-Einstein distribution in case of
bosons showing instead resonant peaks \cite{BoySalgado}.

Recently it has been proved that considering inhomogeneous background for
which the quantum modes that are represented by the mode function can
scatter via their back-reaction on the background, the time evolution of
particle distribution functions display for a finite time interval
approximate thermalization \cite{salle1}.

The mode-function expansion on an inhomogeneous background is needed when
the density matrix describing the initial state of the system does not
commute with the translation. In this case the quantum average will depend
on the space-time position and the mean field is inhomogeneous.

The method of mode-function expansion represents a way of solving the
Eq. (\ref{Eq:homogeneous}) in which the expression
$\langle\varphi_0^2({\bf x},t)\rangle$ need to be renormalised.
We do so by subtracting its value at
$t=0$. Introducing
\be
m^2=m^2_R(T)+{\lambda\over 2}\langle\hat\varphi^2_0({\bf x},0)\rangle
\ee
we obtain
\be
\Bigl[\partial^2 +m^2_R(T)+{\lambda\over 2}\phi^2({\bf x},t)+
{\lambda\over 2}[\langle\varphi^2_0({\bf x},t)\rangle -
\langle\varphi^2_0({\bf x},0)\rangle ]\Bigr ]\varphi_0({\bf x},t)=0.
\label{quant_eq_0}
\ee
The solution of this equation can be constructed by expanding
$\varphi_0({\bf x},t)$ into a series with respect of an appropriately chosen
complete set of time dependent functions, named mode-function with constant 
operatorial coefficients, labelled by "${\bf k}$":
\bea
\label{Eq:modeexpantion}
\varphi_0({\bf x},t)&=&\sum_{ k}\biggl (a_k\psi_k({\bf x},t)+a^+_k
\psi^*_k({\bf x},t)\biggr),\\
\pi ({\bf x},t)&=&\sum_{k}\biggl (a_k\dot\psi_k({\bf x},t)+
a^+_k\dot\psi^*_k({\bf x},t)\biggr ),
\eea
where the creation and annihilation operators obey (by definition)
the usual commutation rules:
\be
[ a_k, a^+_{k'}]=\delta_{k,k'}.
\ee
The standard canonical commutator between $\varphi ({\bf x},t)$ and
$\pi ({\bf x},t)$ requires the relation:
\be
\sum_{k}\biggl [\psi_k({\bf x},0)\dot\psi^*_k({\bf y},0)-\psi^*_k({\bf x},0)
\dot\psi_k({\bf y},0)\biggr ]=i\delta ({\bf x}-{\bf y}),
\ee
which is fulfilled by the choice:
\be
\psi_k({\bf x},0)={1\over \sqrt{2\omega_kV}}e^{-i{\bf kx}},\quad
\dot\psi_k({\bf x},0)=-i\sqrt{\omega_k\over 2V}e^{-i{\bf kx}}.
\label{init_cond}
\ee
Here we choose
\be
\omega_k^2={\bf k}^2+m_R^2(T)+{\bar\phi^2\over 2},\qquad \bar\Phi=
{1\over V}\int d^dx\phi({\bf x},0).
\ee
The $T$ appearing in this relation is the initial temperature, fixed
as part of the initial physical data of the system.

Then an initial state corresponding to thermal equilibrium would
be characterised by the number operator expectation value
\be
\langle a^+_ka_k\rangle_0={1\over e^{\beta \omega_k}-1}\equiv n(\omega_k),
\ee
and all the higher moments expressed in terms of this.
Using the thermal density matrix for the initial moment we find
for the relevant field expectation values
\be
\langle\varphi_0^2({\bf x},t)\rangle =\sum_k|\psi_k({\bf x},t)|^2
(2n(\omega_k)+1),\quad \langle\varphi_0^3({\bf x},t)\rangle =0.
\ee
In the evolution of the order parameter, the contribution from
the second moment at $t=0$ is absorbed partly
into the renormalised squared mass, partly it contributes the temperature
dependence of the  mass. (This is the same renormalization we applied in 
the equation of $\varphi_0$).
In this way we find the effective equation for the
inhomogeneous mean field:
\be
\biggl[\partial^2 +m^2_R(T)+{\lambda\over 2}\sum_k\left(|\psi_k({\bf x},t)|^2-
|\psi_k({\bf x},0)|^2\right)(2n(\omega_k)+1)\biggr]\phi({\bf x},t)+
{\lambda\over 6}\phi^3({\bf x},t)=h.
\label{ren_class_eq}
\ee

The equations for the $k$-modes are found from Eq.(\ref{quant_eq_0}) 
when multiplying it by $a^{\dagger}_p$ and taking its
expectation value:
\be
\biggl [\partial^2 +m_R^2(T)+{\lambda\over 2}\phi^2({\bf x},t)+
\lambda\sum_k \left(|\psi_k({\bf x},t)|^2-|\psi_k({\bf x},0)|^2\right)
(n(\omega_k)+{1\over 2})\biggr]\psi_p({\bf x},t)=0.
\label{ren_mode_eq}
\ee
This equation is the generalisation of the equations proposed in
\cite{boyan95,mottola97} for homogeneous classical background.
The method of mode function expansion in an inhomogeneous background was
applied in numerical simulations in Ref. \cite{salle1}.






\section{\label{s:linresp}Classical linear response theory}

We present now the classical linear response theory of Jakov\'ac and
Buchm\"uller following closely their paper \cite{jako97}. Using this method, 
we can calculate in linear approximation the damping rate of a field.
When we compare in Subsection \ref{ss:dispersion} the analytical 
quantum damping rate with the classical one or the numerical value for the
damping rate with the analytical one as we do in in Section \ref{s:MW}
we need the result for the classical damping rate in the broken phase. Here 
we present for simplicity the derivation of it for the symmetric phase. 
The extension of this method to a classical $O(N)$ scalar theory in the broken 
phase can be found in Appendix \ref{calss}.
 
In a classical theory at finite temperature $T=\beta^{-1}$ defined with the 
Hamiltonian
\be
\label{Eq:class_H}
H = \int d^3x \biggl( {1\over 2} \pi^2 + {1\over 2} (\vec{\partial}\phi)^2
    + {1\over 2} m^2_{cl}(\Lambda) \phi^2 + {1\over 4!}\lambda_{cl} \phi^4 
    + j \phi \biggr)\, ,
\ee
that contains $j(t,\vec{x})$ as an external source, the classical real time
$n$-point functions are defined in the following way
\be\label{Eq:npoint}
\avr{\phi(t_1,\vec{x}_1)\dots\phi(t_2,\vec{x}_2)}_{cl} = {1\over Z} 
\int {\cal D}\pi {\cal D}\phi e^{-\beta H(\pi,\phi)}
\phi(t_1,\vec{x}_1)\dots\phi(t_2,\vec{x}_2)\, ,
\ee
with the partition function
\be\label{eq:Zcl}
Z = \int  {\cal D}\pi {\cal D}\phi e^{-H(\pi,\phi)}\, . 
\ee
In Eq. \ref{Eq:npoint} $\phi(t,\vec{x})$ is the solution of the equations of 
motion
\bea
\dot{\phi}(t,\vec{x}) &=& \pi(t,\vec{x})\, ,\nn
\dot{\pi}(t,\vec{x}) &=& \Delta\phi(t,\vec{x}) - m^2_{cl}(\Lambda) \phi(t,\vec{x})
             - {\lambda_{cl}\over 3!} \phi^3(t,\vec{x}) - j(t,\vec{x})\,  ,
\eea
with the initial conditions
\be
\phi(t_0,\vec{x}) = \phi(\vec{x})\, , \quad
\pi(t_0,\vec{x}) = \pi(\vec{x})\, .
\ee
In Eqs.~(\ref{Eq:npoint}) and (\ref{eq:Zcl}) the integration is performed at 
$t=t_0$ over the space of initial conditions 
$\phi\equiv\phi(\vec{x}),\pi\equiv\pi(\vec{x})$ (averaging over the initial
conditions). 
\rt{From computational point of view the need of relatively large
ensemble averages could be regarded as a drawback of a real time simulation
based on classical dynamics.  As was noted in Ref. \cite{}, in the case of an
$1+1$ dimensional simulation, introducing quantum degrees of freedom through
the method of mode function only apparently make the simulation more
CPU-demanding since the necessity of large ensemble average considerably
diminishes the computational advantage of using classical dynamics.}

As it is well know from the time prior to the discovery of the quantum
theory, a classical theory is not well defined in the UV. As was shown in
\cite{aarts98,Aarts97} for a scalar classical theory of constant temperature
$T$ both for equal time ( or equivalently for static quantities) and unequal
time correlation functions merely a mass renormalization suffices to render
the theory finite.  In Eq. (\ref{Eq:class_H}) the mass term can be split
into the sum of two terms: the renormalized classical mass and the
divergent part of the mass, to be treated as a counterterm:
$m_{cl}^2(\Lambda)=m_R^2+\delta m^2(\Lambda)$. Here $\Lambda$ is an UV
cutoff introduced to regularize the theory, and $\delta m^2(\Lambda)$ is the
sum of two terms, $\delta m_1(\Lambda)$ and $\delta m_2(\Lambda)$,
corresponding to the one and two loop divergent diagrams, as we will see at
the end of this Subsection.

If the classical theory is considered to be the high-temperature
approximation of the quantum theory, then the renormalized classical mass
should be related to the tree-level mass of the quantum theory. Such a
relation between masses can be derived using the dimensional reduction by
matching the result of some quantities calculated in both  quantum and
classical theories. For the exact form of the matching relations (the
masses, the coupling constant and the expressions of $\delta m_1(\Lambda)$
and $\delta m_2(\Lambda)$) I refer to Refs. \cite{matching,aarts98}.

The $j(t,\vec{x})$-dependent solution of the equations of motion 
constructed with the retarded Green's function
\be\label{dcl}      
D_R(t,\vec{x}) = \int {d^3q\over (2\pi)^3} e^{i\vec{q}\vec{x}}
  \Theta(t){1\over \omega_q}\sin(\omega_q t)\, .
\ee
satisfies the integral equation
($t,t'> t_0, x\equiv (t,\vec{x})$),
\be\label{Eq:phj}
\phi(x;j) = \phi_0(x) + \int d^4x' D_R(x-x')\biggl(
  {\lambda_{cl}\over 3!} \phi^3(x';j) + j(x')\biggr)\, .
\ee
$\phi_0$ is the solution of the free (homogeneous) equations of motion which
satisfies the boundary condition. 

As one can see from equation (\ref{Eq:phj}) the classical solution 
depends on the external source in powers of which it can be expressed. 
The idea of linear response theory is to keep only the
term linear in $j(x)$. The linear connection between the solution of the
equation of motion and the external source is given by the retarded response 
function defined through 
\be\label{hrd}
H_R(x-x') = \left.{\delta \phi(x;j)\over \delta j(x')}\right|_{j=0}
\ee
in the form
\be\label{phi}
\phi(x,j) = \int d^4x' j(x') H_R(x-x')\, .
\ee

\begin{figure}[htbp]
  \begin{center}  
  \leavevmode     
     \epsfig{file=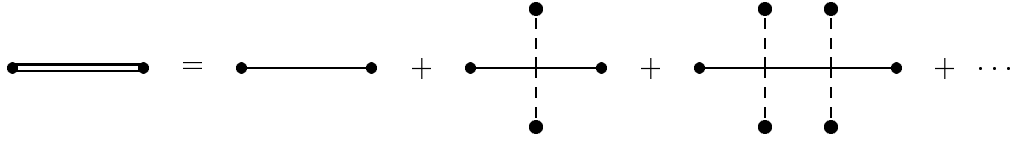}
  \end{center}    
  \caption{The response function $H_R$ expanded in powers of the
     classical field. Full lines denote retarded Green's functions
     $G_R$. The figure is from Ref. \cite{jako97}.}       
\label{Fig:HR}
\end{figure}

The retarded response function depends on the initial conditions 
$\phi(\vec{x})$ and $\pi(\vec{x})$. Substituting
Eqs.~(\ref{Eq:phj}) into (\ref{hrd}) one easily gets the integral equation 
which determines $H_R$,
\be\label{hr}
H_R(x-x') = D_R(x-x') + {1\over 2}\lambda_{cl} \int d^4y D_R(x-y) 
                        \phi^2(y;0)H_R(y-x')\, .
\ee
This equation can be solved iteratively, for example up to order
$\lambda^2_{cl}$
its solution reads:
\bea
\label{Eq:HR}
H_R^{(2)}(x-x')&=&D_R(x-x')+\frac{1}{2}\lambda_{cl}\int d^4y D_R(x-y) 
\phi^2(y;0)D_R(y-x')+\\\nonumber&&
\frac{1}{4}\lambda_{cl}^2\int d^4y\int d^4z D_R(x-y)
\phi^2(y;0)D_R(y-z)\phi^2(z;0)D_R(z-x')+{\mathcal O}(\lambda_{cl}^3),
\eea
and is depicted in Fig. \ref{Fig:HR}.

In order to obtain the finite-temperature retarded response function  
denoted with $D_R^{cl}$, the ensemble average of Eq. (\ref{Eq:phj}) with respect to 
the initial conditions is taken. This yields
\be
\avr{\phi(x)}_{cl} = \int d^4x' j(x') D_R^{cl}(x-x')\, , 
\ee
where 
\be
D_R^{cl}(x-x') = \avr{H_R(x-x')}_{cl} =
 {1\over Z} \int {\cal D}\phi {\cal D}\pi  e^{-\beta H} H_R(x-x')\, .
\ee
$D_R^{cl}$ can be evaluated by first expanding the solution (\ref{hr}) for 
$H_R$ in powers of $\lambda_{cl}$ (cf.~Fig.~\ref{Fig:HR}) and then performing the 
thermal average for each term.

This method means that we have to evaluate
\be
D_R^{cl}(x-x')=\avr{e^{-\beta H_{int}} H_R(x-x')}_{0,\,c},
\ee
using the iterative solution of Eq. (\ref{hr}). In the equation above
$H_{int}$ is the interaction Hamiltonian, the subscript $0$ means thermal
average taken with the free Hamiltonian, and $c$ means that we have to take
only connected graphs.

If the classical theory of interest is only a tool to evaluate the leading
contribution of a quantum theory then, the full result of the classical
theory is irrelevant and one can make the approximation
\be
\avr{e^{-\beta H_{int}} H_R(x-x')}_{0,\,c}\approx\avr{ H_R(x-x')}_{0\,,c}
\ee
because only to leading order in the coupling the results of the classical
and quantum theory are related in the high-temperature limit.

So, what is required is the evaluation of the thermal n-point functions
of the type of (\ref{Eq:npoint}), namely
\be
\langle \phi^2(x_1) \ldots \phi^2(x_{2n})\rangle_{cl}^0 = 2^n
S^2(x_1-x_2) \ldots S^2(x_{2n-1}-x_{2n}) + \mbox{permutations}\, .
\ee
where $(t-t',\vec{x}-\vec{x}')$ is the free two-point function that reads 
\cite{Aarts97,gpa}
\be\label{scl}
S(t-t',\vec{x}-\vec{x}') = \langle \phi(x) \phi(x') \rangle_{cl}^0 \nn
= T \int {d^3q\over (2\pi)^3} e^{i\vec{k}(\vec{x}-\vec{x}')}
    {1\over \omega_k^2} \cos(\omega_k(t-t'))\, .
\ee

From Fig.\ref{Fig:HR} it can be easily seen that the finite temperature 
response function obtained after thermal  averaging with $H_0$ satisfies a 
Dyson-Schwinger equation
\be
\bar{D}_R^{cl}(x-x') = D_R(x-x') + \int d^4y\int d^4y' 
        D_R(x-y)\bar{\Pi}_{cl}(y-y') \bar{D}_R^{cl}(y'-x')\, . 
\ee

$\Pi_cl(x)$ is the classical self energy and by looking at the Eq. (\ref{Eq:HR})
we see that up to ${\cal O}(\lambda_{cl}^2)$ in perturbation theory it is given by:
\bea
\Pi_{cl}^{(1)}(y-y')&=&\avr{\phi^2(y,0)}=
\frac{\lambda_{cl}}{2}S(y-y')\delta(y-y'),\\
\Pi_{cl}^{(2)}(y-y')&=&\avr{\phi^2(y,0) \phi^2(y',0)}=
\frac{\lambda_{cl}^2}{2}S(y-y')D_R(y-y')S(y-y').
\eea

\begin{figure}[htbp]
  \begin{center} 
  \leavevmode 
     \epsfig{file=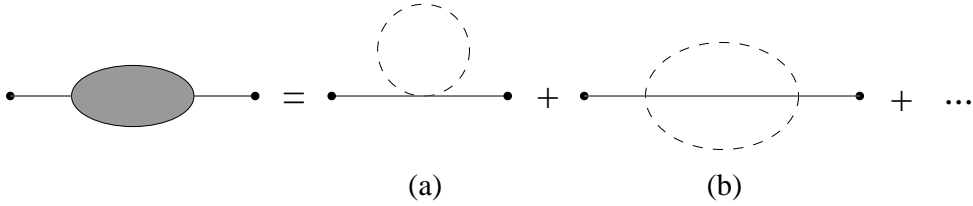, width=13cm} 
  \end{center}
  \caption{Perturbative expansion for the self-energy
     $\Pi$. Dashed lines denote thermal two-point functions $S$.
   The figure is from Ref. \cite{jako97}.}
\label{Fig:Pi}
\end{figure}

These two terms are shown also in Fig. (\ref{Fig:Pi}).

The contribution 
${\cal O}(\lambda_{cl})$ (Fig.~\ref{Fig:Pi}a) is linearly divergent and the
divergence can be removed  by a mass renormalization yielding the counter 
term \cite{Aarts97}
\be
\delta m^2_1 = {1\over 2}\lambda_{cl} S(0,\vec{0}) = {1\over 2}\lambda_{cl} T 
               \int {d^3q\over (2\pi)^3} {1\over \omega_q^2}\, .
\ee
The contribution ${\cal O}(\lambda_{cl}^2)$ (Fig.~\ref{Fig:Pi}b) is logarithmically 
divergent and can be rendered finite by a further mass renormalization
\be
\delta m^2_2 = -{1\over 6}\lambda_{cl}^2\beta\int d^3x S^3(0,\vec{x})
  = -{4\over 3}\lambda_{cl}^2 T^2 \int d\Phi(\vec{0})
{1\over \omega_1\omega_2\omega_3}\, ,
\ee
where $d\Phi(\p)$ is the integration measure,
\be
d\Phi(\p)=\frac{d^3 q_1}{(2\pi)^3 2\omega_1} 
\frac{d^3 q_2}{(2\pi)^3 2\omega_2} \frac{d^3 q_3}{(2\pi)^3 2\omega_3} (2\pi)^3
\delta(\p-\q_1-\q_2-\q_3).
\ee
Contributions of higher order in $\lambda_{cl}$ are all finite.

The damping rate, i.e., the imaginary part of the self-energy is finite. 
Using Eqs.~(\ref{dcl}) and (\ref{scl}) 
one easily obtains for the self-energy to order $\lambda_{cl}^2$,
\bea\label{se}
\bar{\Pi}(\omega,\vec{p})_{cl} &=& \int dt d^3x\ e^{i\omega t} e^{-i\vec{p}\vec{x}}\
                      \bar{\Pi}(t,\vec{x})_{cl} \nn
&=& {1\over 2} \lambda_{cl}^2 T^2 \sum_{\eta_1,\eta_2,\eta_3} \eta_1
\int d\Phi(\vec{p})\ {1\over \omega_2\omega_3}
   {1\over \omega + \eta_1\omega_1 + \eta_2\omega_2 + \eta_3\omega_3 +
i\epsilon}\, ,
\eea 
where $\eta_i = \pm 1$, $i=1,\ldots,3$.

From Eq.~(\ref{se}) one obtains for the imaginary part of the self-energy
($\omega > 0$),
\be\label{gamc}
\bar{\Gamma}(\omega,\vec{p})_{cl}={\pi\over 2}\lambda_{cl}^2 T^2 \int d\Phi(\vec{p}) 
  {1\over \omega_1\omega_2\omega_3} 
  \left[ \omega_1\delta(\omega - \omega_1 - \omega_2 - \omega_3) +
         \omega\delta(\omega + \omega_1 - \omega_2 -\omega_3) \right]\, .
\ee

When comparing the classical damping rate $\bar{\Gamma}_{cl}$ with the
damping rate $\Gamma$ of the full quantum theory it turns out that at very 
high temperatures, i.e., $T\gg m$, where the Bose-Einstein
distribution function becomes
\be
f(\omega) \simeq {T\over \omega} \gg 1\, 
\ee 
to leading order in the coupling the damping rates are given by the classical 
theory.

%% file: scalar1l.tex
\chapter{One-component $\Phi^4$ theory}

The purpose of this chapter is to develop and test the methods of investigation 
of the evolution of out-of-equilibrium systems.

We start in the first Section with a new approach, a kinetic theory, aimed
to substitute the quantum degrees of freedom, coupled to a classical field
theory. This approach can be regarded to be at an intermediate level between
a fully quantum treatment and its high-temperature classical approximation.
This method of treating the short wavelength quantum modes has been used in
gauge theories in numerical works aiming to measure the hot sphaleron rate,
when it was realised that quantum degrees of freedom play an important role
in the determination of this rate.

We present the connection of this way of treating the quantum modes with the
Schwinger-Dyson approach in linear response approximation.

In the second Section we present a numerical investigation of the non-linear
dynamics in a classical $\Phi^4$ theory far from equilibrium.  The system
starts from a metastable state and its evolution is followed as it undergoes
a first order phase-transition to the true vacuum.

\section{Classical kinetic theory of Landau Damping}

Relevance of combined classical field theory and kinetic theory to the high
temperature linear response theory of quantum gauge fields has been
demonstrated by several authors \cite{liu94,iancu94}. These authors have
shown that for high enough temperature the leading non-equilibrium transport
effects of large wave number $|{\bf k}|\gg T$ fluctuations can be reproduced
by a kinetic theory of the corresponding response functions. This approach
has been applied to the study of the QCD plasma \cite{heinz83,heinz86}
and proved successful in reproducing the contribution of hard loops to the
Green's  functions of low $|{\bf k}|\ll T$ modes in Abelian and non-Abelian 
gauge theories.

For gauge fields the central object which allows explicit construction of the 
kinetic Boltzmann equation is the Wong-equation \cite{wong70}, since it 
determines the force exerted on non-Abelian charged particles by external 
fields. 
\rt{The calculation of the appropriate damping coefficients proceeds by 
relating the external fields with the corresponding induced currents, via the 
polarisation tensor. The imaginary part of the polarisation tensor determines 
the damping rate of the field.}

For selfinteracting scalar fields the notion of the external field and of
the force felt by the modes with low wave number is not so obvious. The
physical idea based on the renormalisation group is to represent the high 
wave number fluctuations in form of a gas of massive particles, and study the 
effect of the gas on the dynamics of the slow modes. The derivation of the 
Boltzmann equation for the gas in the low wave number background field was 
presented for the scalar fields in \cite{brandt98} using a non-relativistic 
potential picture. The authors have found the effective action which accounts 
for the loop contribution of high-$k$ fluctuations to the Green's functions of 
the low-$k$ modes. 

The authors of Ref.\cite{brandt98} have assumed positive curvature for the
potential felt by the effective particles at the origin. The above mentioned
gauge investigations have also assumed no scalar background fields and the
restoration of all symmetries. However, certain scalar models with specific
internal symmetries are known where symmetry is not restored up to high
temperatures \cite{weinb74,skage85} and the question of the derivation of a
kinetic equation is still of interest. If the scalar theory is thought to be
part of a non-Abelian gauge+Higgs system, the non-perturbative treatment
\cite{buchm95} of the coupled one-point and two-point Schwinger-Dyson equations
leads to a high temperature solution with non-vanishing vacuum expectation
value for the scalar field. Therefore it is of interest to explore the
physical consequences of the presence of a constant non-vanishing background
from the point of view of the kinetic behaviour of the effective gas.

Motivated by the success of the classical kinetic theory in describing the
effect of the hard quantum modes on the dynamics of the soft (low-k) ones
in the context of a gauge theory in the symmetric phase, we present in
this Section a fully relativistic Lagrangian formalism for the kinetic theory 
of scalar fields for the case of non-zero spatial field average.

The construction starts by proposing a Lagrangian for the effective
particles, representing the high-k modes on the low-k fluctuations.  
\rt{A detailed motivation for this Lagrangian is provided by recalling results 
of earlier investigations.} An advantage of this proposition is that it leads 
to the induced source density of the low-k fluctuations directly, without any
reference to the quantum theory. Evidence for the correctness of the
effective Lagrangian can be presented by comparing its consequences with the
results of the corresponding quantum calculations.  
\rt{In Subsection
\ref{ss:fast} we give a physically transparent simplified rederivation of
the force exerted on an effective particle representing the fast modes in
the background of low-frequency field configurations in the non-relativistic
approximation and making use of the usual wave-particle correspondence.}
In Subsection \ref{ss:kinetic} a fully relativistic rederivation of the
results of Ref. \cite{brandt98} is presented for the kinetic theory of the
fields with large spatial momenta. The emerging correction to the
Bose-Einstein phase-space distribution function of high-k modes is used in
Subsection \ref{ss:slowwaveeq} to compute corrections to the classical field 
equations of the low momentum modes. Finally the Landau damping coefficient
for the off-shell scalar fluctuations is computed. This effect is present
only in the broken symmetry phase of the scalar theory. 
\rt{
Its classical value
is compared with the result of a 1-loop quantum calculation \cite{boya96},
establishing in this way the domain of validity of the proposed classical
treatment.}

\rt{\subsection{Dynamics of the fast modes of the self-interacting scalar field
\label{ss:fast}}}

\subsection{Kinetic theory of particles with background field dependent
mass\label{ss:kinetic}}

The effective gas of high-frequency fluctuations is kicked out of thermal
equilibrium if an inhomogeneous low frequency background fluctuation is
present. This state of the gas induces a source term into the wave equation of 
the low-k modes. In a scalar $\Phi^4$-theory a unified description of the two 
effects can be attempted if in addition to the action $S_{cl}$ describing the 
classical low-frequency dynamics an appropriate action can be introduced for 
a gas particle (that represents the high frequency quantum modes) coupled to
the low-frequency field along its trajectory $\xi_\mu (\tau )$.

We split the $\Phi$-field into two terms: 
\bea
\Phi(x)&=&
\int_{0}^{\Lambda}\frac{d^4k}{(2\pi)^4}\Phi(k)e^{-ik\cdot x}+
\int_{\Lambda}^{\infty}\frac{d^4k}{(2\pi)^4}\Phi(k)e^{-ik\cdot x}
= \phi(x,\Lambda)+\varphi(x,\Lambda), 
\eea
and assuming the spontaneous generation of a non-zero constant average 
background field $\bar\phi$ for the slow modes, we shift
the right hand side of the above expression by $\bar\phi$
\be
\Phi (x)=\bar\phi +\phi (x)+\varphi (x).
\ee
The wave equation for $\phi(x)$ will be modified by  an induced source
density, determined as a statistical average over the phase space
distribution of the gas particles representing the fast modes. 

In this Subsection we give the expression of the force felt by the effective
particles representing the high-$k$ modes in a background. For this we write
down the approximate expression of the Lagrangian quadratic in the fast mode
$\varphi$
\bea
\label{eq:ujL}
\mathcal{L}&=&\frac{1}{2}\partial_\mu\phi\partial^\mu\phi -
\frac{1}{2}m^2(\bar\phi+\phi)^2+
\frac{1}{2}\partial_\mu\varphi\partial^\mu\varphi -
\frac{1}{2}m^2\varphi^2
-\frac{\lambda}{4!}(\bar\phi+\phi)^4-
\frac{\lambda}{4}(\bar\phi+\phi)^2\varphi^2.
\eea
We stress that the $m^2$ is of negative sign.  Leaving out fro Eq.
(\ref{eq:ujL}) the selfinteraction of the $\varphi$ is at the basis of
replacing the parts containing $\varphi$ fields in the Lagrangian with a
gas of massive particle interacting only with the background.

\rt{
From the quadratic Lagrangian the linearised equation of motion for $\varphi$ 
looks like:
\bea
\label{eq:wave}
&\left(\Box-m^2_{eff}[\bar\phi]-\lambda V[\bar\phi,\phi(x)]\right)
\varphi(x)=0,&\\
&\hspace*{-7.3cm}\textrm{where} \hspace{5.5cm}
\label {eq:mass}
m_{eff}[\bar\phi]=\sqrt{m^2+\frac{\lambda}{2}\bar\phi^2},&\\
&\hspace*{-6.67cm}\textrm{and} \hspace{5.5cm} 
V[\bar\phi,\phi(x)]=\bar\phi\phi(x)+\frac{1}{2}\phi^2(x).&
\label{eq:Vdef}
\eea
Using the local dispersion equation suggested by (\ref{eq:wave})
\be
E^2-{\bf P}^2=m^2+\frac{\lambda}{2}\bar\phi^2+\lambda V[\bar\phi,\phi(x)]
\ee
we might say on the basis of wave-particle duality, that the fast modes
can be represented by particles of local $\phi$-dependent mass
\bea
M_{loc}[\bar\phi,\phi(x)]&=&\sqrt{m^2+\frac{\lambda}{2}(\bar\phi+\phi(x))^2},
\eea
while on the average the true mass of the particles is given by 
Eq. (\ref{eq:mass}).
From the non-relativistic approximation to the energy 
\be                                                     
\label{eq:E}                                             
E=\left({\bf P}^2+m_{eff}^2[\bar\phi]+\lambda V[\bar\phi,\phi(x)]\right)^{1/2}
\approx                                                
m_{eff}[\bar\phi]+\frac{\lambda V[\bar\phi,\phi(x)]}{2m_{eff}[\bar\phi]}
+\frac{{\bf P}^2}{2m_{eff}[\bar\phi]},
\ee
one can read off the 
$\phi$-dependent potential felt by the particle locally:
\be
U[\bar\phi,\phi(x)]:=\frac{\lambda}{2m_{eff}[\bar\phi]}V[\bar\phi,\phi(x)].
\ee
From here the non-relativistic expression of the force exerted on the 
particle by the inhomogeneous $\phi$-configurations is
\be
{\bf F}=-grad_\phi U[\bar\phi,\phi(x)]=-\frac{\lambda}{2m_{eff}[\bar\phi]}
\nabla V[\bar\phi,\phi(x)].
\ee
It is easy to suggest intuitively a relativistic generalisation 
of this expression.  Instead, in the next Subsection we shall restart with the 
effective action based on the local mass and find the relativistic expression 
for $dp/d\tau$ applying  the rules of analytical mechanics. 
}

\rt{In view of the preceding subsection} We propose an effective action
describing the system consisting of a massive gas of particles
interacting with a scalar background:
\be
S_{eff}=S_{cl}[\phi]- \sum_i\int\limits_{\sigma_1}^{\sigma_2}
M_{loc}[\bar\phi,\phi(\xi_i(\tau))]d\tau,
\label{effact}
\ee
where $S_{cl}[\phi]$ is simply the action of a $\phi^4$-theory understood
with a cut-off $\Lambda$. The second term is the action for a gas of
relativistic scalar particles superimposed on the background. $\bar\phi$ is
the constant scalar condensate, and $\phi$ is the long wavelength fluctuation 
amplitude on the top of the average field. The local mass is determined by the 
field values along the path $\xi_i(\tau)$ of the i-th particle:
\be
M_{loc}^2[\bar\phi,\phi(\xi_i(\tau))]=m^2+{\lambda\over 2}
(\bar\phi+\phi(\xi_i(\tau)))^2,
\label{Eq:locmass}
\ee
where $m^2$ is the squared mass parameter of the theory, which can be 
negative.

Variation of the second piece of the action in Eq. (\ref{effact}) with respect 
to the field variable $\phi(x)$ should yield the induced source term to the 
wave equation of the low frequency modes. This induced source can be expressed 
with the distribution of the gas particles. The distribution can be derived 
from the solution of a Boltzmann equation describing the gas in the background
field $\phi$.  \rt{This will be done in the following Subsection.}

Variation of Eq. (\ref{effact}) with respect to the particle trajectory provides
the EOM of the particles together with the expression of the force exerted
on the particles by the external field $\phi$. This information is used in
deriving the collisionless Boltzmann-equation.

Concentrating only on one particle, with the usual variational procedure 
\cite{kalman61} one can derive the canonical momentum $P_\mu$ from the 
translational invariance of the action, and the variable rest mass $M$ from 
the invariance of the action with respect to the variation of the proper time 
$\tau$:
\bea
P_{\mu}&=&\left[\frac{\partial L}{\partial\dot \xi_{\nu}}\dot
\xi_{\nu} - L\right]\dot\xi_{\mu}-\frac{\partial L}{\partial\dot
\xi_{\mu}},\\
M & = & \frac{\partial L}{\partial\dot\xi_{\mu}}\dot\xi_{\mu}-L,
\eea
The kinetic momentum is defined through the relation
$p_\mu=M\dot\xi_\mu$
and the force is found from the relation $F_\mu=\frac{dp_\mu}{d\tau}.$
 
The equation of motion of the particle is:
\be
\frac{d}{d\tau}\left[\frac{\partial L}{\partial\dot \xi_{\nu}}-\left( 
\frac{\partial L}{\partial\dot \xi_{\mu}}\dot \xi_{\mu} - L\right)\dot
\xi_{\nu}
\right] =\frac{\partial L}{\partial \xi_{\nu}},
\ee
where $L$ is the sum of the free particle Lagrangian and the interaction 
Lagrangian.

In our case the kinetic momentum coincides with the canonical one and the
EOM can be written in the form:
\be
\label{eq:force}
M_{loc}(\xi)\frac{dp_\mu}{d\tau}=\frac{1}{2}\frac{\partial M_{loc}^2(\xi)}
{\partial\xi_{\mu}}.
\ee
\rt{
The local mass (\ref{Eq:locmass}) is different from the ``average'' or 
``effective'' mass given in Eq. (\ref{eq:mass}).
In Refs. \cite{pisut,plbscikk} following \cite{brandt98} we introduced 
the kinetic  momentum of the particle through the relation
$p_\mu=m_{eff}{d\xi_\mu\over d\tau},$ with the ``effective'' mass given in Eq.
(\ref{eq:mass}). The relation between the canonical and kinetic momentum was
\be
P_\mu=M_{loc}\dot\xi(\tau )={M_{loc}\over m_{eff}}p_\mu,
\label{eq:canmom}
\ee
and then the equation of motion of a particle of mass $m_{eff}$ was obtained
in the form: 
\be
m_{eff}\frac{dp_\mu}{d\tau}=\frac{\lambda}{2}\,
\frac{m_{eff}^2}{M_{loc}^2}\,
\biggl[\partial_\mu V[\bar\phi,\phi(x)] -
{p_\mu\over m_{eff}^2}(p\cdot\partial)V[\bar\phi,\phi(x)]\biggr].
\ee
Apart from a mass renormalization effect due to the second term in the r.h.s., 
we obtained the same result for the induced current as that presented below.
}
The collisionless Boltzmann-equation arises by the application of the
chain rule to the proper-time derivative of the one-particle phase space 
distribution:
\be
\label{Eq:colisionless}
{dn(x,p)\over d\tau}=\biggl[
p\cdot\partial+M_{loc}[\bar\phi,\phi(x)]{dp_\mu\over d\tau}
{\partial\over \partial p_\mu}\biggr]n(x,p)=0.
\ee
In the next Subsection we will show that a statistical average introduce 
a distribution function for relativistic particles.
Appendix \ref{app:partition} shows how the distribution function is introduced 
in the context of relativistic kinetic theory.

Exploiting Eq.(\ref{eq:force}) one writes  the collisionless
Boltzmann-equation for the gas of these particles in the form:
\be
(p\cdot\partial )n(x,p)+
\frac{\lambda}{2}(\bar\phi+\phi(x))\partial_\mu\phi(x)
{\partial n(x,p)\over\partial p_\mu}=0.
\ee

Here one follows the usual perturbative procedure to find corrections 
in the weak coupling limit $\lambda\ll 1$ to  the equilibrium Bose-Einstein 
distribution  by writing the distribution function as a series in powers of 
$\lambda$:
\be
n(x,p)=\tilde n_0(p)+\lambda n_1(\phi (x), p)+\dots, \quad 
\tilde n_0(p)={1\over e^{\beta p_0}-1}\Theta(p_0-\Lambda).
\ee
We have slightly modified the equilibrium Bose-Einstein distribution
function that depends only on the $p_0$, in order to account for the fact,
that the particles represents the degrees of freedom above the
$\Lambda$-scale. 

The formal solution for $n_1(x,p):=n_1(\phi (x), p)$ which is linear in 
$\phi(x)$ is easily found to be
\be
n_1(x,p)=-{\bar\phi\over 2}{1\over (p\cdot\partial)}\partial_\mu \phi(x)
{d\tilde n_0(p)\over dp_\mu}.
\label{Eq:f1}
\ee
In the next subsection we shall calculate corrections to the linearised
field equation for the long wavelength fluctuations. These corrections arise
from the statistically averaged source term emerging through the
$\phi$-dependence of the particle action (\ref{effact}). This completes the
selfconsistent construction of the wave equation for fields interacting with
a relativistic gas of particles.

\subsection{The wave equation of the slow modes\label{ss:slowwaveeq}}

The variation with respect to $\phi$ of Eq. (\ref{effact})
provides the wave equation of the low-$k$ modes
\be
\left(\partial^2+m^2+\frac{\lambda}{6}\bar\phi\right)\phi(x)+
m^2\bar\phi+\frac{\lambda}{6}\bar\phi^3
+\frac{\lambda}{2}\sum_i\int
\frac{\bar\phi+\phi(x)}{M_{loc}[\bar\phi,\phi(x)]}
\delta^{(4)}(x-\xi_i(\tau))d\tau=0.
\label{modwave}
\ee
It contains the source term arising from the gas particles through
the $\phi$-de\-pen\-dence of their local mass. We arrived at this expression
keeping only terms linear in $\phi$.

In order to express the
source term with the help of the one-particle distribution function $n(x,p)$,
we perform a statistical average over the full momentum space and in 
a  small volume around the point $x$. This procedure introduces $n(x,p)$ 
through the relation:
\be
m_{eff}\int {d^3p\over (2\pi )^3p_0}n(x,p)=
\avr{\int d\tau\sum_i\delta^{(4)}({\bf x}-{\bf\xi}_i(\tau ))}.
\ee
Here we introduced $m_{eff}^2=m^2+\frac{\lambda}{2}\bar\phi^2$, the
``average'' or ``effective''  mass square of a particle. With the help 
of it we approximate the dispersion relation of a particle as 
$p_0^2\approx m_{eff}^2+{\bf p}^2$,  
that is the effect of the long wavelength fluctuation is taken into account
only in the force felt by the particle. Also, we keep only the 
${\mathcal O}(\lambda^{0})$ term in the expression $\frac{m_{eff}}{M_{loc}}$ 
arising in Eq. (\ref{modwave}) when the averaging procedure mentioned before
is performed. These two approximations has no effect on the non-local part
of the induced current.  We get the following ``macroscopic'' source
density:
\be
j_{av}(x)={\lambda\over 2}\int {d^3p\over (2\pi )^3p_0}
(\bar\phi+\phi(x))n(x,p).
\label{source0}
\ee

The induced linear response to $\phi (x)$ is found using the perturbative
solution of the Boltz\-mann-equation presented in the previous subsection 
by retaining terms which produce linear functional dependence of $j_{av}$ on 
$\phi$. One finds for the leading terms of the induced source the
following expression 
\be
\label{Eq:jav}
j_{av}(x)=\frac{\lambda}{2}\bar\phi\int{d^3p\over (2\pi )^3p_0}\tilde n_0(p_0)
+j_{av}^{(0)}(x)+j_{av}^{(1)}(x), 
\ee
with
\bea
j_{av}^{(0)}(x)={\lambda\over 2}\phi (x)\int{d^3p\over (2\pi )^3p_0}
\tilde n_0(p_0),\qquad\textrm{and}\qquad
j_{av}^{(1)}(x)={\lambda^2\bar\phi\over 2}\int {d^3p\over (2\pi )^3p_0}
n_1(\phi (x), p).
\eea

The first term in Eq. (\ref{Eq:jav}) which is proportional to the average
field $\bar\phi$ enters in the equation that determines the expectation
value of the $\bar\phi$
\be
m^2+\frac{\lambda}{6}\bar\phi^2+\frac{\lambda}{2}\int{d^3p\over
(2\pi )^3p_0}\tilde n_0(p_0)=0.
\label{Eq:ezisbarphi}
\ee

\rt{In the expression of $j_{av}^{(0)}(x)$ the ${\cal O}(\lambda^2)$ contribution 
is the result of the expansion of $M_{loc}$ in the denominator of 
(\ref{source0}).} This contribution to the thermal mass in the high-$T$ limit 
gives the correct limiting value for the one-component scalar theory, but it
can not account for the $T=0$ renormalisation.

The damping effect arises from the piece of current proportional 
with $n_1(x,p)$ 
\be
j_{av}^{(1)}(x)=
-\frac{\lambda^2}{4}\bar\phi^2\int \frac{d^3p}{(2\pi )^3p_0}
\frac{1}{(p\cdot\partial)}\partial_0\phi (x)\frac{d\tilde n_0}{dp_0}.
\label{resp}
\ee
The imaginary part of the linear response function $\Sigma$ defined through
$j_{av}^{(1)}(x)=\Sigma\phi(x)$ is evaluated in Appendix \ref{app:partition}
by the use of the principal value theorem, when the
$\epsilon$-prescription of Landau is applied to the $(p\cdot\partial )^{-1}$
operator on the right hand side of (\ref{resp}). For an
explicit expression one may assume that $\phi$ represents an off-mass-shell 
fluctuation characterised by the 4-vector: $(k_0 ,{\bf k})$, that is 
$\phi (x)=\phi (k\cdot x)$.

\subsection{\label{ss:MD}Connection with the formalism of 
Danielewicz and Mr{\'o}w\-czy\'nski }
Before explicitly evaluating the linear response function, let us make
connection to the classical kinetic theory for self-interacting scalar fields 
derived first by Danielewicz and Mr{\'o}wczy\'nski \cite{mrow90}, and presented
in the introduction and to present an alternative but equivalent method 
for the derivation of the effect of the quantum (high-k) modes on the EOM of
classical (low-k) modes. 

We sketch the steps followed in the derivation of the effective
equations
for the propagation of a non-thermal signal on a thermalized background:

1. Decomposition of the fields $\phi_i$ into the sum of high-frequency,
thermalized ($\varphi_i$) and low-frequency, non-thermal ($\phi_i$)
components;

2. Derivation of the equations of motion for $\phi_i$ in the background
of the thermal components;

3. Averaging the equations over the thermal background, retaining only
contributions from the two-point functions of the thermalized fields;

4. Approximating the two-point functions resulting from step 3. by
expressions with at most linear functional dependence on the non-thermal
fields, what is sufficient for the calculation of the linear response
function of the theory:
\bea
\nonumber
\avr{\varphi(x)\varphi(y)}&\approx&
\left\langle\varphi(x)\varphi(y)\rangle \right|_{\phi=0}+
\left. \displaystyle\int dz\
\frac{\delta \langle\varphi(x)\phi(y)\rangle}{\delta\phi(z)}
\right|_{\phi=0}\,\phi(z)\\
&\equiv&
\langle\varphi(x)\varphi(y)\rangle^{(0)}
+\langle\varphi(x)\varphi(y)\rangle^{(1)}.
\eea
We start from the Heisenberg equation of motion for the quantum field
\be
(\partial^2+m^2)\hat\Phi+{\lambda\over 6}\hat\Phi^3(x)=0,
\ee
then we split $\hat\Phi(x)$ into the sum of terms 
with low ($p_0<\Lambda$) and high ($p_0>\Lambda$) frequency Fourier-components, 
that is 
$\hat\Phi (x)=\tilde\phi (x)+\varphi (x), \avr{\phi(x)}=0$.
The classical equation of motion for the low frequency component 
$\tilde\phi(x)$ is the following:
\be
(\partial^2+m^2)\tilde\phi (x)+{\lambda\over 6}[\tilde\phi^3(x)+3\tilde\phi
(x)\varphi^2(x)+3\tilde\phi^2(x)\varphi (x)+\varphi^3(x)]=0.
\ee
The effective equation of motion arises upon averaging over the (quantum)
fluctuations of the high frequency field $\varphi(x)$. Dropping the
three-point function one obtains:
\be
(\partial^2+m^2)\tilde\phi (x)+{\lambda\over 6}\left[\tilde\phi^3(x)+
3\tilde\phi(x)\avr{\varphi^2(x)}\right] =0.
\ee
The last term on the left hand side represents the source induced by the
action of the high frequency modes. It is a functional of $\tilde\phi (x)$.
The action of the effective field theory may be reconstructed from this
equation. This approach is closely related to the Thermal Renormalisation
Group equation of D'Attanasio and Pietroni \cite{AP96}.

In the broken phase the non-zero average value spontaneously generated below
$T_c$ is separated from the low-frequency part, $
\tilde\phi (x)=\bar\phi +\phi (x)$.
The expectation value $\bar\phi$ is determined by the effective equation
\be
m^2+{\lambda\over 2}\langle\varphi^2(x)\rangle^{(0)}+{\lambda\over
6}\bar\phi^2=0.
\label{effvev}
\ee
                                                                 
Our present goal is to determine the effective {\it linear} dynamics of the
$\phi$-field, therefore it is sufficient to study the linearised effective
equation for $\phi (x)$:                                         
\be                                                              
(\partial^2+{\lambda\over 3}\bar\phi^2)\phi (x)=-{\lambda\over 2}\bar\phi
\langle\varphi^2(x)\rangle^{(1)}.                                
\ee                                                              
In this equation $\bar\phi$ is the solution of (\ref{effvev}), for this
reason the mass term is
$m^2+\frac{\lambda}{2}\bar\phi^2+\langle\varphi^2(x)\rangle^{(0)}=
{\lambda\over 3}\bar\phi^2=:M^2.$
The linear response theory of is contained in the induced current, determined 
by $\avr{\varphi^2(x)}^{(1)}$.
                                                                 
For the computation of the leading effect of the high-frequency modes in
the low frequency projection of the equation of motion it is sufficient
to study the  two-point function 
$\Delta^{>}(x,y):=\avr{\varphi (x)\varphi (y)}$.
\rt{For its determination we follow the procedure carefully described by
Mr{\'o}wczy\'nski and Danielewicz \cite{mrow90}.}                     

As in Eqs. (\ref{Eq:eom1}),(\ref{Eq:eom2}) we can derive 
\be
\label{Eq:om_most}
\left(\partial_x^2+M^2(x)\right)\Delta^{>}(x,y)=0,
\qquad\qquad
\left(\partial_y^2+M^2(y)\right)\Delta^{>}(x,y)=0,
\ee
where  $M^2(x)=m^2+\lambda\tilde\phi^2(x)/2$.

After adding and subtracting the Wigner transform \footnote{For the
introduction of the Wigner transform see Subsection \ref{ss:kinapprox}.} of
the two equations appearing in (\ref{Eq:om_most}) one arrives at the
following equations for $\Delta (X,p)$
\bea
\label{Eq:massshel}
&&\biggl({1\over 4}\partial_X^2-p^2+M^2(X)\biggr)\Delta (X,p)=0,\\    
&&\biggl(p_\mu\frac{\partial}{\partial X_\mu}+
\frac{1}{2}\frac{\partial M^2(X)}{\partial X_\mu}
\frac{\partial}{\partial p_\mu} \biggr)\Delta (X,p)=0.
\label{Eq:MDboltz}
\eea
The quantity appearing in the induced source is related to the   
Wigner transform of the two-point function through the relation  
\be
\langle\varphi^2(x)\rangle^{(1)}=\int{d^4p\over (2\pi )^4}\Delta^{(1)}
(x=X,p).
\ee
                                                                 
The important limitation on the range of validity of the effective dynamics
is expressed by the assumption that the second derivative with respect
to $X$ is negligible relative to $p^2$ and $M^2$ on the left hand side of
Eq. (\ref{Eq:massshel}). Then this equation is transformed simply into 
a local mass-shell condition, while Eq. (\ref{Eq:MDboltz}) can be 
interpreted as a Boltzmann-equation for the phase-space 
``distribution function'' $\Delta (X,p)$.
Comparing Eq. (\ref{Eq:MDboltz}) to Eq. (\ref{Eq:colisionless}) suggests the 
relation
\be
M(X)F_\mu ={1\over 2}\partial_\mu M^2(X),
\label{force0}
\ee
which supports the form of the particle Lagrangian
\be
L_{particle}=-M[\phi (x)],
\label{particle}
\ee
defined in Subsection \ref{ss:kinetic}\,.

The background-independent solution $\Delta ^{(0)}(p)$ is given by the 
well-known free correlator, slightly modified to account for the lower 
frequency cut appearing in the Fourier series expansion of $\Phi(x)$:      
\be
\Delta^{(0)}(p)=(\Theta (p_0)+\tilde n(|p_0|))2\pi\delta (p^2-M^2),\quad
\tilde n(p_0)={1\over e^{\beta |p_0|}-1}\Theta (|p_0|-\Lambda).  
\label{freecorr}
\ee

We notice here, that in view of Eq. (\ref{Eq:rho_hivat}) or equivalently
because $\bar\phi$ is the solution of Eq. (\ref{effvev}) 
(or Eq. (\ref{Eq:ezisbarphi}) ) we have to use in place of the tree level
mass square $m_{eff}^2$ the one-loop resumed (Hartree)  mass square
$M^2=\frac{\lambda}{3}\phi^2$.
 
An iterated solution of Eq.(\ref{Eq:MDboltz}) starting from (\ref{freecorr}) 
yields
\be
\Delta^{(1)}(X,p)=-{\lambda\over 2}\bar\phi
(p\cdot\partial_X)^{-1}{\partial\phi(X)
\over \partial X_\mu}{\partial\Delta^{(0)}(p)\over \partial p_\mu}.
\label{delta1}
\ee
By taking the Fourier-transform of the induced source
$\displaystyle
j_{ind}(x)=-{\lambda\over 2}\bar\phi\langle\varphi^2(x)\rangle^{(1)}$, and
using the explicit form for $\Delta^{(0)}(p)$ and the relation
$\partial/\partial p_\mu=2 p_\mu\partial/\partial\,p^2$ one easily finds
\be
j_{ind}(k)={\lambda^2\bar\phi^2\over 4}\phi(k)\int{d^4p\over (2\pi)^4}
\frac{k_0}{k\cdot p} 2\pi\delta(p^2-M^2)\frac{d\tilde n(|p_0|)}{dp_0}+
{\lambda^2\bar\phi^2\over 2}\phi(k)\int {d^4p\over (2\pi)^4}
\frac{\partial \Delta^{0}(p)}{\partial p^2}.
\label{Eq:jind2}
\ee

The first term is the non-local contribution. One easily recognises
after performing the $p_0$ integration with help of the $\delta$ function 
that it is exactly the Fourier transform of the r.h.s. of Eq. (\ref{resp}).

The second term calculated explicitly in the Appendix \ref{app:partition}
is a local contribution, and it accounts for the renormalization of the 
selfcoupling $\lambda$.

\subsection{The induced source}

The imaginary part of Eq. (\ref{resp}) or (\ref{Eq:jind2}) determines the rate 
of the Landau-damping. Simple integration steps lead to (see Eq. \ref{Eq:imI1})
\bea
{\rm Im}j(k_0,\k)&=&{\rm Im}\Sigma\phi (k)=
-{\lambda^2\bar\phi^2\over 16\pi}{k_0\over |\k|}\phi (k)
\Theta(|\k|^2-k_0^2)\int_{t_0}^\infty dt{d\hat n\over dt},\nonumber\\   
\hat n(t)&=&{1\over e^{\beta Mt}-1}\Theta (Mt-\Lambda),\quad t_0=
{1/\sqrt{1-(k_0 /|\k|)^2}}.
\eea
This integral is zero if $\Lambda >Mt_0$, but for 
$\Lambda <Mt_0$ it gives
\be
\label{landamp}
{\rm Im}j(k_0,\k)={\lambda^2\bar\phi^2\over 16\pi}
\phi (k){1\over e^{\beta M/\sqrt{1-(k_0 /|\k|)^2}}-1} {k_0\over |\k|} 
\Theta(|\k|^2-k_0^2)
\Theta\Biggl(M-\frac{\Lambda}{\sqrt{1-\frac{k_0^2}{|\k|^2}}}\Biggr)
\ee
independent of the value of $\Lambda$.                           
                                                                 
The result has a very transparent interpretation. In the HTL-limit, when only
the modes with much higher frequencies than any mass scale      
in the theory are taken into account, no Landau-damping arises. {\it The
effective theory is local!} This also means that in a theory with only one
massless degree of freedom, i.e. $M=0$, there is no Landau damping. 
We will see, however, in Chapter \ref{ch:ON} that in the O(N) theory the massless
Goldstone modes ``feel'' the presence of Landau damping due to 
the massive Higgs modes they interact with. In the case of O(N) model the
modes can transform into each other, this is the reason why we give up the 
attempt to describe the model with a kinetic theory. We will use instead a
generalisation of the way presented in Subsection \ref{ss:MD}.
   
Going beyond HTL, ($k_0\ll\Lambda \ll M$) the 1-loop exact   
self-energy contribution, and so the 1-loop exact Landau damping rate  is 
calculated in the next Chapter in Subsection \ref{ss:non_eq_ld}.

Using the formula
\be
\int_a^b n(s) ds=T\ln(1-e^{-\beta b})-T \ln(1-e^{-\beta a})
\ee
one can rewrite the first term from the first line of Eq. (\ref{impis}) 
in the form
\be
{\rm Im}\Sigma_{Landau}=\frac{\lambda^2\bar\phi^2T}{16\pi |\k|}
\ln\frac{1-e^{-\beta\omega_k^+}}{1-e^{-\beta\omega_k^-}}
\Theta (|\k|^2-k_0^2),\qquad
\omega^{\pm}_\k=\left|\frac{k_0}{2}\pm \frac{|\k|}{2}
\sqrt{1-\frac{4M^2}{k_0^2-|\k|^2}}\right|.
\label{dan}
\ee
This coincides with the result previously derived by Boyanovsky {\it et al.}
in Ref. \cite{boya96}. The result of Eq. (\ref{landamp}) holds for small
values of $k_0$ and $\k$ as shown in Subsection \ref{ss:nonli_dy}.

\rt{
Here $M$ is the temperature dependent mass parameter in the finite 
temperature quantum theory. 
For the assessment of the domain of validity of the classical result 
(\ref{landamp}) we compare it to the complete one-loop quantum result
(\ref{dan}).  
With explicit calculation one checks that 
\be
\ln\frac{1-e^{-\beta\omega_k^+}}{1-e^{-\beta\omega_k^-}}\approx
2\,\textrm{sh}\frac{\beta k_0}{2}\,\exp{\left(-\frac{\beta |\k|}{2} 
\sqrt{\frac{4M^2}{|\k|-k_0^2}+1}\right)}\approx\beta k_0 
\exp{\left(-\frac{\beta M}{\sqrt{1-(k_0 /|\k|)^2}}\right)},
\label{Eq:Lapprox}
\ee
provided $\beta |\k|\sqrt{\frac{4M^2}{|\k|-k_0^2}+1}\gg 1$, \,\,
$\beta k_0\ll 1$ and $4 M^2\gg |\k|^2- k_0^2.$ The first condition is needed
for the first step and the second an third one for the second step in formula
(\ref{Eq:Lapprox}).
Both the classical and quantum result goes over into the formula
\be
{\rm Im}\Sigma_{\textrm{approx}}={\lambda^2\bar\phi^2\over 16\pi}\phi (k)
\frac{k_0}{|\k|}\exp{\left(-\frac{\beta M}{\sqrt{1-(k_0 /|\k|)^2}}\right)}
\Theta(|\k|^2-k_0^2)
\Theta\Biggl(M_0-\frac{\Lambda}{\sqrt{1-(k_0 /|\k|)^2}}\Biggr),
\ee
in the limit $\beta k_0\ll 1$;\,\, $\beta M,\, \frac{M}{|\k|}\gg 
\sqrt{1-(k_0 /|\k|)^2}$\, under the assumption that $m_{eff}$ of the
classical theory is identified with $M$.
In the limit $\beta\omega ,\beta |{\bf k}|\ll1$ the conditions above are
satisfied for $k_0\sim |\k|$.
In this way the classical theory provides a convincing argument 
clearly demonstrating that Landau damping is the correct interpretation of 
the result derived from quantum theory.}

\subsection{Discussion}

We have presented two equivalent methods of treating the effective
dynamics of the low-frequency fluctuations of self-interacting scalar fields
in the broken phase of the theory. The dynamics is proved to be non-local
if the effect of fluctuation modes below the mass scale $M$ is also taken
into account. A fully local representation was proposed by superimposing
a relativistic gas with specially chosen field dependent mass on the
original field theory. 

The range of validity of the fully local version of the effective model can
be ascertained only from its comparison with the result of lowering the
separation scale $\Lambda$ in the detailed integration over the fluctuations
with different frequencies. From the comparison one learns that the combined
kinetic plus field theory is equivalent to the effective theory for the
modes with $k_0\ll\Lambda< M$, that is its validity goes beyond the Hard
Thermal Loop approximation.

There is yet another way of constructing a local theory, namely by introducing
an auxiliary field $W(x,{\bf v})$ defined through the equation
\be
v\cdot k \,W(k,{\bf v})=k_0\phi(k),
\ee
where $v=(1,\p/p_0)$. Then by looking at Eq. (\ref{Eq:f1})
one sees that $f_1(k,p)$ the deviation from equilibrium of the one particle
distribution function is given by the equation
\be
f_1(k,p)=-\frac{\bar\phi}{2}\,W(k,{\bf v})\,\frac{d f_0(p_0)}{dp_0}.
\ee
The non-local part of $\Delta^{(1)}(k,p)$ from Eq. (\ref{delta1})is given by
the equation
\be
\Delta^{(1)}(k,p)=-\frac{\lambda}{2}\bar\phi \,W(k,{\bf v})\,2\pi
\delta(p^2-M_0^2) \frac{d\tilde n(|p_0|)}{dp_0}.
\ee

We will see in Section \ref{ch:ON} that the auxiliary field can be
introduced even in a more general term, when we don't use the gradient
expansion and  a Boltzmannian kinetic evolution for these fields doesn't
exist. 

We have also shown that in the broken phase of scalar theories and
in the low-$k_0$ region the Landau-type damping phenomenon occurs
according to an effective classical kinetic theory.  In this context,
the clue to its existence is the presence of the $\sim\bar\phi\phi$ term in 
the fluctuating part of the local mass, which is responsible for the emergence
of a linear source-amplitude relation with non-zero imaginary part.

The extension of our discussion to the $N$-component scalar fields
in the broken phase will be presented in Section \ref{ch:ON}.

\newpage

%% file: scalar1nl.tex
\section{Effective order parameter dynamics and the decay of a metastable
vacuum state far from equilibrium} 

The reaching  of equilibrium from a metastable state involves a large number of
interesting effects from instabilities observed in the mixed phase of first
order phase transitions \cite{nucl} to the inflation in the early Universe
\cite{Boyan99}. The final state is reached in an irreversible process.

The decay of metastable states is usually discussed in the framework of the
nucleation scenario \cite{Binder76}. It has been implemented in the form of
saddle point expansions in classical \cite{Langer67}, and quantum systems
\cite{Coleman80}. This large amplitude instability is 
\removetext{, however, only} the first of the possible instabilities, 
suggested by mean  field analysis. It 
consists of the creation of a bubble of the true vacuum
larger than the critical size, embedded into the false one. 

Another possibility, the instability against fluctuations with infinitesimal
amplitude leads to the spinodal phase separation.  A recent observation made
it clear that soft fluctuations of these inhomogeneous unstable modes
generate in equilibrium the Maxwell construction by their tree-level
contribution to the renormalization group flow \cite{Alex99}. The fluctuation 
induced flatness of the effective potential suggests the dominance of spinodal
phase separation in equilibrium. 

Since the type of instability one observes, might depend essentially on the
time scale of the observation, a detailed investigation of the time
evolution can separate the effects of the two kinds of instabilities. This
is made possible by large scale computer simulations of the thermalisation
process in closed systems.

Whether the relevant mechanism for a first order phase transition is the
formation of bubbles of the new phase, as described by thermal
nucleation theory, or the gradual change of a large region of the sample,
due to small amplitude spinodal instabilities described by spinodal
decomposition is also an intriguing question in heavy ion physics where the
actual expansion rate of the plasma may favour one or the other scenario
\cite{fraga}.

Another question, left open by the Maxwell construction in equilibrium
\cite{Callaway83}, concerns the structure of the vacuum with spontaneous 
symmetry breaking. In fact, the effective potential is related to the
probability distribution of the order parameter and the Maxwell-cut applied
to the former suggests that the latter is also degenerate in the mixed
phase. Either we accept that the vacua with spontaneously broken symmetry
correspond overwhelmingly to the mixed phase or a dynamical mechanism is
seeked to eliminate the mixed phase from among the final states of the time
evolution.

In cosmology different slow-roll scenarios of inflation are being
considered. Recent studies of the dynamics of inflaton fields with large
number of components (large $N$ limit) displayed for a first time a
dynamical version of the Maxwell construction \cite{Boyan99a}.  The Hartree
type solution of the quantum dynamical equations leads to the conclusion
that the order parameter might get rest with finite probability at any value
smaller than the position of the stable minimum of the tree level effective
potential, corresponding to the stabilisation of a mixed state.

Detailed investigations of the thermalisation phenomenon were performed
also in noisy-damped systems, coupled to an external heat bath
\cite{Alford93,Gleiser94a,Gleiser94b,Berera98}. The relaxation to thermal 
equilibrium of the space averaged scalar field (the order parameter) 
starting from metastable initial values has been thoroughly investigated. 
In these simulations the damping coefficient is treated as an 
external control parameter. Using the numerical solution of
the corresponding Langevin-equations the validity range of the analytical 
results for the homogenous nucleation mechanism has been explored. 

We  focus on an alternative description of the decay process
of the metastable vacuum state. The process is described exclusively in
terms of the homogenous order parameter (OP) mode.  The evolution of the OP
is studied in interaction with the rest of the system as described by the
reversible dynamical equations of motion of the full system.  Careful
analysis of its dynamics allows us to explore the effects of both kinds of
the above mentioned instabilities. The transition of the order parameter
from the metastable to the stable vacuum is induced by a homogenous external
``magnetic'' field, whose strength is systematically reduced. No random
noise is introduced to represent any external heat bath, the friction
coefficient of the effective order parameter dynamics is determined
internally.
 
Our model, a spacelike lattice regulated \at{classical} scalar $\Phi^4$
field theory in its broken symmetry phase is introduced in Subsection
\ref{ss:th_on_lat}. In Subsection \ref{ss:time-history} we describe the 
evolution of the system starting from order parameter values near a metastable 
point which relaxes first to this state under the combined effect of parametric
resonances and spinodal instabilities. The second stage of the transition to 
the stable ground state is the actual focus of our discussion.

Characteristic intervals of the observed order parameter trajectory are
reinterpreted as being the solutions of some effective point-particle
equation of motion, which displays dissipation effects explicitly. Our
approach can be understood also as the real-time version of the lowest mode
approximation used for the estimation of finite size dependences in
Euclidean field theory \cite{Zinn94,Binder84}. In this sense our approach
can be considered also as the numerical implementation of a real time
renormalisation group strategy.

One of our principal goals is to reconstruct the thermodynamics of the
\at{classical} ``OP-ensemble'' on the (meta)stable branches of the
OP-trajectory (Subsection \ref{ss:metastable}). Its dissipative dynamical 
equations near equilibria will be established. On the transition trajectory 
we shall elaborate on the presence of the Maxwell construction in the effective
dynamics describing the motion after nucleation (Subsection \ref{ss:jump}). The
statistical aspects of the approach to the equilibrium are established for 
reference and comparison in  Subsection \ref{ss:nucl}. The conclusions of this 
investigation are summarised in Subsection \ref{ss:polonyi_concl}.

\rt{The results of this study can be usefully compared with classical
investigations of metastability and nucleation in the kinetic Ising model
\cite{Binder74}. This system has first order dissipative dynamics 
by its definition.  Still several relaxation features of the kinetic Ising 
model are comparable to our findings, since the ``numerical experiment''
performed in the two models are essentially the same.  Especially, the
relaxation function of \cite{Binder74} is simply related to the order
parameter we focus our attention. In both cases in the mechanism of the
bubble growth the aggregation of spontanously generated small size regions
of the true ground state to its surface plays important role.}

\subsection{\label{ss:th_on_lat}Classical cut-off $\Phi^4$-theory on lattice}
The energy functional of a classical system in a two-dimensional box of size
$L_d$ coupled to an external magnetic field of strength $h_d$ is of the 
following form:
\be
E_d=\int d^2x_d\left[{1\over 2}\left({d\Phi_d\over dt_d}\right)^2+{1\over 2}
\bigl (\nabla_d\Phi_d\bigr )^2+{1\over 2}m^2\Phi_d^2+{1\over 24}
\lambda\Phi_d^4-h_d\Phi_d\right ].
\label{dimfulenergy}
\ee
The index $d$ is introduced to distinguish the dimensionfull quantities from
the dimensionless ones, defined by the relations (for $m^2<0$):
\bea
t=t_d|m|,\qquad x=x_d|m|,\qquad
\Phi =\sqrt{\lambda\over 6}{1\over |m|}\Phi_d,\qquad h=\sqrt{\lambda\over 6}
{1\over |m|^3}h_d.
\eea
For the spatial discretisation one introduces a lattice of size
$L_d=Na_d=Na{1\over |m|}.$

The energy functional of the lattice system can be written as
\bea
E\equiv {\lambda\over 6|m|^2}E_d={a^2\over a_t^2}\sum_{\bf n}\left [{1\over 2}
\left(\Phi_{\bf n}(t)-\Phi_{\bf n}(t-a_t)
\right)^2+{a_t^2\over 2a^2}\sum_{\bf i}\left(\Phi_{\bf n+\hat i}-\Phi_{\bf n}
\right)^2\right.\nonumber\\
\left.
-{a_t^2\over 2}\Phi_{\bf n}^2+{a_t^2\over 4}\Phi_{\bf n}^4-a_t^2h\Phi_{\bf n}
\right].
\label{dimlessenergy}
\eea
(Here we have introduced the dimensionless time-step $a_t$, which should be
chosen much smaller than $a$, and $\bf n$ denotes the lattice site vectors.) 
The equation of motion to be solved numerically is the following:
\bea
\Phi_{\bf n}(t+a_t)+\Phi_{\bf n}(t-a_t)-2\Phi_{\bf n}(t)-{a_t^2\over a^2}\sum_i
(\Phi_{\bf n+\hat i}(t)+\Phi_{\bf n-\hat i}(t)-2\Phi_{\bf n}(t))
\nonumber\\
+a_t^2(-\Phi_{\bf n}+\Phi_{\bf n}^3-h)=0.
\label{latteq}
\eea

The initial conditions for Eq.(\ref{latteq}) were chosen as
\be
\dot\Phi_{\bf n}(t=0)=0,\qquad \Phi_{\bf n}(t=0)=\Phi_0+\xi\Phi_1.
\label{white}
\ee
The random variable $\xi$ is distributed evenly on the interval $(-1/2,1/2)$. 
Therefore the
starting OP-value is $\Phi_0$. The energy density $E/Na^2$ is controlled 
through the magnitude of $\Phi_1$. In this study we have chosen $\Phi_0=0.815$
and $\Phi_1=4/\sqrt{6}$. The latter corresponds to a temperature value 
$T_i=0.57$ in the
metastable equilibrium (from Eq.(\ref{tempeq})). This value is much below 
the critical temperature of the system ($T_c\simeq 1.5T_i$). 
It has been checked that at this energy density all other
choices of $\Phi_0>0$, for fixed $h$, find a unique metastable equilibrium. 

Eq.(\ref{latteq}) was solved with $a=1$ and with typical $a_t$ values in the
range (0.01-0.09). It has been checked that the statistical characteristics
of the time evolution is not sensitive to the variation of $a_t$, though the
``release'' time (the moment of the transition from metastability 
to the true ground state) 
in any single run with given initial conditions might change 
considerably under the variation of $a_t/a$. Three lattice sizes were 
systematically explored: $N=64,128,256$. Several single runs were performed
also for $N=512$ and $N=1000$ with the aim to analyze in more detail
some self-averaging
physical quantities on different portions of the trajectory under the
assumption of the ergodicity of the system. The magnetic field $h$ 
inducing the transition was
varied in the range $h\in -(0,0.08)/\sqrt{6}$. For the reconstruction of the
effective potential felt by the OP also positive values were 
chosen up to $h=0.5/\sqrt{6}$. The smaller the value of $|h|$ was fixed, 
the longer the ``release'' times have grown on the average. For this reason 
also the runs were prolongated with decreasing $|h|$, 
and for the smallest $|h|$ the length of a run reached up to 
$(10^6-10^7)|m|^{-1}$ until the transition took place. 

\rt{Table \ref{runstatistics} shows the transition event statistics for
each $(h,N)$ pair, used in this analysis. On smaller lattices, for a few
$h$ values very large number of transitions were observed, in order to
clarify the nature of the corrections to the nucleation mechanism.}
\at{For a careful comparison of our transition statistics with the
generally used statistical nucleation theory, and also for
understanding the systematics of its change when $h$ has been
diminished a large number of ($h$, $N$) pairs were used in this
analysis. Altogether \hbox{24~422} transition events have been
recorded (for \hbox{$N=64$:~~16~908}, \hbox{$N=128$:~~2~903},
\hbox{$N=256$:~~4~411}).}
For the largest systems at the smallest $h$ the event rate was 
1-2/day/ 400 MHz-processor.
 

\subsection{\label{ss:time-history}Time-history of the order parameter}
\begin{figure}[htbp]                                       
\includegraphics[width=8.25cm]{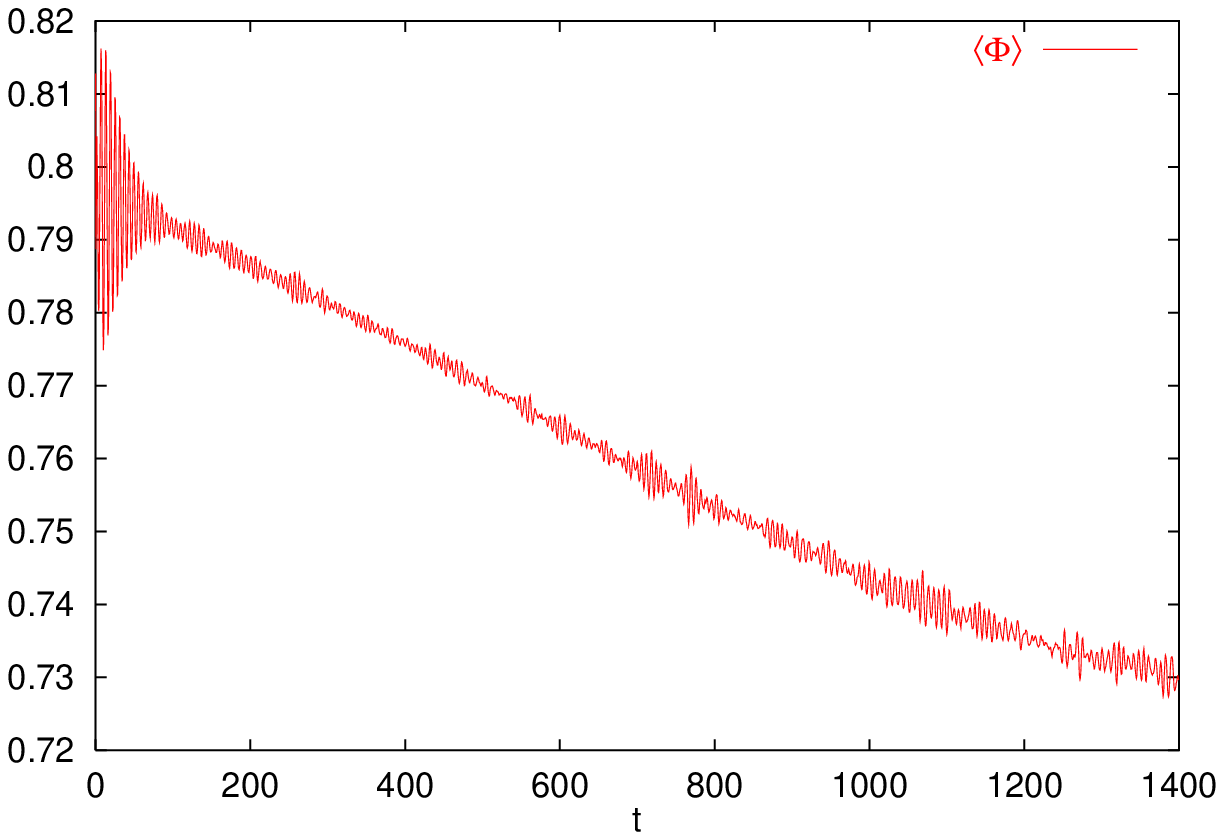}
\includegraphics[width=8.25cm]{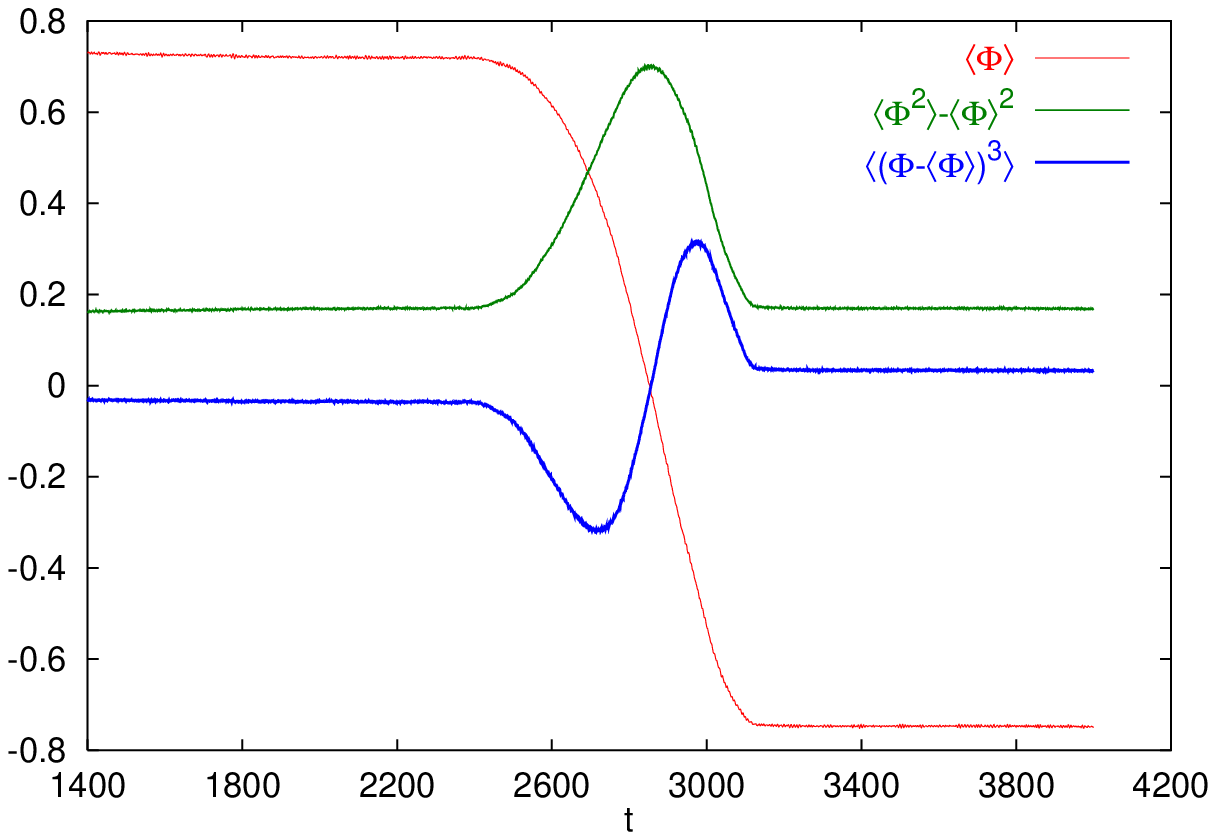}
\caption{\label{time_hist}The time evolution of the order parameter,
its MS fluctuation and the third moment.
The example is selected from runs on a $N=512$ 
lattice with $h=-0.04/\sqrt{6}$ external source strength.}
\end{figure}

\begin{figure}[htbp]
\begin{center}
\includegraphics[width=8.5cm]{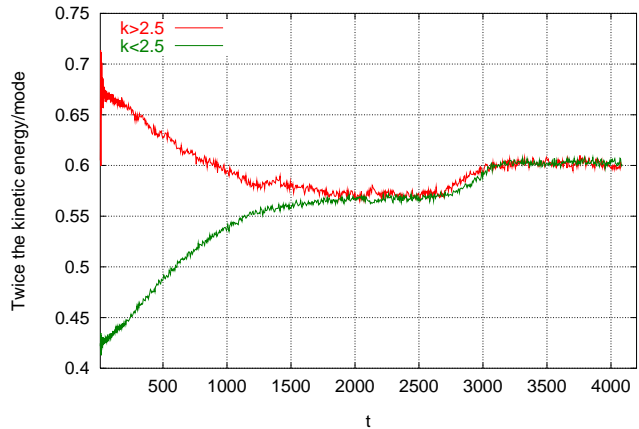}
\end{center}
\vspace*{-1cm}
\caption{\label{k_hist}The time evolution of the kinetic energy content of
the $|{\bf k}|>2.5$ and $|{\bf k}|<2.5$ regions averaged over the
corresponding $|\bf{k}|$-intervals.
The example is the same as in Fig. 1.}
\end{figure}

A typical OP-history is displayed in Fig.\ref{time_hist}. 
In the same figure we show also the history of the OP mean square 
(MS)-fluctuation ($\langle\Phi^2\rangle -\langle\Phi\rangle^2$) and
of its third moment ($\langle (\Phi -\langle\Phi\rangle )^3\rangle $). 
The evolution
of the non-zero $\bf k$ modes is demonstrated in Fig.\ref{k_hist}, where
the averaged kinetic energy content of the $|{\bf k}|<2.5$ and $|{\bf k}|>2.5$
regions is followed. Although the separation value is somewhat arbitrary, 
namely it divides into two nearly equal groups the spatial frequencies
available in the lattice system,
the figure demonstrates the most important features of the evolution of the 
power in the low-$|{\bf k}|$  and high-$|{\bf k}|$ modes.

In general, five qualitatively distinct parts of the trajectory can be 
distinguished, although some of the first three might be missing for 
some initial configurations and/or magnetic field strengths.

\begin{figure}[htbp]
\begin{center}
\includegraphics[width=8.25cm]{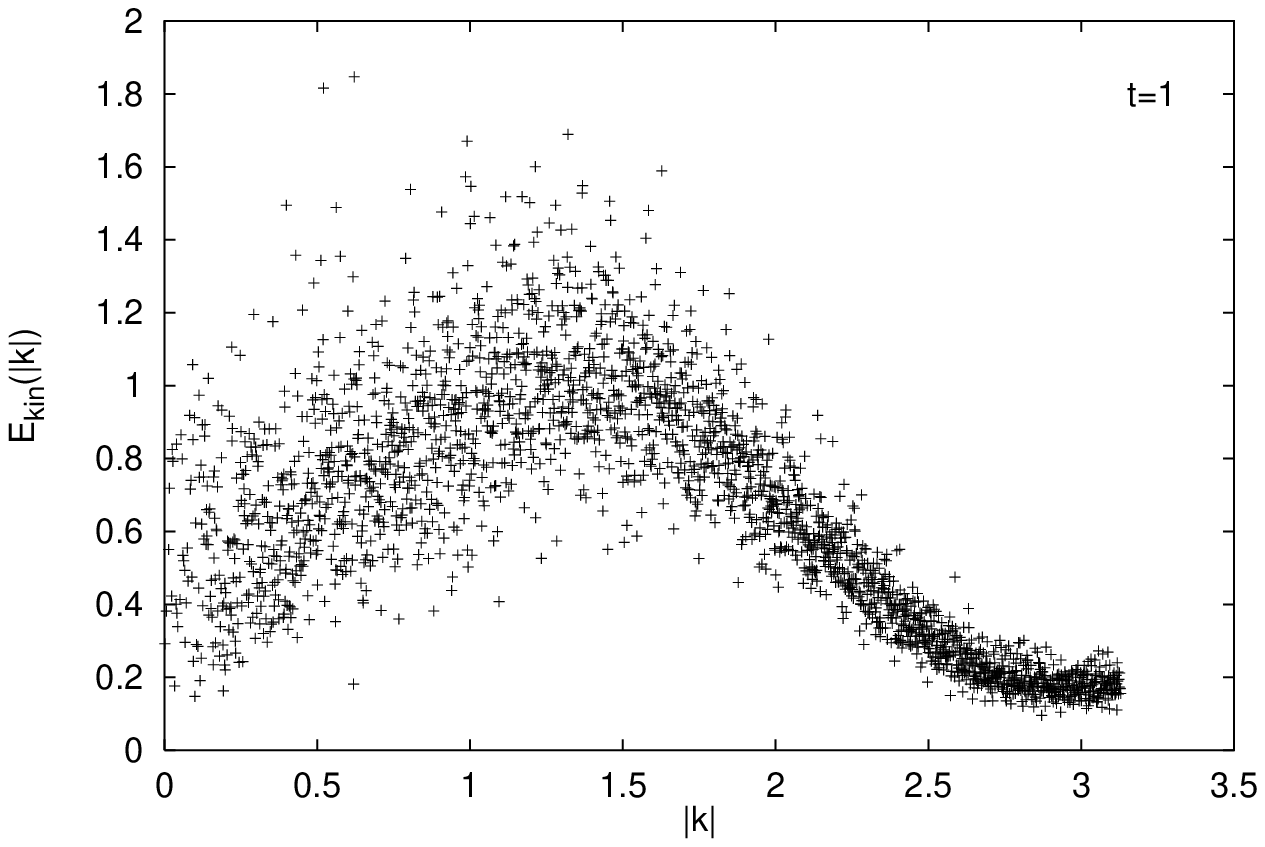}
\includegraphics[width=8.25cm]{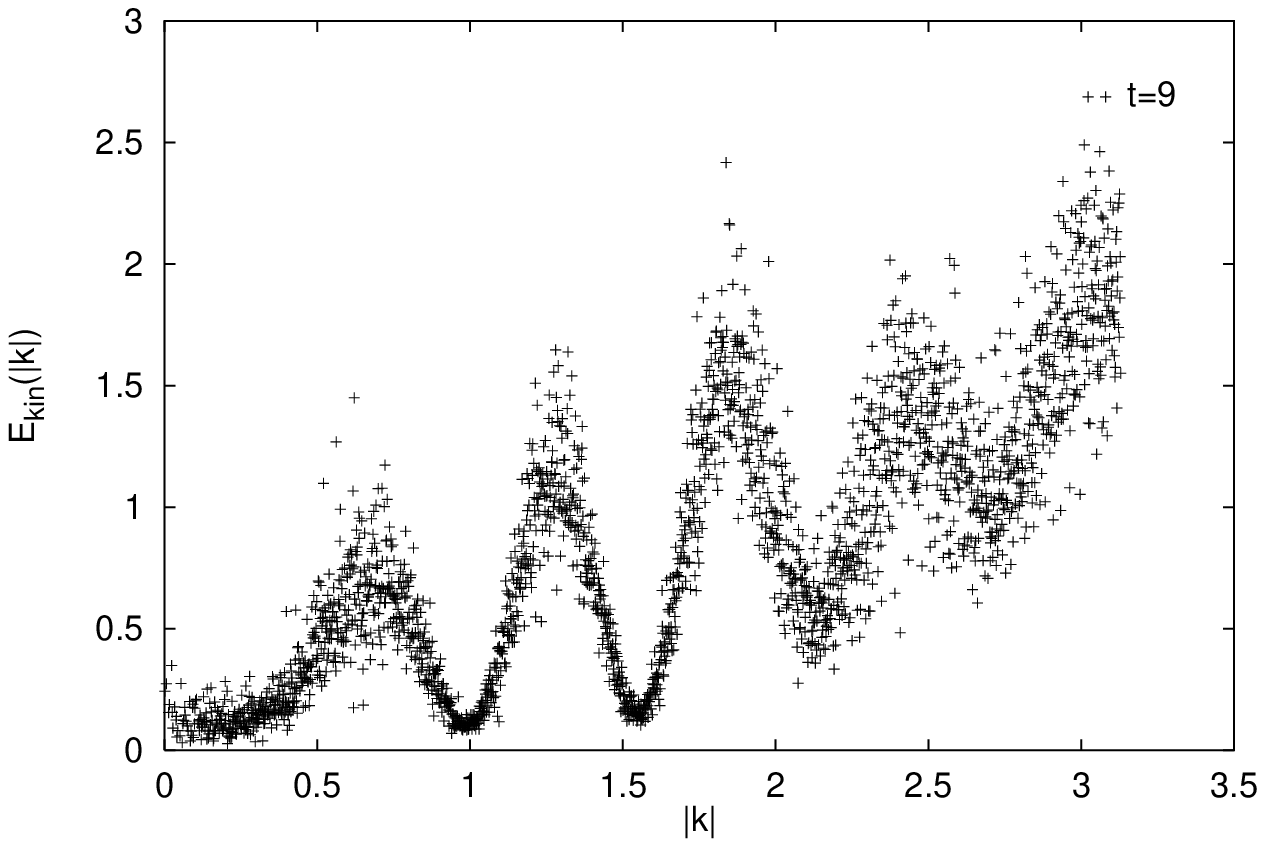}
\end{center}
\vspace*{-0.5cm}
\caption{Early time power spectrum.}
\label{Fig:peaks}
\end{figure}

The OP-motion usually starts with large amplitude damped oscillations.
The ``white noise'' initial condition of Eq.(\ref{white}) corresponds to a 
$\bf k$-independent Fourier amplitude distribution, therefore the initial 
distribution of the kinetic energy is $\sim\omega^2(|{\bf k}|)$.
During this period, in the power spectrum of the kinetic energy, first 
a single sharp peak shows up at a resonating $|{\bf k}|$-value ($|{\bf k}|
\sim 1.5$), which breaks up into several peaks ($|{\bf k}|<1$) at later times 
due to the non-linear interaction of the modes, see Fig. (\ref{Fig:peaks}). 
At the end of the first period the whole $|{\bf k}|<1$ range 
gets increased power, the $|{\bf k}|>1.5$ part of the power spectrum does
not seem to change.

Next a slow, almost linear (modulated) decrease of the OP follows.  At the
same time its MS fluctuation increases linearly.  In Fig.\ref{k_hist} an
energy flow towards the low-$|{\bf k}|$ part can be observed, which proceeds
through the excitation of single modes in this part. As a result the minimum
of the effective potential is continously shifted to smaller $\Phi$-values,
as if the temperature would gradually increase (see the left part of
Fig.\ref{time_hist}). The full (microscopic) kinetic energy density shows in
this period less than 5\% variation. In view of the picture based on the
Maxwell construction it is rather surprising that independently of the
initial conditions the OP converges towards a well-defined absolute value,
depending only on the total energy density.

On the third portion the average value of the OP and its moments stay
constant (see the right picture in Fig.\ref{time_hist}). The average energy
content of the low and high-$|{\bf k}|$ part of the spectra is nearly the
same. This suggests the establishment of a sort of thermal (meta)equilibrium.

In terms of the terminology introduced for the inflation, the first period
leading to this (quasi) -stationary state can be called {\it preheating},
and the second {\it reheating}.  Directly before the moment of the
transition to the true vacuum a peak appears in the power spectrum in the
narrow neighbourhood of ${\bf k}=0$ with varying position in time. On the
snapshots of the real space configurations a set of randomly distributed
small bubbles of the true ground state appear with a radius increasing in
time, until one of the bubbles exceeds the critical size.

The fourth portion of the motion is the transition itself. The value of the
OP MS-fluctuation increases by about a factor of three and the temporal
width of this transient increase measures very well the transition time. The
transition time decreases on larger lattices, the height of the jump in the
OP-fluctuation is not sensitive to the lattice size.  Also the third reduced
moment shows a characteristic variation.  An increase of the temperature
proceeds smoothly during the transition of the OP to its stable value
(Fig.\ref{k_hist}). The slight separation of the two curves in
Fig.\ref{k_hist} gives a feeling on the degree of uniformity of the
temperature variation of the different modes.

The last portion of the trajectory represents stable (thermalised)
oscillations around the true ground state.  Here a complete equilibration of
the power spectrum can be observed (see Fig.\ref{k_hist}) corresponding to a
somewhat increased temperature $(T\sim 0.6)$.

\subsection{\label{ss:metastable}Motion near the (meta)\-stable point}
Our analysis of the motion around the (meta)stable value of the order
parameter explores the consequences of assuming the ergodicity hypothesis for
\at{a sufficiently long} finite time interval,  \at{after the system has 
already reached the equilibrium}. \at{ The equilibrium is characterised by a 
limiting probability density in the configuration space. The averaging with
this density should provide the same value as the one yielded by a single
long time evolution when subsequent configurations are used in constructing
the statistics. Our ``statistical system'' is now a single degree of
freedom, the order parameter of the lattice system, interacting with all
other ($k\ne0$) modes.}

\at{
The effective equation of motion can be thought to result from the
application of a ``molecular dynamical renormalization group'', to our
microscopical equations.  The blocking in space is performed by projecting
the field configuration $\Phi({\bf x},t)$ on the OP $\Phi (t)$. It represents 
the infrared (IR) end point of such a blocking whose effective theory is now
reconstructed from the actual time dependence found numerically. Combining
ergodicity of the full system with the renormalization group (RG) concept we
arrive to the conclusion that ensemble averages of any OP-function coincide
with time averages of the same function.
} 
\rt{
In case of the motion around metastable minima we attempt to detect
signatures of the metastability in the period directly preceding the start
of the transition of the order parameter to the stable position.}

\at{
In view of the RG concept we look for an effective equation of motion where
the value of OP is determined exclusively by its values preceding in time.
Assuming the absence of long memory effects a ``gradient'' expansion in time
can be envisaged, leading to a local differential equation.}
\at{
We introduce into this phenomenological ``Newton-type'' equation a term
violating time-reversal invariance. We are not able to derive it from the
original system, we only wish to test its presence. The sustained motion of
OP, however, requires, in this case the presence of a random ``force'' term,
too.  The deterministic part of the ``force'' is expected to be related to
the equilibrium effective potential, since this object determines the
stationary probability distribution for the OP near equilibrium.}

\at{The above considerations lead us to write down the  equation of
motion, which represents a linear relation between the acceleration and the
velocity of the order parameter:}
\be
\ddot{\Phi}_d+\eta_d(\Phi_d)\dot{\Phi}_d-h_d
+{dV_{\textrm{eff}}(\Phi_d)\over d\Phi_d}=\zeta_d,
\label{eqfriction}
\ee
where $\zeta_d$ is a noise term. In the corresponding dimensionless equation
of motion the following new rescaled quantities will appear:
\be
\eta_d=|m|\eta, \qquad \zeta_d=\zeta |m|^3\sqrt{6\over\lambda}.
\ee
The fitting procedure for the coefficients on the left hand side of
Eq.~(\ref{eqfriction}) was the following.
For a given interval of time $I_t$ the region of the order parameter space
visited by the system $\left\{\Phi(t)|t\in I_t\right\}$ was divided into small 
bins. Having defined the time set \linebreak 
\hbox{$T_{\Phi_b}=\left\{t\in I_t|\Phi_b<\Phi(t)<\Phi_b+\Delta\Phi\right\}$} 
corresponding to a given bin, the linear relation
$\ddot\Phi(t_b)=-\eta(\Phi_b) \dot\Phi(t_b)-f(\Phi_b)$
was fitted with the method of least squares
using the $\dot\Phi(t_b)$, $\ddot\Phi(t_b)$ data measured at time moments 
$t_b$ belonging to $T_{\Phi_b}$. We have obtained in this way the coefficient 
functions $\eta(\Phi_b), f(\Phi_b)$. Once the functions $\eta(\Phi)$ and
$f(\Phi)\equiv-h+V_{\textrm{eff}}'(\Phi )$ are determined, we evaluate 
for each time $t$ the expression 
$\ddot\Phi(t)+\eta(\Phi(t))\dot\Phi(t)+f(\Phi(t))$. Its actual value
determines the random noise function $\zeta(t)$, whose statistical
features (autocorrelation) should be extracted from the data.

The {\it effective force} $f(\Phi )$ calculated from the time-average of the
oscillatory motion around the equilibrium, is expected to
agree with the force coming from the theoretically determined
finite temperature effective potential calculated perturbatively
in the cut-off two-dimensional field theory for some appropriately chosen
value of the temperature \cite{Kapusta89}. With one-loop accuracy the
expected equality reads:
\be
f(\Phi )_{measured}=-h+{d\over d\Phi}\left[V(\Phi )+{T\over 8\pi}V^{''}
(\Phi )\left(1+\log{\Lambda^2\over V^{''}(\Phi )}\right)\right]_{theory}
+{\cal O}(T^2).
\label{fforce}
\ee
The expressions on the right hand side of this equation are connected
to the dimensionfull quantities of the original one-loop computation in the
following way:
\be
T={\lambda\over 6|m|^2}T_d,\quad 
V(\Phi )=-{1\over 2}\Phi^2+{1\over 4}\Phi^4={\lambda\over 6|m|^2}V_d,
\quad \Lambda ={\pi\over a}.
\ee
The dimensionless temperature is defined by the time-average of the 
kinetic energy based on the assumption that in the effective theory of the
OP it has the usual expression in terms of $\dot\Phi (t)$:
\be
\lim_{t\rightarrow\infty}{1\over t}\int_0^tdt'{1\over 2}
(\dot\Phi (t',x))^2\equiv{1\over 2}\overline{\dot\Phi^2}^{t}={T\over 2}.
\label{tempeq}
\ee
One should note that this definition in the dimensionfull
version implies a ``Boltzmann-cons\-tant'', $|m|^2$ multiplying $T_d$.

By measuring the order parameter average for different values of the
external field $h$ we obtain the magnetisation curve of the system. In the
numerical work we restarted the computation at different values of the
external magnetic field but one might as well change $h$ adiabatically and
measure the force law at each (quasi)equilibrium point. The results should
agree in the stable regime, up to the phenomenon of the hysteresis. The
resulting curve can be viewed as the numerical Legendre transformation by
identifying the external source $h$ with the derivative of the effective
potential,
\be
-h+V'_{\textrm{eff}}(\la\Phi\ra_{h,measured})=0.
\label{equilib}
\ee
It was found that $f(\Phi )_{measured}$ is always vanishing at 
$\Phi=\la\Phi\ra_{h,measured}$,
thus the force acting on the order parameter at the
equilibrium position $\la\Phi\ra_h$ induced by the external 
source $h$ is indeed always $-V'_{eff}(\la\Phi\ra_h)$. 

\begin{figure}[htbp]
\begin{center}
\includegraphics[width=8.5cm]{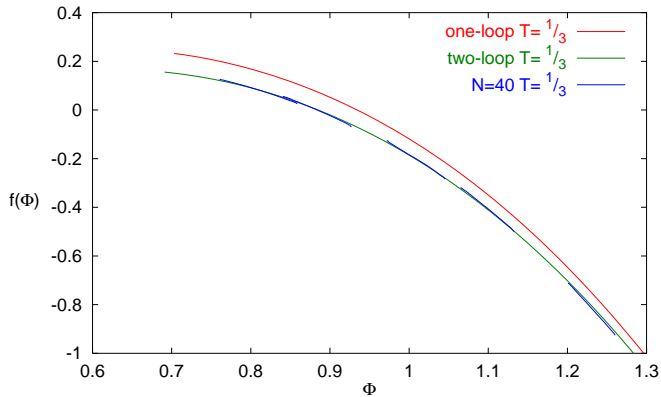}
\end{center}
\vspace*{-0.8cm}
\caption{\label{fphi} The force $f(\Phi)$ as the function of the order
parameter on $40\times40$ lattices with
 $T=1/30$ and $1/3$. The force is measured by shifting the center of motion
to different $\Phi$ values by applying appropriate $h$ fields to the system.
The finite size effects are negligible, see 
text. For comparison we display the force arising from the derivative 
of the one-loop and for $T=1/3$ also of
the two-loop expressions of the effective potential.}
\end{figure}

The two sides of the relation (\ref{fforce}) are shown in Fig.\ref{fphi}.
To the values of the external field used in preparing Fig.\ref{fphi}
($h\sqrt{6}=-0.5, -0.02, 0, 0.5, 1, 2)$ 
single runs were selected by the requirement
that the system stayed in the metastable vacuum up to the time $10^6$.
The force shown in Fig.\ref{fphi} demonstrates that not only the
equilibrium positions but also the fluctuations of the OP
are governed by the {\it static} effective potential. Similar measurements,
performed on $100\times 100$ lattices showed no finite size effects
in the stable regime, $h>0$. The error of
$f(\Phi )$ is not shown in the Figure, since the typical values
($\sim 0.004$ for $T=1/3$ and $\sim 0.002$ for $T=1/30$) are too small to be
displayed.

Another piece in the effective equation of motion \eq{eqfriction}
is the {\it friction term} whose presence indicates the dynamical
breakdown of the time reversal symmetry. The friction coefficient
$\eta(\Phi)$ proves clearly non-vanishing and shows only weak 
dependence on the actual value of the OP around its equilibrium position. 
The breakdown of the time-reflection symmetry in a closed system must arise 
only in presence of
infinitely many degrees of freedom. Till then only statements on the 
Poincar{\'e}-time can be made. Thus our non-zero results for $\eta$
require further clarifications. 

The point is that there are two types of infinities, controlled by the {\it
temporal} UV and the IR cutoffs, respectively. The spontaneous symmetry
breaking is driven by the IR modes, and the non-trivial minima of the
potential energy arise from the presence of infinitely many degrees of
freedom in the IR (thermodynamical limit). On the contrary, the dynamical
symmetry breaking \cite{dynam} is the result of the effects of the derivative 
terms in the action and the infinitely many UV modes (continuum limit). 
The breakdown of the time reversal, being related to a linear time-derivative 
term in the effective equation of motion, should come from the UV, the short 
time behaviour of the system.

In fact, one expects no friction when the UV cutoff, $a_t$ is \at{so} small,
\at{that not enough energy can be dissipated during such a short time. In
quantitative terms one should have}
\be\label{tau}
a_t>\tau={2\pi\over\sqrt{{8\over a^2}+2}}+{\mathcal O}(T), 
\ee
where $\tau$ is the time scale of the fastest mode in the system. 
The right hand side relation of Eq.\eq{tau} gives \at{an estimate of} the 
maximal frequency from the free dispersion relation 
$p_0^2=4(\sin^2p_xa/2+\sin^2p_ya/2)/a^2+M^2(T)$
of the lattice Hamiltonian system for fixed $a$ in the limit $a_t\to0$.
The fast modes absorb energy from the OP
in a single time step for $a_t>\tau$, and the friction term should appear 
in the effective equation of motion.

\begin{figure}
\begin{center}
\includegraphics[width=8.5cm]{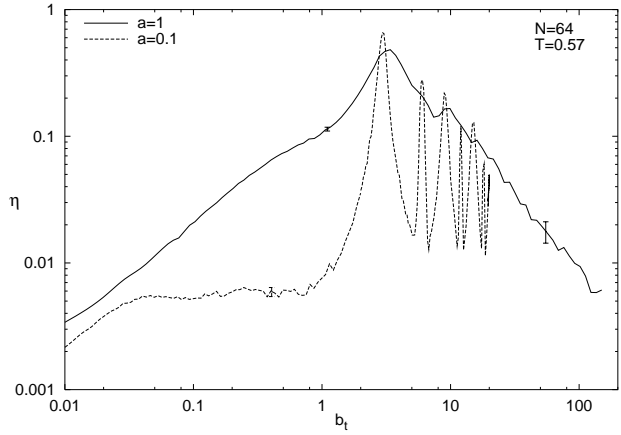}
\end{center}
\vspace*{-0.8cm}
\caption{\label{atdep}
The friction $\eta (\la\Phi\ra_{h=0})$ as the
function of the timelike lattice spacing $b_t=na_t$. The time scale of the
fastest
mode is $\tau\approx2$ and $0.22$ for $a=1$ and 0.1, respectively,
according to the perturbation expansion. Typical error bars
appear on the curves both for the IR and UV regimes.}
\end{figure}   

With help of the temporal blocking
\be\label{tblo}
\Phi(t)\to\Phi_{n}(t)={1\over n}\sum_{k=-n/2}^{n/2-1}\Phi(t+ka_t)
\ee
one can construct the trajectories $\Phi_{n}(t)$
corresponding to larger values of the time
cut-off, $b_t=na_t$. Such blocking
was performed in time up to $n=2000$ and a discrete time
 equation of motion of the form
Eq.(\ref{eqfriction}) was reconstructed for the blocked trajectory
\at{$\Phi_n(t)$}. 
 The non-trivial dependence of $\eta$ on the temporal cutoff $a_t$ is shown
in Fig. \ref{atdep}. 

One can distinguish two regimes separated by a crossover apparently
independent of the lattice size, located at $b_t\approx0.2\tau$. A scaling
behavior is observed on the UV side, where $\eta(b_t)$ tends to zero with a
critical exponent close to 1 and the dissipative force being constant. This
result should be independent of the actual blocking details. In the other
regime, on the IR side $\eta$ goes over first into a $b_t$-independent
regime, corresponding to the saturation of the energy transfer from the OP
and giving a stable, microscopic definition of the friction coefficient.
Finally, in the far IR part a qualitatively different oscillatory behaviour
sets in.

The location of the crossover from the UV scaling regime to the plateau can
be understood by writing the fluctuation dissipation theorem in the
corresponding discretised form: $\langle\zeta^2\rangle =2\eta T/b_t$.
The linearly increasing regime of $\eta (b_t)$ implies constant second moment 
for the noise.  When $\eta (b_t)$ reaches the plateau the second moment of
the noise decreases like $1/b_t$. The crossover therefore is located at the
autocorrelation time scale of the noise.

A qualitative interpretation of the oscillatory IR regime can be based on
the observed small amplitude beating phenomenon in the OP trajectory. This
can be recognized by closer inspection of the left side of
Fig.\ref{time_hist}, which persists further also on the right side. It is
reflected in the OP autocorrelation function, too, since in the course of
the blocking an interference effect occurs on the right hand side of
Eq.\eq{tblo} due to this regularity. This feature is relevant to the value
of $\eta (b_t)$, responsible for the decay of all kinds of fluctuations.

The appareance of peaks in $\eta(b_t)$ at both the
maximal destructive and constructive interferences can be modeled
semi-quantitatively by
identifying the  beating part of the OP-motion 
${\textrm{Re}}\delta\Phi={\textrm {Re}}\delta\Phi_0\exp(i\omega t)$
with the stationary solution of a single weakly damped
driven harmonic oscillator,
$\ddot{\delta\Phi}+\eta\dot{\delta\Phi}+\omega_0^2\delta\Phi=
f_0\exp(i\omega t)$.
The blocking acts on the trajectory $\delta\Phi(t)$
as $\delta\Phi(t)\to u\delta\Phi(t)$, where 
$u=(\exp(i\omega b_t)-1)/i\omega b_t$.
It does not change the relative phase of the driving force and 
of the blocked oscillation amplitude, leading to a relation
between the parameters of the original and the blocked equation of motion:
\be
-\omega^2+i\eta\omega+\omega_0^2=-\omega^2u^2+i\bar\eta\omega u+\bar\omega_0^2.
\ee

The friction coefficient for the blocked trajectory turns out to be
$\bar\eta=(\eta+\omega \textrm{Im}(u^2))/\textrm{Re}(u)$. It is easy to see
that $\textrm{Re}(u)$ is vanishing at maximal constructive and destructive 
interferences. This provides singularities in $\bar\eta$.  Whenever
$\textrm{Re}(u)=0$ we have $\textrm{Im}(u^2)=0$ and numerator changes sign
in the vicinity of the singularity. Thus
$\bar\eta>0$ apart for a short time interval around the singularities where
the non-harmonic features should stabilise the fluctuations and
keep $\bar\eta>0$, as observed in our simulation. 

The $a$ dependence appearing in Fig.\ref{atdep} arises from the following
two effects.  One is that for larger $a$ the maximal oscillation frequency
is smaller and the time resolution of the system becomes cruder. Another is
that larger $a$ represent bigger physical volume, many more soft modes and
less harmonic system, which tends to invalidate the simple picture based on
a single harmonic mode. As a result the effects of oscillatory nature will
be smeared, as one clearly recognizes in the figure.

The quality of any proposed deterministic equation of motion (e.g. equations
similar to Eq.(\ref{eqfriction}) with zero on the right hand side), can be
judged by the amplitude and the autocorrelation of its error term, {\it the
noise term} $\zeta$. The amplitude of the noise was found at least two-three
order of magnitude below the average level of the force as fitted to Eq.
\eq{eqfriction}.
The autocorrelation function of the noise of Eq. \eq{eqfriction} appeared to
be local, approximately of the form
$\la\zeta_d(t)\zeta_d(t')\ra\approx\delta''(t-t')$.  The status of the
fluctuation-dissipation theorem will be investigated for more complex field
theoretical systems in future investigations.

\begin{figure}[htbp]
\begin{center}
\includegraphics[width=8.5cm]{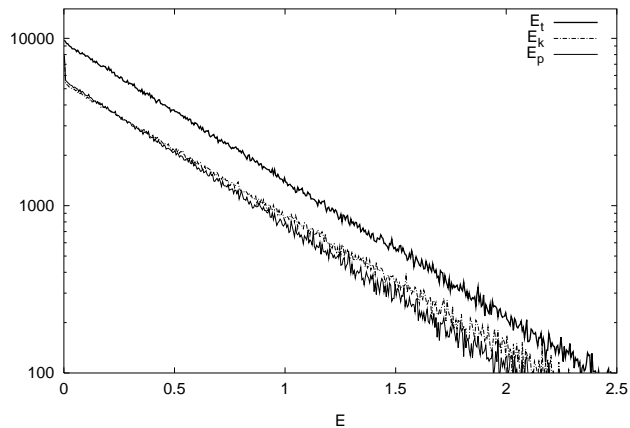}
\end{center}
\vspace*{-0.8cm}
\caption{\label{th} The histogram of the energies $E_t$, $E_k$
and $E_p$ of Eq.(\eq{ens}) on a $256\times256$ lattice with $h=2/\sqrt{6}$.}
\end{figure}

The selfconsistency of the definition of the temperature in Eq.(\ref{tempeq})
and also the establishment of the thermal equilibrium
can be tested further by plotting histograms for the following quantities:
\bea\label{ens}
E_k&=&\frac{1}{2}\left({d\Phi\over dt}\right)^2,\nonu
E_p&=&{\mu^2\over2}(\Phi-\langle\Phi\rangle_h)^2,\\
E_t&=&E_k+E_p,\nonumber
\eea
where $\mu^2$ is the slope of the force as the function of the order 
parameter, determined numerically. 
Typical results for the energy-histograms 
are shown in Fig.\ref{th}. 
It shows perfect agreement of all slopes and good agreement with the 
expectations based on equipartition of the energy between the kinetic and
the potential parts. We have checked for several temperatures, that the 
temperature
determined by these histograms and from the $|{\bf k}|$-spectra of the kinetic 
energy agree. This piece of information is parallel to the recent detailed
investigations of the thermalisation in $(1+1)$-dimensional $\Phi^4$-theories
\cite{parisi97,aarts00}.

It is worthwhile noting that the relaxation from a given initial
condition to the thermally distributed state like in this figure takes at
least one order of magnitude longer time for systems in the symmetric phase
compared to the spontaneously broken case. At higher energy density
(temperature) one would expect larger collision frequency, therefore shorter
thermalisation time. The opposite result hints to the importance of slow,
soft modes, whose presence is due to the symmetry breaking mechanism. We
shall argue in the next Subsection that these modes are responsible for the
realization of the Maxwell-cut in the potential term of the effective
equation of motion for the OP. What we find remarkable is that these modes
are present not only in the mixed phase but also near (meta)stable
equilibria, among the dynamical fluctuations around the ordered vacuum in
contrast to the massive perturbative excitation spectrum in the equilibrium.

We complete the analysis of this subsection with the remark, that after the
transition to the stable vacuum shortly a thermalised distribution is
recovered for the OP at somewhat increased temperature which agrees with the
value given by the equipartition.

\subsection{\label{ss:jump}Jump from the metastable to the stable vacuum}
As it has been emphasized in the Introduction, there are (at least)
two different descriptions of the transition from the
metastable state to the stable one. \at{In the first approach an}\rt{The}
expansion around the
lower pass, the \rt{``sphaleron''}\at{critical bubble (bounce)}
configuration yields a detailed
space-time picture and the transition rate by means of the 
analytic continuation of the potential experienced by the OP near the
(meta)stable position \cite{Langer67,Coleman80}.

Another possibility is to provide for the OP a probabilistic description,
obtained by the elimination of all other degrees of freedom in some kind of
blocking procedure.  This description is based usually on some underlying
Master equation for the probability distribution of the OP and the resulting
Fokker-Planck equation \cite{vankampen}. The transition to the stable state
appears in this approach formally as a tunneling solution of the
Fokker-Planck equation. The probabilistic feature of the dynamics of the OP
is supposed to arise from near the (meta)stable equilibrium assuming 
\at{a statistical ensemble of}\rt{averaging over} the initial conditions.

In our simulation we find results analogous with the predictions of the
probabilistic description by analysing the OP-motion starting from a single,
well defined initial condition.  One might wonder at this point if it is
possible to understand the probabilistic tunneling of the OP by following
the system from a unique initial condition. The self averaging in time can
not be used for this argument since such a transition occurs only once
during the evolution in a (quasi)irreversible manner.
\at{We have to develop a third approach to the \hbox{metastable$\to$stable}
transition.}

In general, the effective equation of motion for $\Phi(t)$ reflects the
typical landscape of the microscopic potential energy functional around the
actual point $\Phi({\bf x},t)$ in the configuration space.  The classical
origin of what appears as a tunneling on the Fokker-Planck level must be the
arrival of $\Phi({\bf x},t)$ to the vicinity of some narrow valley opening
up towards the stable vacuum. In traversing this valley the landscape
changes and the typical fluctuations will be different from those felt in
the metastable regime. The constants parametrising the effective equation of
motion must reflect this change.

Our goal in this Subsection is to construct \rt{such a ``classical tunneling''}
\at{an effective} description of the transition to the stable state by
carefully tracing the time evolution of the OP.  This will be achieved by
projecting the microscopic equation of motion onto the homogeneous mode and
phenomenologically parametrising it similarly to Eq.(\ref{eqfriction}).

\begin{figure}   
\begin{center}
\includegraphics[width=8.5cm]{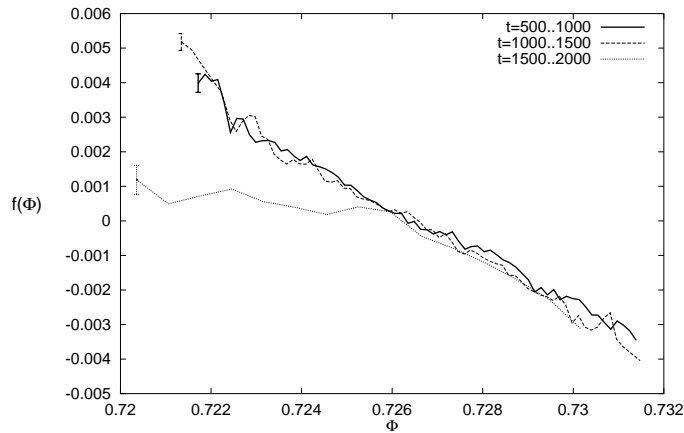}
\end{center} 
\vspace*{-0.8cm}              
\caption{\label{Fig:phi} The force $f(\Phi)$ as the function of the order
parameter on $100\times100$ lattice with $h=-0.04/\sqrt{6}$. 
Each curve corresponds to time intervals of length $500|m|^{-1}$.
The interval corresponding to the continous curve ends $1000|m|^{-1}$
before the transition, the long dashed curve refers to the next
time interval and the short dashed curve is the force measured during the last 
time interval preceding the transition. One clearly observes the bending
down left from the central OP-value in the metastable regime. Typical
error bars are shown on the leftmost points of the three curves.}
\end{figure}

As long as the system is far from the narrow valley of the instability
the force is time independent and agrees with the force derived from the
perturbative effective potential according to the part of Fig.~\ref{fphi}
corresponding to $\Phi>0.9$. When \rt{the external magnetic field brings}
the system \at{arrives} close\rt{r} to the entrance of the unstable valley,
$(\Phi\approx0.8)$,  the soft modes start to be important. This is reflected 
in the slight glitch in the leftmost piece of the measured force law in 
Fig.~\ref{fphi}. A sequence of glitches results in a situation depicted in 
Fig.~\ref{Fig:phi}.  It shows the force acting on the OP in three successive time 
intervals preceding the event of tunneling for  $h=-0.04/\sqrt{6}$. The 
$f(\Phi )$ curves determined using Eq.~(\ref{eqfriction}) in the disjoint 
time subintervals coincide within error bars for all, but the last one.  In
this last interval preceding directly the transition towards the direction
of negative $\Phi$ values, the fitted force bends down and its average
becomes (a small positive) constant. This is characteristic feature of the
instants when the system finds the entrance into the unstable potential
valley. The \rt{flattening of the potential, the} vanishing of the force is
an indication for the dynamical realization of the Maxwell-cut. The OP moves
fast through the valley and the method of fitting the trajectory to
Eq.~(\ref{eqfriction}) for finding the force fails due to the insufficient
statistics.

The fluctuation moments depicted in Fig. \ref{time_hist} tell a bit more
about this region. The increased values of the moments, the renormalized
coupling constants in Wilsonian sense at vanishing momentum, indicate the
enhanced importance of the soft interactions as the OP tunnels through the
mean field potential barrier. This softening makes the OP fluctuating with
larger amplitude. The valley of instability is in a surprising manner
flatter than the typical landscape around equilibrium.  The flatness along
the motion of the OP (the mode ${\bf k}=0$ in momentum space) comes from the
Maxwell-cut. The average curvature of the potential in the transverse 
directions, i.e. for modes with  $|{\bf k}|\not=0$, can be estimated by the 
second functional derivative of the  two dimensional effective action with 
cutoff $|{\bf k}|$, which  can be taken as ${\bf k}^2+V''_{\bf k}(\Phi)$.
The increased MS fluctuation of the OP corresponds to a decrease in
$V''_{{\bf k}=0}(\Phi)$ in the valley. It pushes down
$V''_{\bf k}(\Phi)$ also for small non-vanishing $|{\bf k}|$, in the
low momentum regime which is expected to be the most influenced by the 
changing landscape.

The lesson of Fig. \ref{Fig:phi} is that the potential itself should be considered
as a fluctuating quantity. This helps to translate into more quantitative 
terms the above qualitative points. We discuss the projection  of the
microscopic equation of motion \eq{latteq} onto the
zero momentum sector,
\be
0=\ddot\Phi-\Phi+\Phi^3+3\overline{\varphi^2}^V\Phi
+\overline{\varphi^3}^V-h\equiv\ddot\Phi+V_{\textrm{inst}}'(\Phi),
\label{eq_OP}
\ee
where the symbol $\overline{\varphi^n}^V$ means the space average of 
$\varphi^n$, ($\overline{\varphi}^V=0, \Phi (t,{\bf x})=\Phi (t)+
\varphi ({\bf x},t)$). The instant potential introduced in the second line 
contains a {\it deterministic} piece, which is the sum of the tree-level 
potential
and of the slowly varying part the second and third moments
(eg. $(\overline{\varphi^n}^V)_{det}$). This last 
feature clearly appears graphycally in Fig.\ref{time_hist} and a simple
model based on a two-phase model of the transition period will be constructed 
below to account for it. The remaining oscillating pieces of the moments 
provide the probabilistic fluctuating contribution to the instant potential:
\be\label{inst}
V_{\textrm{inst}}(\Phi)=-h\Phi-\frac{1}{2}\Phi^2+{1\over4}\Phi^4
+(\overline{\varphi^3}^V)_{det}\Phi
+{3\over2}(\overline{\varphi^2}^V)_{det}\Phi^2+\zeta_0\Phi+{3\over 2}
\zeta_1\Phi^2.
\ee
The additive ($\zeta_0$) and the multiplicative ($\zeta_1$) noises are given 
by the differences
\bea
\zeta_0(t)&=&\overline{\varphi^3}^V(t)-(\overline{\varphi^3}^V)_{det},\nonu
\zeta_1(t)&=&\overline{\varphi^2}^V(t)-(\overline{\varphi^2}^V)_{det}.
\label{zetadef}
\eea
\at{Note that no friction terms can be introduced in a natural way into
the system \hbox{(\ref{eq_OP}-\ref{zetadef})}. Therefore we face the intriguing
question, how irreversibility is realised in such a system. The
time-correlation matrix $\overline{\zeta_i(t)\zeta_j(t+\tau)}^t$
appears to us to be the key object for its investigation, to which we
plan to return in the future.}
\rt{
Note that the breakdown of the time reversal invariance is expected to result
from the cross-correlation between the two noises rather than the effective 
equation of motion \eq{eqfriction} with a single noise and a friction term.}

In order to gain more insight how this works we build into 
Eq.\eq{eq_OP} the consequences of the mixed two-phase picture 
of the phase transformation. The microscopical basis for this picture
is provided by the thermal nucleation whose quantitative discussion is
given  in the Subsection \ref{ss:nucl}.

More specifically, we assume that the space can be splitted into sharp 
domains (neglecting the thickness of the walls in between), where the field is 
the sum of the constant background values 
$\Phi_{0\pm}$ and the fluctuations $\tilde\varphi_{\pm}$ around it,
\be
\Phi_\pm({\bf x},t)=\Phi_{0\pm}+\tilde\varphi_\pm({\bf x},t).
\ee
We assume local equilibrium in both phases, based on the smooth evolution
of the temperature as displayed in Fig.\ref{k_hist}.

The actual value of the order parameter is determined by the surface
ratio $p(t)$
occupied by the stable phase:
\be
{\Phi(t)}=p(t)\Phi_{0-}+(1-p(t))\Phi_{0+}.
\ee
Simple calculation then yields
\bea
\label{bubble_moments1}
(\overline{\varphi^2}^V)_{det}(t)&=&
\frac{\Phi_{0+}-\Phi(t)}{\Phi_{0+}-\Phi_{0-}}
\left(\overline{\Phi_-^2({\bf x},t)}^V-\overline{\Phi_+^2({\bf x},t)}^V\right)
+\Phi_{0+}^2
-\Phi^2(t)+\overline{{\tilde\varphi_{+}}^2}^V,\\\nonumber
(\overline{\varphi^3}^V)_{det}(t)&=&
\frac{\Phi_{0+}-\Phi(t)}{\Phi_{0+}-\Phi_{0-}}
\left(\overline{\Phi_-^3({\bf x},t)}^V-\overline{\Phi_+^3({\bf x},t)}^V\right)
\\&&
+\overline{\Phi_+^3({\bf x},t)}^V-3\Phi(t)(\overline{\varphi^2}^V)_{det}(t)-
\Phi^3(t),
\label{bubble_moments2}
\eea
where the volume averages should be read off the corresponding 
equilibria on the two sides of the transition. If one takes the values of 
$\Phi_{0\pm},\overline{\tilde \varphi^n_{\pm}}^V$ 
from the respective equilibria determined in the same 
simulation, a quite accurate description of the shape of the
two fluctuation moments arises in the whole transition region and its close 
neighbourhood \at{using $\Phi(t)$ to parametrize their $t$-dependence. 
(Note that $\overline{\Phi^n_\pm(x,t)}^V$ does not depend on time.)
The deterministic part of the moments are therefore well-defined functions
of $\Phi$.} In this way, a simple explicit construction can be given for the
effective noisy equation of the OP-motion, if the above pieces are 
supplemented by the correlation characteristics of the two kinds of noises.

The RMS (root mean square) fluctuations found numerically on 
the transition
part of the trajectory and its close neighbourhood are the following:
\be
\sqrt{\overline{\zeta_0^2(t)}^t}=0.0063(10),\quad \sqrt{\overline
{\zeta_1^2(t)}^t}=0.0054(10), \qquad \textrm{for}\, N=64
\ee
(its magnitude increases with the size of the system).
The magnitude of their equilibrium cross correlation was found $\sim 10^{-5}$.

As a corollary of this construction one can demonstrate the absence of the
deterministic part of the acceleration of the order parameter in the 
transition period. 

Substituting Eqs.(\ref{bubble_moments1}), (\ref{bubble_moments2})  into 
Eq.(\ref{eq_OP}), one finds for the deterministic part of the force,
\be
f\left(\Phi\right)=\left(-1+\frac{\vavr{\Phi_+^3({\bf x},t)}-
\vavr{\Phi_-^3({\bf x},t)}}
{\Phi_{0+}-\Phi_{0-}}
\right)\Phi+\frac{\Phi_{0+}\vavr{\Phi_-^3({\bf x},t)}-
\Phi_{0-}\vavr{\Phi_+^3({\bf x},t)}}{\Phi_{0+}-\Phi_{0-}}-h.
\label{phieom}
\ee
The average of the equations of motion in the respective equilibria,
\be
\la\Phi_\pm^3({\bf x},t)\ra-\Phi_{0\pm}-h=0
\ee
implies the vanishing of the deterministic force in
Eq.~(\ref{phieom}), when exploiting the equality of the volume and the
ensemble average in this case.
Eq.~(\ref{eq_OP}) transforms into
\be
\ddot\Phi(t)+\zeta_0(t)+3\zeta_1(t)\Phi(t)=0.
\ee
This is the dynamical realization of the Maxwell-construction holding when
the mixed phase model with local equilibrium is valid.

If the force is calculated from the 
full instant potential, $V(\Phi)_\textrm{inst}$ in Eq. \eq{inst},
it depends parametrically on the 
moments $\overline{\varphi^2}^V(t)$ and $\overline{\varphi^3}^V(t)$. The
approximate trajectory
$\Phi_\textrm{inst}(t)$, defined by minimising $V(\Phi)_\textrm{inst}$
with respect to $\Phi$, where the moments are taken from the
numerically determined time evolution, reproduced accurately
the observed OP, $\Phi(t)$ with the following RMS/unit time:
\be
\sqrt{\overline{(\Phi (t)-\Phi_{\textrm {inst}}(t))^2}^t}=0.0014(5),
\qquad \textrm{for}\, N=64.
\ee
This construction interpretes the OP-trajectory 
as a continous deformation of the instant potential with the OP ``sitting''
permanently in the actual minimum. Notice that such motion is possible 
only if the OP continously undergoes some sort of dissipation.

The vanishing of the acceleration was checked by comparing the computer
generated OP trajectory in the transition region with a ballistic motion in
a viscous medium. In particular it was tested that the ratio $h/\dot\Phi$ is
nearly time and $h$ independent for small enough $h$.  The dynamical
friction measured by the above ratio tends to a constant with decreasing $h$
at fixed lattice size. Although the error for \at{its} value in each
individual transition is rather small, the central values obtained in
different runs fluctuate quite strongly, which leads eventually to a large
error of the mean calculated as an average of runs with different initial
configurations having the same energy density.

It is worth to note that the order of magnitude of these ``renormalized'',
i.e.~IR determined values of the friction coefficient agree with the peak
value in Fig.~\ref{atdep} for $N=64$. The ``running'' friction coefficient
determined by the blocking in time, however, drops as
$b_t$ is increased and takes extremely small values at $b_t\approx20$,
the average time length of the ballistic fits for the transition. 
This serves an example
for the dependence of the renormalization group flow on the details
of the blocking. It remains to be understood whether the agreement
between the peak value and the ballistic fit is an accident
or follows from the internal dynamics.

\subsection{\label{ss:nucl}Connection with the nucleation picture}

In this subsection we analyse the transition using the more conventional
statistical approach of thermal nucleation theory.

From the study (on lattices up to $N=512$) of the detailed microscopical field 
configurations it turned out that the phase transformation starts by the 
nucleation of a single bubble of the stable ($\Phi <0$) phase. 
Late-coming further large bubbles are aggregated to it as
well as the small size (consisting of $n_{bubble}<50-60$ joint sites) 
bubbles. Following the nucleation the expansion rate of the large bubble 
governs the rate of change of the order parameter.
To a very good approximation each individual transition could have been 
characterised by a constant value of $\dot\Phi$ in this interval. Two 
mechanisms are known to lead to this behavior. The first is scattering of hard
waves (``particles'') off the bubble wall, while in the second
the expansion velocity is limited by the diffusive 
aggregation of smaller bubbles \cite{aggreg}. 
The latter process seem to be the dominant in the kinetic Ising model 
\cite{Binder74}.

\begin{figure}[htbp]
\begin{center}
\includegraphics[width=8.5cm]{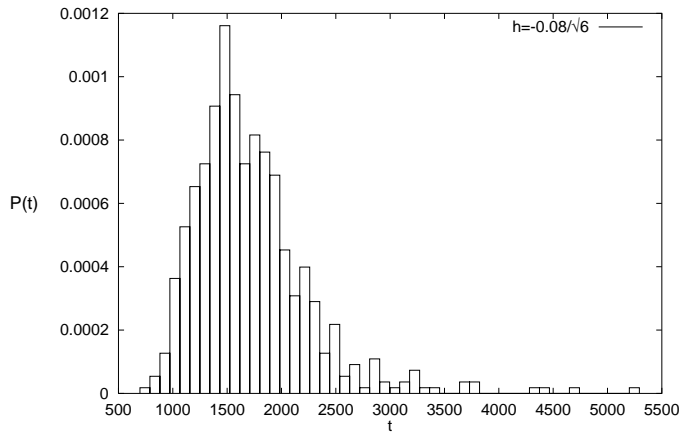}
\end{center}
\vspace*{-0.8cm}
\caption{\label{release_hist} The normalised
``release'' time histogram counting the frequency of a certain time
moment-bin when the order parameter first takes a negative value. The figure
represents the statistics based on 750 events at $h=-0.08/\sqrt{6}$ on an
$N=64$ lattice.}
\end{figure}

The statistics of the ``release'' time of the supercritical bubble from the
metastable state shows at first sight rather peculiar characteristics.
The binned histogram for larger values of $h$ shows an asymmetric
peaked structure which apparently deviates from the exponential distribution
characterising the thermal nucleation scenario (Fig.\ref{release_hist}).
The very early transitions ($t<t_{max}$) seem to be suppressed for small
values of $h$. We expect they correspond to transitions, which happen before
the system reaches the metastable state.

The fall-off for $t>t_{max}$
starts nearly exponentially, but the histogram develops a very long tail
when $h\rightarrow 0$. The bigger sample is used for estimating 
the probability distribution, the more suppressed is the weight of these 
events in the normalised distribution. In practice this means a longer 
time interval where a good linear fit can be obtained to the 
log-linear histogram. Eventually, the separation of a clean 
exponential signal is possible, and the slope can be compared with 
predictions of the nucleation theory. 

Knowing the frequency histogram of the time $t$ that elapsed 
before the system has escaped from the metastable region by nucleating a 
growing bubble of the stable state, we can construct the nucleation rate per 
unit area: $\Gamma(h)$. Indeed, for a lattice size $L$ we can read off 
$\Gamma$ by fitting to the expected form $P_h(t)\sim exp(-t\Gamma(h) L^2)$. 

\rt{We didn't measured the elapsed time 
from the begining of the metastable region but we have thrown out those points
that didn't correspond to a nucleation from a metastable state. In the
figure we can see the $h$ dependence of the $\ln \Gamma(h)$ obtained by
fitting to the measured data. This $h$ dependence is what we have to
explaine theoretically.}

\begin{figure}[htbp]
\begin{center}
\includegraphics[width=8.5cm]{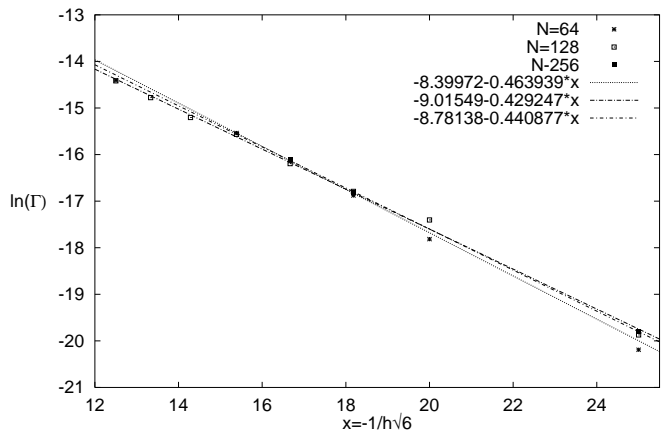}
\end{center}
\vspace*{-0.8cm}
\caption{\label{t_size}
Nucleation rate for unit lattice surface
versus the reciprocal of the magnetic field for different
lattice sizes. The error bars are smaller than the size of the symbols
representing the data points.}
\end{figure}

The nucleation rate is proportional to the volume of the system $L$. In
Fig.\ref{t_size} we see evidence for the size independence of the
transition rate per unit surface. The logarithm of this quantity can be
estimated following the standard nucleation theory
\cite{Langer67,Coleman80,Gleiser93} where the nucleation rate per unit
area is given by:
\be
\label{eq:nucl_rate}
\Gamma\sim e^{-S_2/T},\qquad
\textrm{with}\qquad
S_2=S(\Phi_B)-S(\Phi_f(h)),
\ee
$\Phi_B$ is the bounce solution and $S$ the Euclidean action 
$S=\int d^2x\left[\frac{1}{2}(\partial_\mu\Phi)^2+U(\Phi)\right]$
with $U(\Phi)$ defined below in Eq. (\ref{Eq:Udef}).

In the so called thin wall approximation, when the energy density
difference beween the true and the false vacuums is tiny the coefficient
$S_2$ can be calculated analiticaly and in the relation
\be
\ln \Gamma =K-{S_2\over T}.
\ee
$S_2$ is determined approximately by the action of the bounce solution,
connecting the two vacua. In the investigation below we have studied at fixed 
temperature the $h$-dependence of $S_2/T$, assuming the $h$-independence 
of $K$, although this might be not the case.

The bounce is a solution of the Euclidean Euler-Lagrange equation of a
theory defined by the action $S(\Phi)=\int
d^2r\left(\frac{1}{2}(\nabla\Phi)+\frac{1}{2}m^2\Phi^2+
\frac{\lambda}{24}\Phi^4+h\Phi\right)$
such that (see fo exemple \cite{Coleman80})
\begin{enumerate}
\item
$\Phi_B$ approaches the false vacuum $\Phi_f(h)$ at the infinity,
\item
$\Phi_B$ is not a constant,
\item
$S(\Phi_B)$ is the minimal amoung the solutions obeying
1. and 2. \,.
\end{enumerate}

Since the nucleation proceeds through approximately spherical bubbles we need
an $O(2)$-invariant solution of the equation of motion i.e. the solution of
\be
\frac{d^2\Phi}{dr^2}+\frac{1}{r}\frac{d\Phi}{dr}=m^2\Phi+
\frac{\lambda}{6}\Phi^3+h,
\label{eq:O2bounce}
\ee
with the boundary conditions ensuring that the field be regular at the origin 
($\Phi'_B(0)=0$) and in the false vacuum at spatial infinity 
($\Phi_B(\infty)=\Phi_f(h)$).

As it is well-known, in the thin wall
approximation the differential equation  (\ref{eq:O2bounce})  reduces to
\be 
\frac{d^2\Phi}{dr^2}=\frac{dU_+(\Phi)}{d\Phi},
\label{eq:soliton}
\ee
where $U_+(\Phi)$ is the a symmetric function of $\Phi$, with minima at 
$\Phi_{f,t}=\pm\sqrt{\frac{6}{\lambda}}|m|$, obtained by replacing $h$ by $0$
in the tree level potential of Lagrangian  and dropping the constant term:
\be
\label{Eq:Udef}
U(\Phi)=\frac{1}{2}m^2\Phi^2+\frac{\lambda}{24}\Phi^4+h\Phi=
\underbrace{\frac{\lambda}{24}\left(\Phi^2+6\frac{m^2}{\lambda}\right)^2}_
{U_+(\Phi)} +h\Phi-\frac{3}{2}\frac{m^4}{\lambda}
\ee 
Multiplying equation (\ref{eq:soliton}) by $\Phi'$ one can see that it admits a 
first integral 
$\left[\frac{1}{2}(\Phi')^2-U_+\right]'=0,$
whose value is determined by the condition that $\Phi(\infty)=\Phi_f$
and $\frac{d\Phi}{dr}\Bigl|_{\infty}=0$ because of the finiteness of the energy.
So, we have 
\be
\frac{1}{2}(\Phi')^2-U_+(\Phi)=-U_+(\Phi_f)=0.
\ee
This equation determines a monotonic $\Phi$ which goes from $\Phi_t$ to
$\Phi_f$. The solution depends on a single integration constant $\bar r$, 
a point at which the $\Phi$ takes the average of its two extreme values 
$\Phi_t$ and $\Phi_f$:
\be
\int_{0}^{\Phi}\frac{d\bar\Phi}
{\sqrt{2\left[U_+(\bar\Phi)-U_+(\Phi_f)\right]}}=r-\bar r.
\ee
Performing the integral  we obtain the well-known kink solution 
\be
\Phi(r)=|m|\sqrt{\frac{6}{\lambda}}\tanh\left[\frac{|m|}{\sqrt2}(r-\bar r)
\right].
\ee
Expressed analytically the thin wall approximation of the bounce looks like:
\be
\Phi_{B}(r) = \left\{
\begin{array}{ll}
\Phi_{t},  & 0 < r < \bar r - \Delta r \\
\Phi_{wall}(r),  & \bar r - \Delta r < r < \bar r + \Delta r\\
\Phi_{f},     & r > \bar r + \Delta  r
\end{array}
\right.
\label{eq:bubble}
\ee
which describes a bubble of radius $\bar r$ of the stable phase $\Phi_{t}$
embedded in the metastable phase $\Phi_{f}$. 
$\Phi_{wall}(r)=|m|\sqrt{\frac{6}{\lambda}}\tanh\left[\frac{|m|}{\sqrt2}
(r-\bar r)\right]$ describes the bubble wall separating the two phases. 

The unknown parameter $\bar r$, the radius of the ball, can be determined by 
computing the action in the above approximation and demanding that it be
stationary under variations of $\bar r$. We have to calculate the action in
three distinct regions according to the solution (\ref{eq:bubble}).\\
Outside the wall, $\Phi=\Phi_f$ hence, ${S_2}_{outside}=0$.\\
Inside the wall, $\Phi=\Phi_t$ hence, 
${S_2}_{inside}=2\pi \frac{\bar r^2}{2}h\left(\Phi_t-\Phi_f\right)=
2\pi\bar r^2h\Phi_t-= -2\pi\bar r^2h\sqrt{\frac{6}{\lambda}}|m|$.\\
Within the wall, in the thin-wall approximation,
\bea
\nonumber
{S_2}_{wall}&=&2\pi\bar r\int dr\left[\frac{1}{2}{\Phi'}_{wall}^{2}+
U_+(\Phi_{wall})-U_+(\Phi_f)\right]\\
&=&2\pi\bar r\int_{\Phi_t}^{\Phi_f}d\Phi_{wall}\sqrt{2U_+(\Phi_{wall})}=:
2\pi\bar r S_1. 
\eea
So, we can conclude that the action asscociated with the bounce in the
thin-wall approximation is:
\be
\label{Eq:kink}
(S_2)_d=-2\pi\bar r^2h\sqrt{\frac{6}{\lambda}}|m|+2\pi\bar r S_1,
\ee 
where the $S_1$ is the energy of the kink
\be
S_1=\int dr
\left[\frac{1}{2}{\Phi'}_{wall}^{2}+U_+(\Phi_{wall})\right]=
3\frac{|m|^4}{\lambda}\int_{-\infty}^{\infty}dr
\frac{1}{\cosh^4\left(\frac{r|m|}{\sqrt2}\right)}=4\sqrt2\frac{|m|^3}{\lambda}.
\ee
The first term on the right hand side of Eq. (\ref{Eq:kink}) corresponds to the 
volume energy of a bubble of radius $\bar r$,
the second one to its surface energy. 

From the stationarity of $S_2$ with respect of the variation of $\bar r$ 
($\frac{dS_2}{d\bar r}=0$) we can calculate $\bar r$:
\be
\bar r=\frac{S_1}{2h|m|}\sqrt{\frac{\lambda}{6}}=
\frac{2}{h\sqrt3}\frac{|m|^2}{\sqrt{\lambda}}.
\ee
For the critical bubble size a very simple expression is obtained for the 
exponent of the rate in terms of dimensionless quantities:
\be
{S_2(\textrm{thin~wall})\over
T}={16\pi\over\sqrt{6}}{|m|^5\over\lambda^{3/2}}
{1\over h_dT_d}={4\pi\over9}{1\over hT}.
\ee

\rt{
With dimensionless quantities we have the relation
\be
T_r=6\,0.576/\lambda_r.
\ee
Using this and the relation between dimensionfull and dimensionless
quantities we obtain: 
\be
\frac{B}{T}=\frac{8\pi}{3}\frac{|m|^2}{\lambda T}\frac{1}{h_r}=
\frac{8\pi}{3}\sqrt6\frac{1}{\lambda_r T_r}\frac{1}{h_{polonyi}}
=\frac{5.99}{h_{polonyi}}.
\ee}

\rt{
The condition for the validity of our approximation is that $\bar r$ is
large compared to the length scale on which the $\Phi$ varies significantly.
This means that $|m|\bar r \gg 1$ i.e. 
$\frac{3}{\sqrt2}\frac{h}{|m|^3}\sqrt\frac{\lambda}{6} \ll1$. On the lattice
this inequality is
\be 
\frac{3}{\sqrt2}h_r\ll1 \qquad\textrm{or} \qquad 
\frac{\sqrt3}{2}h_{polonyi}\ll1,
\ee
so, the thin-wall approximation seems resonably in the range of the
parameter $h$ used.  Neverthless, let's verify the validity of the
approximation by a direct lattice calculation of the $O(2)$ symmetric
bounce.}

\rt{
In the thin wall approximation the following expression is used
based on the tree level potential:
\be
(S_2)_d=-2\pi\bar r^2h\sqrt{\frac{6}{\lambda}}|m|+8\sqrt2\pi\bar r
\frac{|m|^3}{\lambda},
\ee
}
The linear dependence on $1/h$ is fulfilled in our numerical calculations
very well, but the predicted action
is more than one order of magnitude larger than what can be derived from the
slope of Fig.\ref{t_size}: $S_2/T (\textrm{measured})\approx 0.1/(hT)$.
If one relaxes the
thin wall approximation and solves numerically the two-dimensional bounce
equation directly for several $h$, one finds 
$S_2(\textrm{bounce})/T=0.78/(hT)$.

\rt{
On the lattice the equation of the bounce and the action associated with it 
looks like:
\be
\frac{d^2\Phi}{dx^2}+\frac{1}{x}\frac{d\Phi}{dx}=-\Phi+\Phi^3+h
\label{eq:latO2}
\ee
and
\be
S(\Phi)=\pi\sum \Delta x x\left[
\frac{1}{2}\left(\frac{d\Phi}{dx}\right)^2-
\frac{1}{2}\Phi^2+\frac{1}{4}\Phi^4+h\Phi\right]
\label{eq:Slat}
\ee
respectively.
The 2-nd order DE can be written as a system of two first order coupled DE:
\bea
\frac{d\Phi}{dx}&=&Y(x)\\         
\frac{dY}{dx}&=&-\Phi+\Phi^3+h-\frac{Y(x)}{x}
\eea}
We mention here that
for finding the bounce solution \rt{of (\ref{eq:latO2}) all} we have to 
\rt{do is to} 
choose an initial value for $\Phi>{\Phi}_{min}(h)$ evolve it and see what's 
happening at a later $x$. If $\Phi>{\Phi}_{max}(h)$ then the solution 
overshoots and if $\frac{d\Phi}{dx}<0$ then the solution undershoots. 
We have to tune the initial value of $\Phi$ and the steps of the fourth order 
Runge-Kutta algorithm we used, in order to achieve that the solution stays very 
close to  ${\Phi}_{max}(h)$, as long as possible. \rt{The numerical solution 
is depicted in Fig. \ref{Fig:bounce}}. Given the bounce solution we can simply evaluate its
action by performing a numerical integration \rt{of Eq. (\ref{eq:Slat})}.

\rt{
\begin{figure}
\begin{center}
\includegraphics[width=8cm]{bounce12.eps}
\end{center}
\vspace*{-0.8cm}
\caption{\label{Fig:bounce}
The numerical bounce solution.}
\end{figure}
}

Further improvement can be obtained by applying the temperature corrected
effective potential in the bounce equation. We have determined the
parameters of the $T$-dependent potential directly from our numerical 
calculation in the following way. The restoring force was measured for a 
certain $h$ around both the stable and the metastable minima as described in 
Subsection \ref{ss:metastable}. Next an interpolating fit has been constructed of the form: 
$h_{\textrm{eff}}+m^2_{\textrm{eff}}\Phi+\lambda_3\Phi^2+
\lambda_{\textrm{eff}}\Phi^3/6$. 
It turned out that in the best fit $\lambda_3\approx 0$ is fulfilled always,
while the values of the other coefficients are not too far from their tree
level values. Then a bounce solution can be built on the corresponding
(real!) potential which includes the temperature corrections. The final
result is quite close to the measured value of the rate logarithm:
$S_2(T,\textrm{bounce})/T=0.29/(hT)$. The same real interpolation can be built
on the neighbourhood of the minima of the 2-loop $T$-dependent 
effective potential, leading to $S_2(T, \textrm{bounce})/T=0.2/(hT)$.
By common experience in the surface tension simulations a factor of 2-3 
difference in $S_2$ is expected to arise relative to the mean field theory.

We conclude, that the finite temperature corrections are important for
quantitative treatment of the nucleation rate within the thermal nucleation
theory.

\begin{figure}                       
\begin{center}                
\includegraphics[width=8.5cm]{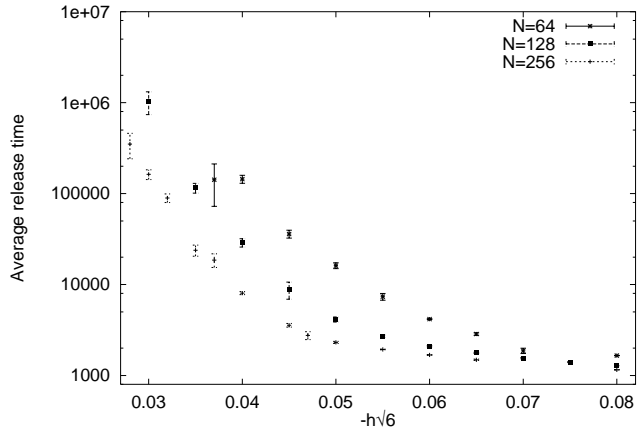}
\end{center} 
\vspace*{-0.8cm}
\caption{\label{t_release} The average ``release'' time as a function
of $h$ measured on $N=64,128,256$ lattices.} 
\end{figure}

Next we turn to the discussion of the possible nucleation threshold at small
$h$. The average ``release time'' increases for fixed lattice size with
decreasing $h$ (Fig.\ref{t_release}).  On a lattice of fixed size we found
small $h$ values for which we could not detect any transition, what makes
very probable the existence of a threshold value of the external field
$h_{th}(N)$. We did not attempt to locate this value beyond the simple
hyperbolic fits to the few largest $\langle t_{release}\rangle$ values. The 
value of $h_{th}(N)$ remains stable when the smallest $h$ is excluded from the
fit, therefore we conclude that a threshold magnetic field exists similar to 
the case of the kinetic Ising model \cite{Binder74}. The value of $h_{th}$ 
decreases when the lattice size is increased. Intuitively we expect 
$h(\infty )=0$, but with the three lattices studied by us this conjecture 
cannot be demonstrated.

\subsection{\label{ss:polonyi_concl}Conclusions} 
In this Section we have investigated in detail the decay of the false vacuum
in a classical lattice field theory. This serves as a demonstrative example
of the nonlinear relaxation of a far out of equilibrium classical cut-off
field theory. Our investigation was based dominantly on the effective theory
of the order parameter. Two versions of the effective theory were
reconstructed from its trajectory derived from the microscopical equations
of the theory. The first refers to the (meta)stable branch of the motion.
The second one which takes into account the existence of a mixed phase
during the transition period describes very well the transition together
with its neighbourhood. The first equation has the form of conventional
mechanical motion taking place in a dissipative noisy environment. The
dissipation, the dynamical breakdown of the time reversal symmetry was found
only for times longer than the minimal microscopic time scale of the system,
the autocorrelation time of the noise.

Our results offer a``dualistic'' resolution of the competition between the
nucleation and the spinodal phase separation mechanisms in establishing the
true equilibrium. On the one hand, we find that the statistical features of
the decay of the false vacuum agree with the results obtained by expanding
around the critical bubble. The microscopic mapping of the field
configurations during the relaxation supports the bubble creation scenario.
Alternatively, the effective OP-theory displays the presence of soft modes
and produces dynamically a Maxwell-cut when the time dependence of the
transition trajectory is described in the effective OP theory.  We find that
the larger is the system the smaller is the external field which is able to
produce the instability.  For infinite systems an infinitesimal field pushes
the system through the Maxwell-cut, where no force is experienced by the OP.
Therefore it will not stop before reaching the true homogenous ground state
passing by the mixed states with constant velocity.

%% file: scalar2l.tex
\chapter{The damping of the Goldstone modes}

This Chapter is devoted to the dynamics of the Goldstone modes.  First we
derive a set of effective equations for the long-wavelength modes of an
$O(N)$ symmetrical scalar field theory. Next, we analytically determine the
damping rate of the Goldstone modes as a function of the wave number $\bf k$.
Next, in an $O(2)$ symmetrical scalar field theory we determine the same
rate in the presence of a small explicit symmetry breaking parameter $h$.
We study the behaviour of the damping rate as $h\rightarrow 0$.

By performing a $2+1$ dimensional numerical simulation in an $O(2)$
symmetric classical field theory, we check the large time asymptotic power
law of the relaxation of an arbitrary field configuration. The manifestation
of Mermin--Wagner theorem is also discussed.

\section{Effective theory for the soft modes in O(N) $\Phi^4$
model\label{ch:ON}}

\rt{
In the study of dynamical phenomena with the participation of
long-wavelength Higgs and Goldstone-modes one can account for the effect
of the high frequency modes most conveniently by deriving a number
of effective equations of motion. In a second step one might express
 the correction terms of the original equations with
help of a set of auxiliary fields, allowing local representation of
some nonlocal effects. A remarkable feature of this approach is that
for pure gauge theories a simple kinetic interpretation can be given
to the dynamics of the auxiliary fields
\cite{heinz83,blaizot99,bodeker98,arnold99,litim99}.}

\rt{
Recently we have shown that a similar interpretation is possible for
fluctuations of the one-component self-interacting scalar field in the
broken symmetry phase \cite{plbscikk}. 
The evolution of the modes with wave number $|{\bf k}|$ was considered under 
the assumption $M<T$; that is the mass scale set by the spontaneous
symmetry breakdown is smaller than the temperature.  Since 
$M\sim {\sqrt\lambda}\langle\Phi\rangle$, 
this condition relates the vacuum expectation value of the field to the
temperature and is actually met in systems with second order phase transition
not much below the transition point. This condition actually was used when 
the comparison of the quantum results with the classical dynamics was made.}

\rt{
The high frequency modes have a non-trivial impact on the dynamics of the
low frequency modes if the scale characterising the low-frequency
fluctuations is much below the mass scale of the symmetry breaking. In this
sense one has to go beyond the usual hard thermal loop approximation, since
the effect of loops with momentum $p\simeq M<T$ should be taken into
account. The classical statistical mechanical system proposed in
\cite{plbscikk} reproduces the linear source-amplitude response computed in
quantum theory \cite{boya96}. }

In this section we extend our discussion to the broken phase dynamics of the
N-component scalar field theory with $\Phi^4$ self-interaction. This model
is relevant to the dynamics of the $\pi -\sigma$-system $(N=4)$ 
\cite{rajagopal93} and is actively investigated in connection with the
phenomenon of the disoriented chiral condensate
\cite{boyanovsky95,boyanovsky97,biro97, rischke98}.  The appearance, the
evolution and the damping of the low frequency fluctuations and
instabilities of the chiral order parameter are carefully studied in these
papers. Less attention is paid to the damping of the Goldstone-modes. Though
in the realistic case the explicit breaking of the $O(N)$ symmetry leads to
massive ``Goldstone''-bosons, it is of interest to see what is the intrinsic
dynamics of these excitations in the ``ideal'' spontaneously broken case. It
turns out that the damping of the Goldstone-fluctuations with frequency
$k_0$ is essentially different in the respective domains
\be
k_0\ll{M^2\over T}\ll M<T,\qquad \textrm{and}\qquad 
{M^2\over T}\ll k_0\ll M<T.
\ee
Our conclusion is that both the on-shell dissipation time and the large time
asymptotics of the Goldstone fields is qualitatively different in the first
region in comparison to the fluctuations of the order parameter.

The physical picture is very appealing. Radial fluctuations along the order
parameter relax fairly quickly, what freezes fast the length of the vacuum
expectation value of the $\Phi$ field. The second stage of the relaxation
consists of the slow rotation of the order parameter, which is described by
the relaxation of collectively excited low momentum Goldstone fluctuation
modes.
  
The high-temperature one-loop quantum dynamics of scalar models has been
investigated very actively recently partly as a kind of theoretical
laboratory for developing powerful calculational methods, partly with
the aim to provide answers to important questions of inflatory
cosmology and the physics of heavy ion collisions. Our work was
substantially influenced by References \cite{mrow90,
boyanovsky96b,boyanovsky98,bodeker95,greiner97}.

We first derive in Subsection \ref{ss:eff_eq_slow} the effective equations
for the low frequency modes.  The coefficients of the terms correcting the
classical equations are determined by various $n$-point functions of the
fast modes.  For the effective dynamics we work out the modification of the
linear part of the equations (the self-energy operator), which is fully
determined by the two-point function of the high-frequency fields. A detailed 
study of this quantity appears in Subsection \ref{sec:2pointfunction}. Stated 
in more technical terms, we compute one-loop contributions to the two-point
functions with full $T=0$ propagators and restrict the temperature dependent
contribution to the $p_0>\Lambda$ modes. This approach follows the finite 
temperature renormalization group transformation scheme of D'Attanasio and
Pietroni \cite{AP96}.  In Subsection \ref{ss:non_eq_ld} we shall discuss, in 
particular, whether the linear response of the high frequency modes to long 
wavelength fluctuations allows a classical kinetic theory interpretation.  
Using the results obtained for the two-point function of the theory, in
Subsection \ref{ss:nonli_dy} we present the explicit effective field equations, 
with help of auxiliary fields introduced to handle the nonlocal nature of the 
effective dynamics. Our results are summarised in Subsection \ref{ss:On_concl}.
In Appendix \ref{app:pertth} some results of the main text are rederived
with help of the conventional perturbation theory in a form explicitly
showing its equivalence to the Dyson-Schwinger treatment. Appendix \ref{calss} 
presents the iterative solution of the dynamics of the classical $O(N)$-model. 
Some relevant integrals are explicitly evaluated in Appendix \ref{sec:appcalc}.

\subsection{The effective equations of motion for the slow modes
\label{ss:eff_eq_slow}}

The Lagrangian of the system is given by
\be
L={1\over 2}(\partial_\mu{\bf\Phi}_a)^2-{1\over 2}m^2({\bf\Phi}_a)^2-
{\lambda\over 24}(({\bf\Phi}_a)^2)^2.
\label{Lagr_dens}
\ee
The aim of this investigation is to integrate out the effect of $p_0>
\Lambda$ high frequency fluctuations. From the point of view of the
final result it turns out to be important whether $\Lambda <M$ or
$\Lambda >M$, where $M$ is the mass scale spontaneously generated in
the broken symmetry phase. In the spirit of the renormalization group
the value of $\Lambda$ will be lowered gradually and the importance of
passing the scale $M$ will become evident in this process. Our actual
interest will concentrate on the temporal variation of the lowest
frequency fluctuations $k_0\ll\Lambda$.

We separate in the starting Lagrangian the high-frequency modes
($\varphi_a(x)$ with frequency $\omega>\Lambda$) and the low-frequency modes
($\tilde\phi_a (x)$ with frequency $\omega<\Lambda$)
\be
{\bf\Phi}_a(x)\rightarrow \tilde\phi_a (x)+\varphi_a (x).
\ee
Averaging over the high frequency fluctuations gives $\langle
\varphi_a(x)\rangle=0$ which yields also 
$\tilde\phi_a (x)=\avr{\Phi_a(x)}$. This separation leads to
\bea
\nonumber
L&=&{1\over 2}(\partial\tilde\phi_a)^2+{1\over 2}(\partial\varphi_a)^2-
{m^2\over 2}\Bigl [(\tilde\phi_a)^2+(\varphi_a)^2\Bigr ]
-{\lambda\over 24}\Bigl [(\tilde\phi_a)^2(\tilde\phi_b)^2+
(\varphi_a)^2(\varphi_b)^2\Bigr ]\\
&&-{\lambda\over 24}\Bigl [4(\tilde\phi_a\varphi_a)(\tilde\phi_b\varphi_b)+
4(\tilde\phi_a\varphi_a)(\varphi_b)^2+4(\tilde\phi_a\varphi_a)(\tilde\phi_b)^2+
2(\tilde\phi_a)^2(\varphi_b)^2\Bigr ].
\label{shiftedL}
\eea
From Eq.(\ref{shiftedL}) the equations for the slow modes can be
derived:
\bea
\nonumber
&&(\bbox +m^2)\tilde\phi_a+{\lambda\over 24}
\left[8\varphi_a(\tilde\phi_b\varphi_b)+
4\tilde\phi_a(\tilde\phi_b)^2+4\varphi_a(\varphi_b)^2\right.\\
&&\left. 
\qquad\qquad\qquad\qquad\qquad+4\varphi_a(\tilde\phi_b)^2+
8\tilde\phi_a(\tilde\phi_b\varphi_b)+4\tilde\phi_a(\varphi_b)^2\right]=0.
\eea
The effect of the high-frequency fluctuations on the slow ones is
obtained by averaging the equations with respect to their
statistics. At one-loop level accuracy $\langle
\varphi_a\varphi_b\varphi_c\rangle=0$ and one arrives at
\be
(\bbox +m^2)\tilde\phi_a
+{\lambda\over 6}\tilde\phi_a(\tilde\phi_b)^2+
{\lambda\over 3}\tilde\phi_b\langle\varphi_a\varphi_b\rangle+
{\lambda\over 6}\tilde\phi_a\langle(\varphi_b)^2\rangle =0.
\ee
We introduce at this point the conventional notation:
\be
\Delta_{ab}(x,y)\equiv\langle\varphi_a(x)\varphi_b(y)\rangle .
\ee
Below no summation will be understood when repeated indices appear without
the explicit summation symbol. We shall also use specific pieces extracted
from the above two-point functions defined as
\be
\Delta^{(0)}_{ab}(x,y)\equiv\Delta_{ab}(x,y)|_{\tilde\phi =0},\qquad
\Delta^{(1)}_{ab}(x,y)\equiv\int dz{\delta\Delta_{ab}(x,y)\over
\delta\tilde\phi_c (z)}|_{\tilde\phi =0}\cdot\tilde\phi_c(z).
\ee

In the broken symmetry phase one has to separate the nonzero average
value from the slowly varying field, and write the equation only for
the fluctuating part:
\be
\tilde\phi_a\rightarrow\bar\phi\delta_{a1}+\phi_a,
\label{breaking}
\ee
where we have chosen the direction of the average to point along the
$a=1$ direction. We shall analyse the resulting equations for the
$a=1$ and the remaining $a=i\neq 1$ components separately.  Since the
main effect of the high frequency modes we are interested in is the
modification of the mass term (self-energy contribution), therefore we
shall restrict our study to the linearised equation of $\phi_a(x)$.
One has to take into account that the two-point functions
$\langle\varphi_a\varphi_b\rangle$ depend on the background $\phi_a(x)$. For
the linearised equations it is sufficient to compute them only up to
terms linear in the background.

The two equations are 
\bea
&&
(\bbox +{\lambda\over 3}\bar\phi^2)\phi_1(x) + \frac{\lambda\bar\phi}2
\phi_1^2(x) + \frac{\lambda\bar\phi}6 \sum\limits_{i\neq1} \phi_i^2(x) +
\frac\lambda6 \phi_1(x) \sum\limits_a\phi_a^2(x) + J_1(x)=0,\nonumber\\
&&
\bbox \phi_i(x) + \frac{\lambda\bar\phi}3 \phi_i(x)\phi_1(x) +
\frac\lambda6 \phi_i(x) \sum\limits_a\phi_a^2(x) + J_i(x)=0,\qquad (i\neq1)
\label{effeqs}
\eea
with the induced currents
\bea
J_1(x) &=& {\lambda\over 2} \bar\phi\Delta^{(1)}_{11}(x,x) + {\lambda\over
  6}(N-1)\bar\phi\Delta^{(1)}_{ii}(x,x),\nonumber\\
J_i(x) &=& {\lambda\over 3}\left(
\Delta^{(0)}_{ii}(x,x)-\Delta^{(0)}_{11}(x,x)\right)
\phi_i+{\lambda\over 3}\bar\phi\Delta^{(1)}_{i1}(x,x).
\label{j_ind}
\eea
We have simplified the equations using the fact $\Delta_{i,b\neq i}$ has no
$\phi$-independent piece, since with zero background the correlators are 
diagonal. The equations (\ref{j_ind}) express the linear response of the fast 
modes to the presence of a low frequency background producing an effective 
source term to their classical dynamical equations. Also one has not to forget 
that $\bar\phi$ is now the solution of the constant part of the equations:
\be
m^2\bar\phi+{\lambda\over 6}\left(
3\Delta^{(0)}_{11}(x,x)+(N-1)\Delta^{(0)}_{ii}(x,x)\right)\bar\phi+
{\lambda\over 6}\bar\phi^3=0.
\label{avphi}
\ee

\subsection{The two-point function of the fast modes}
\label{sec:2pointfunction}

Varying Eq.(\ref{shiftedL}) with respect to $\varphi (x)$ one arrives at
the equations of motion of the fast modes in the background of
$\tilde\phi (x)$:
\bea
\nonumber
&&(\bbox +m^2)\varphi_a+{\lambda\over 24}\left(
8\tilde\phi_a(\tilde\phi_b\varphi_b)+
4\varphi_a(\varphi_b)^2+4\tilde\phi_a(\tilde\phi_b)^2\right.\\
&&\qquad\qquad\qquad\qquad\qquad\qquad\left.
+8(\tilde\phi_b\varphi_b)\varphi_a
+4\tilde\phi_a(\varphi_b)^2+4\varphi_a(\tilde\phi_b)^2\right)=0.
\eea
The equations of motion linearised in the high frequency fields
can be written in the form
\be
\left\{(\bbox+m^2)\delta_{ab}+{\lambda\over 6}\left[
2\tilde\phi_a\tilde\phi_b+(\tilde\phi_c)^2\delta_{ab}\right]\right\}\varphi_b=
-{\lambda\over 6}\tilde\phi_a(\tilde\phi_b)^2.
\label{lineq}
\ee
We introduce the notation
\be
m_{ab}^2(x)=m^2\delta_{ab}+{\lambda\over 6}\left[
2\tilde\phi_a(x)\tilde\phi_b(x)+ (\tilde\phi_c(x))^2\delta_{ab}\right]
\label{massmatr}
\ee
and apply Eqs.(\ref{lineq}) and (\ref{massmatr}) to the fields appearing
in the definition of the two-point function $\Delta_{ac}(x,y)$:
\be
\bbox_x \Delta_{ac}(x,y)=-m^2_{ab}(x)\Delta_{bc}(x,y),\qquad
\bbox_y \Delta_{ac}(x,y)=-m^2_{cb}(y)\Delta_{ab}(x,y).
\ee
In the derivation of these homogeneous equations we have exploited that
$\langle\tilde\phi_a(x)(\tilde\phi_c(x))^2\varphi_b(y)\rangle\linebreak =0$.
After the Wigner-transformation
\be
\Delta (X,p)=\int d^4ue^{ipu}\Delta (x,y),\qquad u=x-y
\ee
one arrives at the exact linear equations for the Wigner
transforms. (The use of the Latin letters $k,p,q$ is reserved for the 
4-momenta and $p\cdot k, kq$, etc. denotes their Minkovskian scalar products.) 
For the two-point functions diagonal in $O(N)$ indices one
finds in the broken phase to linear order in the background (after
using Eq.(\ref{breaking}))
\bea
&&
\hspace*{-1cm}
\biggl[{1\over 4}\bbox_X-ip\cdot\partial_X-p^2+ M^2_a\biggr]\Delta_{aa}(X,p)+
\lambda_a\bar\phi\int{d^4q\over (2\pi)^4}\phi_1(q)\Delta_{aa}(X,
p-{q\over 2})e^{-iqX}=0,\nonumber\\
&&
\hspace*{-1cm}
\biggl[{1\over 4}\bbox_X+ip\cdot\partial_X-p^2+M^2_a\biggr]\Delta_{aa}(X,p)+
\lambda_a\bar\phi\int{d^4q\over (2\pi)^4}\phi_1(q)\Delta_{aa}
(X,p+{q\over 2})e^{-iqX}=0,
\label{MDeq}
\eea
where for the Goldstone bosons one has
\be
M_{i}^2=0,\qquad \lambda_i={\lambda\over 3},
\ee
while for the heavy (``Higgs'') mode
\be
M^2_1={\lambda\over 3}\bar\phi^2,\qquad \lambda_1=\lambda.
\ee
Here the tree level mass relations were used, as they correspond to the
actual order of the perturbation theory. This procedure  can be improved 
using Eq.(\ref{avphi}) for $\bar\phi$, what is equivalent to the
resummation of the perturbative series.

After performing the Fourier-transformation also with respect to the
center-of-mass coordinate and subtracting the resulting two equations
one arrives at
\bea
2p\cdot k\Delta_{11}(k,p)&=&-\lambda\bar\phi\int{d^4q\over (2\pi )^4}\phi_1(q)
\left[\Delta_{11}(k-q,p+{q\over 2})-\Delta_{11}(k-q,p-{q\over 2})\right],
\nonumber\\
2p\cdot k\Delta_{ii}(k,p)&=&-{\lambda\over 3}\bar\phi\int{d^4q\over (2\pi )^4}
\phi_1(q)
\left[\Delta_{ii}(k-q,p+{q\over 2})-\Delta_{ii}(k-q,p-{q\over 2})\right].
\label{diageq}
\eea

For the mixed two-point function $\Delta_{i1}$ in the same steps a
similar equation is derived:
\be
\biggl(2p\cdot k+{\lambda\over 3}\bar\phi^2\biggr)\Delta_{i1}(k,p)=
-{\lambda\over 3}\bar\phi \int{d^4q\over (2\pi )^4}
\left[\Delta_{ii}(k-q,p+{q\over 2})-
\Delta_{11}(k-q,p-{q\over 2})\right]\phi_i(k).
\label{offdiageq}
\ee

In the kinematical region, where the center-of-mass variation is very
slow one can assume $q\approx k$. (If there would be no $X$-dependence
one would find $\Delta(k-q,P)\sim \delta (k-q)$). If in addition one
restricts the variation of $p$ in the high frequency fields by the
relation $|k_\mu |\ll\Lambda \simeq |p_\mu |$, one can approximate the integrals
in these equations by a factorized form. If the functions
$\Delta_{aa}(k,p-{k \over 2})$ are replaced by the first two terms of
their power series with respect to $k/2$, one recovers the drift
equations proposed originally by Mr{\'o}wczy\'nski and Danielewicz
\cite{mrow90}. In case of the diagonal two-point functions
these equations are identical to the collisionless Boltzmann equation
for a gas of scalar particles, whose masses
$(M_1^2+\lambda\bar\phi\phi_1(x)$ and $M_i^2+
\lambda\bar\phi\phi_1(x)/3$) are determined by the $\phi_1$ field:
\be 
p\cdot\partial_X\Delta_{aa}(X,p)+{\lambda_a\over
  2}\bar\phi\partial_X\phi_1 \partial_p\Delta_{aa}(X,p)=0.
\label{kineq}
\ee
Below we shall discuss an alternative to this expansion, which does
not exploit the assumption of slow $X$-variation. At the end we will
be in position to assess the validity of the assumption which led to
the Boltzmannian kinetic equations (\ref{kineq}).

At weak coupling one can attempt the recursive solution of Eqs.
(\ref{diageq}) and (\ref{offdiageq}). The starting point of the
iteration is the assumption that {\it the high frequency modes are
close to thermal equilibrium}, therefore one has for the starting
2-point functions \cite{LeBellac}
\bea
&&
\nonumber
\Delta_{ii}^{(0)}(p)=2\pi\delta (p^2)
\left(\Theta (p_0)+\tilde n(|p_0|)\right),\\
&&
\Delta_{11}^{(0)}(p)=2\pi\delta (p^2-M_1^2)
\left(\Theta (p_0)+\tilde n(|p_0|) \right),
\nonumber\\
&&
\tilde n(x)=n(x)\Theta (x-\Lambda )={1\over e^{\beta x}-1}\Theta (x-\Lambda).
\label{propagator}
\eea
Since the starting distributions are independent of $X$, the integrals
in Eq. (\ref{MDeq}) factorize exactly and one has for the first
corrections the following explicit expressions:
\bea
\Delta^{(1)}_{aa}(k,p)&=&-{1\over 2p\cdot k}\lambda_a\bar\phi\phi_1(k)
\biggl[\Delta_{aa}^{(0)}(p+{k\over 2})-\Delta_{aa}^{(0)}(p-{k\over 2})\biggr],
\nonumber\\
\Delta_{i1}^{(1)}(k,p)&=&-{1\over 2p\cdot k+M_1^2}{\lambda\over 3}\bar\phi
\phi_i(k)\biggl[\Delta^{(0)}_{ii}(p+{k\over 2})-\Delta^{(0)}_{11}(p
-{k\over 2})\biggr].
\label{corr1eq}
\eea
The retarded nature (ie. forward time evolution) will be taken into
account with the Landau prescription $k_0\to k_0+i\epsilon$
(see Appendix \ref{app:pertth} for the equivalent formulas in
the conventional perturbation theory).

\subsection{Non-equilibrium linear dynamics of the slow modes
\label{ss:non_eq_ld}}

The quantum improved, linearised equations of motion are
(see eq.~(\ref{effeqs}))
\begin{equation}
  (\bbox + M_a^2)\phi_a(x) + J_a(x) =0.
\label{lineq1}
\end{equation}
The (retarded) self-energy function is introduced through the
 relation
\begin{equation}
  J_a(x) = \int\!d^4y\,\Pi_a(x-y)\,\phi_a(y),
\end{equation}
where, after Fourier-transformation one obtains, (see eq.~(\ref{corr1eq}))
\begin{eqnarray}
& \Pi_1(k) = \displaystyle\int\frac{d^4p}{(2\pi)^4}\Biggl[ &\!\! \left(
  \frac{\lambda\bar\phi} 2 \right)^2  \frac1{p\cdot k}
  \left(\Delta_{11}^{(0)}(p-\frac k2) - \Delta_{11}^{(0)}(p+\frac k2)
  \right) \nonumber\\
&& + \left(\frac{\lambda\bar\phi} 6 \right)^2 (N-1) \frac1{p\cdot k}
  \left(\Delta_{ii}^{(0)}(p-\frac k2) - \Delta_{ii}^{(0)
}(p+\frac k2)
  \right)\Biggr], \nonumber\\
& \Pi_i(k) = \displaystyle\int\frac{d^4p}{(2\pi)^4}\Biggl[ &\!\! \left(
  \frac{\lambda\bar\phi} 3 \right)^2  \frac1{2p\cdot k + M_1^2}
  \left(\Delta_{11}^{(0)}(p-\frac k2) - \Delta_{ii}^{(0)}(p+\frac k2)
  \right)  \nonumber\\
&& - \frac\lambda 3 \left(\Delta_{11}^{(0)}(p) - \Delta_{ii}^{(0)}(p)
  \right)\Biggr].
\label{pis}
\end{eqnarray}
These integrals can be calculated (or at least reduced to 1D
integrals). Some details of the calculations can be found in Appendix
~\ref{sec:appcalc}, here we just state the results:
\begin{eqnarray}
&& \Pi_1(k) = \left[ \left(\frac{\lambda\bar\phi} 2 \right)^2\,
  R_1(k,M_1) + \left(\frac{\lambda\bar\phi} 6 \right)^2 (N-1)
  \,R_1(k,0) \right],\nonumber\\
&& \Pi_i(k) = \left(\frac{\lambda\bar\phi}3\right)^2 R_i(k,M_1).
\label{pis1}
\end{eqnarray}
The explicit form of their real parts is the following:
\begin{eqnarray}
&\hspace*{-1.25cm} 
\textrm{Re} R_1(k,M)=& \int\limits_M^\infty\!d\omega\, (1+2\tilde
n(\omega))\,\left[{\cal A}(\frac{2|\p||\k|}{2\omega k
_0+k^2}) + {\cal
    A}(\frac{2|\p||\k|}{-2\omega k_0+k^2}) \right], \nonumber\\
&\hspace*{-1.25cm}
 \textrm{Re} R_i(k,M)=& \int\limits_M^\infty\!d\omega\, \left[ (1+\tilde
  n(\omega))\,{\cal A}(\frac{2|\p||\k|}{2\omega k_0+k^2+M^2}) + \tilde
    n(\omega) {\cal A}(\frac{2|\p||\k|}{-2\omega k_0+k^2+M^2}) \right] 
  \nonumber\\ 
&&\hspace{-1.85cm} +\int\limits_0^\infty\!d|\p|\, \left[ (1 + \tilde n(|\p|))\, 
{\cal
    A}(\frac{2|\p||\k|}{-2|\p| k_0+ k^2- M^2}) + \tilde n(|\p|) {\cal
    A}(\frac{2|\p||\k|}{2 |\p| k_0+k^2-M^2}) \right],
\label{repis}
\end{eqnarray}
where $|\p|^2=\omega^2-M^2$ and
\begin{equation}
  {\cal A}(x)=\frac1{4\pi^2|\k|}\textrm{arth}(x) = \frac1{8\pi^2|\k|} \ln
  \left|\frac{1+x}{1-x}\right|.
\end{equation}
The expressions of the imaginary parts look as follows:
\begin{eqnarray}
& \textrm{Im} R_1(k,M)=&\frac{-1}{4\pi|\k|}\Biggl[\Theta(-k^2)\!\!\!\!\!
  \int\limits_{P_c-k_0}^{P_c}\!\!\!\!\!dP\,\tilde n(P) \,+\,
  \Theta(k^2\!-\!4M^2)\int\limits_{k_0/2}^{P_c} \!dP\,(1 + \tilde
  n(k_0\!-\!P) + \tilde n(P))\Biggr],
  \nonumber\\ 
& \textrm{Im} R_i(k,M)=&\displaystyle{\frac{-1}{16\pi|\k|}} \Biggl[\,
  \Theta(M^2-k^2)\left[ \int\limits_{Q_+}^{Q_++k_0}\!\!\!  dP\,\tilde
  n(P) + \int\limits_{|Q_--k_0|}^{|Q_-|}\!\!\! dP\,\tilde n(P) \right]\,
  \nonumber\\
&& + \Theta(k^2-M^2)\int\limits_{Q_+}^{Q_-} \!dP\,(1 + \tilde n(k_0 - P)
  + \tilde n(P))\Biggr],
\label{impis}
\end{eqnarray}
where
\begin{equation}
  P_c=\frac {|\k|}2\sqrt{1-\frac{4M^2}{k^2}} + \frac{k_0}2,\quad
  Q_+=\left|\frac{k^2-M^2}{2(k_0+|\k|)}\right|, \quad 
  Q_-=\frac{k^2-M^2}{2(k_0-|\k|)}.
\end{equation}

The real part is divergent at zero temperature which cancels against
the coupling constant counterterm contribution. Care has to be taken,
however, when implementing the regularization, to maintain the Lorentz
invariance for the pieces containing $T=0$ parts of the propagators
$\Delta^{(0)}$, which is manifest in the original form in
Eq.~(\ref{pis}).

It is remarkable that the usual domain of Landau damping ($|\k
|^2>|k_0|^2$) is apparently extended up to $M_1^2+|\k |^2>|k_0|^2$. An
analogous situation has been noted and interpreted recently for the
propagation of a light fermion in heavy scalar plasma
\cite{boyanovsky98b}. Below we find for the damping of soft on-shell
Goldstone-modes the same interpretation.
 
The linearised equations of motion can be analysed from several points
of view. One is the determination of the dispersion relations. This
describes the quasiparticles, and physically corresponds to the
``dressing'' of an (external) particle passing through the thermal
medium. The other point of view is the field evolution
\cite{boyanovsky96b}, when one follows the solution of the field
equations developing from given initial conditions (history).

\subsection{\label{ss:dispersion}Dispersion relations for on-shell waves}

The position of the poles is determined by the equation
\begin{equation}
  k^2-M_a^2 -\Pi_a(k) = 0.
\end{equation}
We split $k_0$ into real and imaginary parts: $k_0=\omega-i\Gamma$,
and assume $\Gamma\ll\omega$. Then the perturbative solution can be
written as
\be
  \omega^2 = \omega_0^2 + \textrm{Re}\Pi(\omega_0,\k),\qquad
  \Gamma =  - \frac{\textrm{Im}\Pi(\omega_0,\k)}{2\omega_0},
\ee
where $\omega_0^2=M_a^2+\k^2$. The corresponding time dependence for
fixed  wave vector \k\ and given initial amplitudes ($\partial_t\phi
(t=0,\k )=P({\bf k}),\phi (t=0,\k )=F({\bf k})$) is found using Eq.
(\ref{init_problem}) and taking into account only the poles
\begin{equation}
  \phi(t,\k)= \biggl[P(\k) \frac{\sin\omega t}{\omega} + F(\k)
  \cos\omega t \biggr]\,e^{-\Gamma t}.
\label{disprel}
\end{equation}

\paragraph{Goldstone modes.}

The tree level dispersion relation has no mass gap. The second
equation of Eq. (\ref{pis}) shows that $\Pi_i(k=0)=0$, that is no mass gap
is created radiatively neither at finite temperature; this is the
manifestation of the Goldstone theorem. The on-shell imaginary part,
as Eq.~(\ref{pis1}) shows, comes from the continuation of the
Landau-damping extending up to $k^2<M_1^2$. Going back to Appendix
\ref{sec:appcalc} (Eq.(\ref{gh_diff})), one finds that the imaginary part 
receives contribution from the collision of a hard thermal Goldstone particle
with distribution $n(p_0)$ with the soft external Goldstone wave of
momentum $k$ producing a hard Higgs-particle minus the inverse
reaction, when a hard thermal Higgs with momentum distribution
$n(p_0+k_0)$ decays into a soft and a hard Goldstone. The two
contributions can be combined into a single integral leading to the
following expression for the damping rate:
\begin{equation}
  \Gamma_i(\k) = \left(\frac{\lambda\bar\phi}3\right)^2\,\frac1{32\pi
  \k^2} \int\limits_{M_1^2/4|\k|}^{M_1^2/4|\k| + |\k|}\!\! dp\,\tilde n(p).
\end{equation}
If $|\k |\ll M_1$ we can write with a good approximation
\begin{equation}
 \Gamma_i(\k) = \left(\frac{\lambda\bar\phi}3\right)^2\!\frac1{32\pi
  |\k |} \,\tilde n(\frac{M_1^2}{4|\k |}).
\label{goldstone_damping}
\end{equation}
This contribution survives the IR cut, if $\Lambda<M_1^2/4|\k|$. 

The result (\ref{goldstone_damping}) is interesting from several
points of view. First, in HTL approximation, where we neglect all the
masses, we would found $\Gamma_i \sim \tilde n(0) =0$. On the other
hand the classical approximation corresponds to the substitution
$n(x)\to T/x$ in Eq.(\ref{goldstone_damping}), which results 
(see Appendix \ref{calss}) in
\begin{equation}
  \Gamma_i^{{\rm class}} = \frac{\lambda T}{24\pi}.
\label{goldstone_class}
\end{equation}
However, for very small momenta ($|\k |\ll M_1^2/4T$) the correct result
is exponentially small, deviating considerably from the classical
result. The purely classical simulations therefore cannot reproduce
the correct Goldstone dynamics in the long wavelength region, they
would overestimate the damping rate.

The most important consequence of the form of the Goldstone damping is
the generation of a new dynamical length scale $M_2^{-1}=4T/M_1^2$.
The components of the Goldstone condensate with longer wave-length can
not (exponentially) decay, they survive for a longer time.

To get more insight in the dynamics of the Goldstone modes we have calculated 
the damping rate $\Gamma (|{\bf k}|)$ for classical
on-shell Goldstone modes of the $O(2)$ symmetric scalar fields propagating 
in a thermal medium of the broken symmetry phase taking into account
the effect of the explicit symmetry breaking. We have specialised the
discussion of the {\it linear response function} to the $O(2)$ case
just to simplify some formulae. The conclusions appear to be of
intrinsic validity for the dynamical effect of the Goldstone phenomenon.

The study of the Goldstone-propagation through thermal medium is of
particular interest for the interpretation of the pion-sigma dynamics in
heavy ion collisions. Several recent field theoretical studies have dealt
with this problem \cite{boyan95,rischke98,csernai99} mainly concentrating
on the evolution of the heavy (sigma or Higgs) component. Though it has 
been observed in \cite{rischke98} that homogeneous pion condensates do not
decay, the origins of this statement have not been fully explored. The
effect of explicit symmetry breaking was taken into account by some authors
by treating the pions as effectively massive degrees of freedom, without
exploiting the consequences of the approximate symmetry 
\cite{boyanovsky96b,mrow97}.

The Lagrangian of the $O(2)$ symmetric scalar field theory with explicit 
symmetry breaking differs from the one presented in Eq. (\ref{Lagr_dens})
only by an additional  $h\phi_1$ term. We will follow what happens with the 
damping rate $\Gamma (|{\bf k}|)$ as the strength of the explicit symmetry 
breaking goes to zero.

The difference in the linearised equations of motion for $\phi_2$ Eq.
(\ref{avphi}) is that due to a modified equation for the equilibrium (static)
value of the vacuum expectation, $\bar\phi$ one obtains
\be
m^2\bar\phi +\frac{\lambda}{6}\bar\phi^3-h+{\lambda\over 2}\bar\phi\
\avr{\varphi_1^2}^{(0)}+
{\lambda\over 6}\bar\phi\langle\varphi_2^2\rangle^{(0)}=0.
\ee
Also it induces the following mass parameters for the Higgs and Goldstone modes:
\be
M_1^2={\lambda\over 3}\bar\phi^2+{h\over\bar\phi},\qquad
M_2^2={h\over \bar\phi}.
\label{masses}
\ee

The effective linear wave equation for the field $\Phi_2$ looks like:
\bea
\lp-k^2+{h\over\bar\phi}\rp\phi_2(k)&=&
{\lambda\over 3}\phi_2(k)\displaystyle\int \frac{d^4p}{(2\pi )^4}
\lp\Delta_{11}^{(0)}(p)-\Delta_{22}^{(0)}(p)\rp\nonumber\\
&&+\lp\frac{\lambda}{3}\bar\phi \rp^2\phi_2(k)
\int\frac{d^4p}{(2\pi )^4}\,
\frac{\Delta_{22}^{(0)}(p+k/2)-
\Delta_{11}^{(0)}(p-k/2)}
{2p\cdot k+M^2_1-M^2_2}.
\label{eff_eq}
\eea
For $k=0$ the right hand side of this equation vanishes, what reflects
the effect of the Goldstone-theorem on the self-energy function of $\phi_2$.

The decay rate of the $\phi_2$-field is determined by the imaginary part of 
the self-energy function i.e. the coefficient of $\phi_2$  appearing in the 
right hand side of (\ref{eff_eq}).

The kinematical analysis of the range of variables contributing to 
this integral is much more involved than in the case when the symmetry
breaking happens exclusively spontaneously but it goes along the same lines
with the presentation in Appendix \ref{sec:appcalc}. After a straightforward 
but tedious procedure we find for the imaginary part of the self-energy 
function:
\bea
{\rm Im\ }\Pi_2(k_0,|{\bf k}|)&=&\lp{\lambda\over 3}\bar\Phi
\rp^2\frac{1}{16\pi |{\bf k}|}\times\nonumber\\
&&\left[\Theta (-k^2)
\int\limits_{\alpha_{-}}^\infty ds\ \left(n(s)-n(s-k_0)\right)
+\Theta (-k^2)
\int\limits_{\alpha_{+}}^\infty ds\ \left(n(s+k_0)-n(s)\right)\right.
\nonumber\\
&&+\Theta (k^2)\Theta (M_2^2-M_1^2)\Theta \lp(M_1-M_2)^2-k^2\rp
\int\limits_{\alpha_{-}}^{\alpha_{+}}ds~(n(s)-n(s-k_0))\nonumber\\
&&+\Theta (k^2)\Theta (M_1^2-M_2^2)\Theta \lp(M_2-M_1)^2-k^2\rp
\int\limits_{\alpha_{+}}^{\alpha_{-}}ds\ \lp n(s+k_0)-n(s)\rp\nonumber\\
&&-\left.\Theta (k^2)\Theta \lp k^2-(M_1+M_2)^2\rp
\int\limits_{\alpha_{-}}^{\alpha_{+}}ds\ \lp 1+n(s)+n(k_0-s)\rp\right],
\label{Eq:forthterm}
\eea
with
\be
\alpha_{\pm}=\left|
\frac{1}{2k^2}\left(k_0\left(k^2-M_1^2+M_2^2\right)\pm
|{\bf k}|\sqrt{(k^2-M_1^2+M_2^2)^2-4k^2M_2^2} \right)\right|.
\ee

In case of on-shell propagation ($k_0^2-|{\bf k}|^2=M_2^2$) and
$M_2<M_1$, only the fourth term of Eq. (\ref{Eq:forthterm})
contributes. With its help for the decay rate one finds
\be
\Gamma ={(\lambda\bar\phi /3)^2\Theta(M_1-2M_2)\over
32\pi|{\bf k}|\sqrt{|{\bf k}|^2+M_2^2}}
\left[ \int\limits_{\alpha_{+}}^{\alpha_{+}+k_0}ds\ n(s)
-\int\limits_{\alpha_{-}}^{\alpha_{-}+k_0}ds\ n(s) \right].
\label{rate}
\ee

For $M_2=0$ and $k_0\neq 0$ it reproduces the result of 
Eq. (\ref{goldstone_damping}) for the damping rate of a Goldstone wave 
without explicit symmetry breaking.

For $M_2\neq 0$, the damping rate is a continuous function of $|{\bf k}|^2$
and can be expanded around $|{\bf k|}=0$:
\bea
\Gamma (|{\bf k}|=0)&=&{\lambda^2\bar\phi^2\over 288\pi }
\Theta \lp M_1-2M_2\rp
\frac{M_1}{M_2^3}\sqrt{M_1^2-4M_2^2}\times\nonumber\\
&&
\lp\exp \lp-\frac{\beta M_1^2}{2M_2}\rp 
-\exp \lp -\frac{\beta (M_1^2+M_2^2)}{2M_2}\rp
\rp\times\nonumber\\
&&
\lp\exp \lp-\frac{\beta (M_2^2+M_1^2)}{2M_2}\rp-1\rp^{-1}
\lp\exp \lp-\frac{\beta M_1^2}{2M_2}\rp-1\rp^{-1}.
\label{ratexp}
\eea
In particular, it can be evaluated for $T \gg M_2,M_1$, with a result which
exactly coincides with the classical expression for the decay rate,
arising from Eq.(\ref{rate}) with the replacement $n(s)\rightarrow
n_{cl}(s)=T/s$. 

However, for $M_2\rightarrow 0$  the expression (\ref{ratexp}) does not
follow the classical theory, but vanishes non-analytically as 
$\exp (-\beta M_1^2/2M_2)$. This result demonstrates that not only the 
self-energy function, but {\it also the decay rate vanishes for $|{\bf k}|=0$, 
when the strength of the explicit symmetry breaking goes to zero}.

The vanishing with $h$ of the decay rate of a homogeneous Goldstone
modulation of the vacuum is in conformity with the expectation according to
which the equivalence of the vacuum states in a system with spontaneous
symmetry breaking should be manifest also in the dynamical evolution of
fluctuations and external signals: when an external signal superimposed on
the actual vacuum realizes the transition over to an equivalent vacuum state
no relaxation to the initial vacuum should occur (absence of the restoring
force).

The quantity $\Gamma (h, |{\bf k}|)$ arises at this order of the
perturbation theory as the difference of the rates of two processes: the
transformation of the Goldstone-wave into a Higgs-wave with the absorption
of a (high-frequency) thermal Goldstone-particle and its inverse. From the
macroscopic point of view our result presents the rate of transformation of
a Goldstone-signal into Higgs-signal when propagating through the
thermalised medium. In the approximation when one assumes that a single act
of transformation is not followed by any further interaction with the
thermal bath, it gives also the inverse life-time of the Goldstone-wave.
However, if one anticipates further multiple transformations between the
Goldstone and the Higgs forms of propagation, then the damping rate of the
original Goldstone signal will considerably increase for small, but finite
$h$ due to the more efficient damping of the $k_0\approx T$ Higgs-waves.
However, for $h=0$ the full rate still vanishes, since the
Goldstone-to-Higgs transformation rate itself becomes zero. An analysis of
A. Jakov\'ac \cite{jako99b} presents a systematic way for taking into
account the effect of the larger Higgs-width in the Goldstone-propagator.

\paragraph{Higgs modes.}

From Eq.~(\ref{pis}) one can easily read off the on-shell
($k_0^2=\k^2+M_1^2$) value of the imaginary part of the
self-energy. This is entirely due to Higgs scattering into a soft +
hard Goldstone-pair, and leads to
\begin{equation}
  \Gamma_1=\frac{-\textrm{Im} \Pi_1}{2k_0} =
  \frac{\lambda M_1^2}{96\pi k_0|\k|} (N-1) 
  \int\limits_0^{|\k|/2} \!dP\,(1 + \tilde n(\frac{k_0}2 - P) + \tilde
  n(\frac{k_0}2 + P)).
\end{equation}
In the $k\ll M_1$ limit it simplifies to
\begin{equation}
  \Gamma_1= \frac{\lambda M_1}{96\pi} (N-1) \left(\frac12 +\tilde
  n(\frac{M_1}2) \right).
\end{equation}
This survives the cut if $2\Lambda<M_1$. 

Since $M_1<T$ we can perform a high temperature expansion. 
This means at the same time that the classical approximation is applicable. 
We find
\begin{equation}
  \Gamma_1 = \frac{\lambda T}{48\pi} (N-1),
\end{equation}
which is $(N-1)/2\,\Gamma_i^{{\rm class}}$.

\subsection{Solution of the initial value problem of the slow fields}

Equation(\ref{lineq1}) describes an integro-differential equation. For its
solution we have to know the complete past history of the fields. 
\rt{The previous subsection provides a specific way to look at the problem.}
In general we can introduce 
$z_a(x)=\int_{y_0<0}\!d^4y\Pi_a(x-y)\phi_a(y)$, then for $x_0>0$
\begin{equation}
  \phi_a(x)= \int\frac{d^4k}{(2\pi)^4}\,e^{-ikx}\,\frac{z_a(k)}
  {k^2-M^2-\Pi_a(k)}.
\label{init_problem}
\end{equation}
Since all singularities are in the lower $k_0$ half plane, $\phi_a(x)$
defined by this relation vanishes for $x_0\equiv t<0$.  $z(x)$ can be
used for setting the initial conditions \cite{boyanovsky96b}, as it
can describe jumps in the field as well as in its time derivative. For
example $\phi_a(x)=0$ for $x_0<0$ and $\phi_a(t=0)=F_a$, $\partial_t
\phi_a(t=0)=P_a$ corresponds to
\begin{equation}
  z_a(x) = -P_a({\bf x}) \delta(x_0) - F_a({\bf x})
  \delta^{\prime}(x_0), \qquad z_a(k)= ik_0 F_a({\bf k}) - P_a({\bf
    k}).
\end{equation}
Direct substitution of Eq. (\ref{init_problem}) into Eq. (\ref{lineq1})
shows that it satisfies the effective homogeneous wave equation. 
By evaluating the $k_0$ integral for $t=0$ (for details, see below) one 
also can demonstrate that it fulfils the initial conditions set above.

\begin{figure}[t]
\begin{center}
\includegraphics[width=14cm]{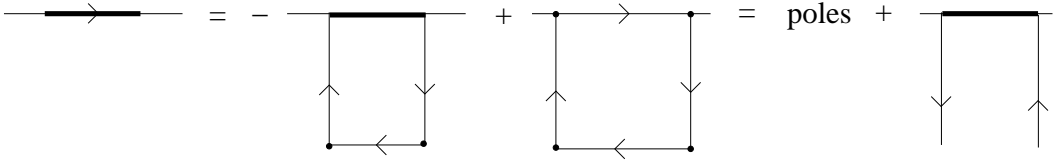}   
\end{center}
\vspace*{-0.8cm}
\caption{This figure shows how to pick up the contribution of a cut
(thick line).}
\label{Fig:cut_contur}
\end{figure}

In general, we are faced with the computation of Fourier transforms of
functions analytic on the upper complex plane. The procedure of
extracting their large time asymptotics was investigated already in
Refs.~\cite{boya96, boyanovsky96b}.  For $t>0$ the $k_0$
integration contour can be closed with an infinite semi-circle in the
lower half-plane and we pick up the contribution of the cuts and poles
inside the closed integration path:
\begin{equation}
  f(t)= \int\limits_{-\infty}^\infty \!\frac{dk_0}{2\pi}\, f(k_0)\,
  e^{-ik_0 t} = \sum\limits_{\omega\in\textrm{poles}}\!\!\!\!
  (-iZ(\omega)) e^{-i\omega t} + \sum\limits_{I\in\textrm{cuts}}
  \int\limits_I\!  \frac{dk_0}{2\pi i}\, \rho(k_0)\, e^{-ik_0 t},
\end{equation}
where $Z(\omega)$ is the residuum of the physical poles of the $f$-function
\footnote{The poles below a cut do not contribute to this formula,
they lie on the unphysical Riemann sheet.}, $\rho=i\,\textrm{Disc} f$
is the discontinuity along the cut, $I=[\omega_1,\omega_2]$ is the
support of the cut.  The second term can be computed again by
completing the original integration interval to a closed contour 
(see Fig. \ref{Fig:cut_contur}). The
discontinuity itself may have poles (but no cuts!) which contribute in
the same way as the ``normal'' poles. After these poles are
``encircled'', there are two straight contours - parallel to the
imaginary axis - left, starting at the two ends of the cut and running
in the interval $[\omega,\omega-i\infty]$, where $\omega$ is the end
(or starting) point (threshold). After this analysis one arrives at
the generic form
\begin{equation}
  f(t)= \sum\limits_{\omega\in\textrm{poles}}\!\!\!\! (-iZ(\omega))
  e^{-i\omega t} + \!\!\!\! \sum\limits_{\omega\in\textrm{thresh.}}
  \!\!\!\! (\mp) e^{-i \omega t} \int\limits_0^\infty \!
  \frac{dy}{2\pi}\, \rho(\omega -iy)\, e^{-yt},
\end{equation}
where the $-$ sign is to be applied for the starting, the $+$ sign for
the end point of the cut, and we have to take into account all poles,
also the ones on the unphysical Riemann sheet. Expanding $\rho$ around
$\omega$ into power series of $y$, the $y$ integration can be
performed.  The term $\sim y^n$ of the expansion contributes to the
time dependence $t^{-n-1}$. The large time behaviour is {\em dominated
by the lowest power term of the expansion}. If $\rho$ cannot be power
expanded then the damping for large times is faster than a power-law.

In our case the position of the poles is determined by the dispersion 
relations (see previous subsection), and for given initial values we find
for the pole contributions
\begin{equation}
  \phi_{pole}(t,\k)= Z(\k)\, \biggl[P(\k) \frac{\sin\omega
    t}{\omega} + F(\k) \cos\omega t \biggr]\,e^{-\Gamma t}.
\end{equation}
This expression coincides with eq.~(\ref{disprel}) apart from the wave
function renormalization $Z(\k )$. The time dependence of the
quasi-particle pole contribution was analysed before.

For the cuts the factorized form can be used for the spectral
function: $\rho_a = z_a \rho^G_a$, where $\rho^G_a$ is the spectral
function of the propagator:
\begin{equation}
  \rho^G_a = \frac{-2\textrm{Im}\Pi_a} { (k^2-M^2 -\textrm{Re} \Pi_a)^2
    + (\textrm{Im}\Pi_a)^2}.
\end{equation}
The thresholds and the leading large-time behaviour are determined by
$\textrm{Im}\Pi$, Eq.~(\ref{impis}). The imaginary part of the
Goldstone self energy is non-analytic only at $k^2=0$, but also there
the non-analytic piece vanishes as $\sim\exp(-M^2/(2T|k_0-|\k||))$. This
finally leads to a damping faster than any power which can be seen after
a saddle point analysis
\begin{equation}
  \phi_{cut,i} \sim \frac{ |\k|}{\lambda M_1^2} \left(\frac{\beta
      M_1^2}{t^3}\right)^{1/4} \, e^{-2\sqrt{\beta M_1^2t}}\, \left[ P(\k)
    \frac{\sin |\k|t }{|\k|} + F(\k) \cos |\k|t \right].
\end{equation}
The cut contribution for the Higgs damping, on the other hand, is
similar to the zero temperature case
\begin{eqnarray}
  &\phi_{cut,1}(t,\k)= &-\frac{\lambda T}{24 M_1}\, \frac1{(\pi M_1
  t)^{3/2}}\,\left[ P(\k) \frac{\sin (2M_1 t-\pi/4) }{2M_1} + F(\k) \cos
  (2M_1 t -\pi/4) \right]  \nonumber \\
  && + \frac{\lambda |\k|}{96\pi M_1}\, \frac1{\pi M_1 t}\,\left[ P(\k)
  \frac{\sin (|\k|t+\pi/2)}{|\k|} + F(\k) \cos(|\k|t +\pi/2) \right].
\end{eqnarray}
The first term comes from the threshold of the Higgs pair production,
the second from the threshold of the Goldstone pair production or their
Landau damping. The Landau damping of the Higgs particles does not 
contribute to the power law decay.

The terms decaying as some powers of time will dominate the time
evolution after the period of the exponential decay for the Higgs
bosons
($|\k|^{-1}\ll M_1^{-1}$). Similar is the case for small Goldstone
domains ($|\k|^{-1}\ll M_2^{-1}$), there the exponential decay is
followed by a $\sim\exp(-\sqrt{t})$ time evolution. In case of large
Goldstone domains ($|\k|^{-1}\gg M_2^{-1}$), however, because of the
exponentially small damping rate, the situation is reversed: the
$\sim\exp(-\sqrt{t})$ behaviour will be dominant for intermediate
times, while the amplitude is reduced by a factor of $Z^{-1}$. Only
for very long times will become the exponential damping term, arising
from the Goldstone-pole, relevant. Its action will erase completely
the large size domains.

\subsection{Non-local dynamics of the slow modes \label{ss:nonli_dy}}

The calculation of the effective non-local evolution of the
low-frequency modes can be based on Eq.(\ref{effeqs}) using the
explicit expressions for the Fourier transforms of the induced
currents defined in Eq.(\ref{j_ind}):
\bea
J_1(k) & = & {\lambda^2\bar\phi^2\over 4}\int{d^4p\over (2\pi )^4}
{1\over p\cdot k}
\phi_1(k)\left[\Delta_{11}^{(0)}(p-k/2)-\Delta_{11}^{(0)}(p+k/2)\right]
\nonumber\\
&&
+(N-1){\lambda^2\bar\phi^2\over 36}\int{d^4p\over (2\pi )^4}{1\over p\cdot k}
\phi_1(k)\left[\Delta_{ii}^{(0)}(p-k/2)-\Delta_{ii}^{(0)}(p+k/2)\right],
\nonumber\\
J_i(k) & = & -{\lambda\over 3}\int{d^4p\over (2\pi )^4}
{2p\cdot k\over 2p\cdot k+M_1^2}
\phi_i(k)\left[\Delta_{11}^{(0)}(p-k/2)-\Delta_{ii}^{(0)}(p+k/2)\right].
\label{expl_source}
\eea

Implicitly in all Wigner-transforms $\Delta^{(0)}$ the Bose-Einstein
factor is understood with a low frequency cutoff, well separating
the modes treated classically from the almost thermalised high
frequency part of the fluctuation spectra.

For the purpose of numerical investigations the above non-local form
of the induced currents is difficult to use. In this subsection we
discuss the introduction of auxiliary fields making the numerical
solution of the non-local dynamics easier to implement.  We shall not
take into account the nonlinear piece of the source induced at
one-loop level, since in weak coupling the leading nonlinear effect
comes from the tree-level cubic term. The consistent inclusion of the
higher power induced sources will not lead to qualitatively new,
leading effects unlike the linear source, which is responsible for
damping. We leave this extension for future investigations.

We concentrate first on the induced current $J_1$. By appropriate
shifts of the integration variable $p$ it can be rewritten as
\bea
J_1(k) & = & {\lambda^2\bar\phi^2\over 2}\int {d^4p\over (2\pi )^4}\left[
{1\over 2p\cdot k+k^2}-{1\over 2p\cdot k-k^2}\right]\Delta_{11}^{(0)}(p)
\phi_1(k)
\nonumber\\
&&
+(N-1){\lambda^2\bar\phi^2\over 18}\int{d^4p\over (2\pi )^4}\left[
{1\over 2p\cdot k+k^2}-{1\over 2p\cdot k-k^2}\right]\Delta_{ii}^{(0)}(p)
\phi_1(k).
\eea
After explicitly performing the $p_0$ integration one arrives at the
following expression which has a clear interpretation as a specific
statistical average:
\bea
J_1(k) & = & \phi_1(k)\left[{\lambda^2\bar\phi^2\over 2}\int{d^3p\over
(2\pi )^3}{1\over 2p_0}(1+2\tilde n(p_0))\left({1\over 2p\cdot k+k^2}-
{1\over 2p\cdot k-k^2}\right)\Big|_{p_0=
({\bf p}^2+M_1^2)^{1/2}}\right.\nonumber\\
&&\left.
+(N-1){\lambda^2\bar\phi^2\over 18}\int{d^3p\over (2\pi )^3}{1\over 2p_0}
(1+2\tilde n(p_0))\left({1\over 2p\cdot k+k^2}-{1\over 2p\cdot k-k^2}\right)
\Big|_{p_0=|{\bf p}|} \right].
\label{statav}
\eea
The temperature independent piece corresponds to the $T=0$
renormalization of $\lambda$, fixed at the scale $k^2$. It is absorbed
into the mass term of $\phi_1$, therefore we retain in the integrand
only the terms proportional to the cutoff Bose-Einstein factors.
 
We introduce two complex auxiliary ``on-shell'' fields $W^a(x,{\bf
p})$, $a=1,i$. They correspond to the two different mass-shell
conditions appearing in the above equation, and fulfil the equations:
\be
(2p\cdot k-k^2)W^a(k,{\bf p})=\phi_1(k).
\label{auxeq}
\ee
Clearly, $J_1(x)$ can be expressed as a well defined combination of
the thermal averages of these fields:
\be
-J_1(x)=\lambda^2
\bar\phi^2\Bigl (\langle W^1(x,{\bf p})\rangle
+\langle W^1(x,{\bf p})^{*}\rangle 
+{N-1\over 9}(\langle W^i(x,{\bf p})\rangle+
\langle W^i(x,{\bf p})^{*}\rangle )\Bigr ).
\label{j1_ind}
\ee
Thermal averages are defined by the usual formula 
\be
\langle W^a(x,{\bf p})\rangle =\int{d^3p\over (2\pi )^3}{1\over 2p_0}\tilde
n(p_0)W^a(x,{\bf p})\Big|_{p_0=({\bf p}^2+M_a^2)^{1/2}},\qquad a=1,i.
\ee
For the relevant combination one can use in place of Eq. (\ref{auxeq}) the
equations arising after the combination $W^a(x,{\bf p})-W^{a}(x,{\bf p})^*$ is
eliminated:
\be
\left[(2p\cdot k)^2-(k^2)^2\right]\left(W^a(k,{\bf p})+W^a(-k,{\bf p})^*
\right)=2k^2\phi_1(k).
\label{auxsymeq}
\ee
Equations (\ref{effeqs}), ({\ref{j1_ind}) and (\ref{auxsymeq})
represent that form of the non-local dynamics which is best adapted
for numerical solution.

For very small values of $k_0$ a simplified form can be derived which
coincides with the result of the conventional kinetic treatment of the
Higgs-modes presented in Subsection \ref{ss:MD}.  One arrives at this
approximate expression of $J_1(k)$ if one performs an appropriate
three-momentum shift ${\bf p}\rightarrow {\bf p\mp k}/2$ on the ${\bf p}$ 
variable in Eq. (\ref{statav}).  One finds the following expressions for the
denominators, accurate to linear order in $k_0$:
\bea
2p\cdot k\pm k^2&\rightarrow& 2k_0\sqrt{({\bf p\mp k}/2)^2+M_a^2}-
2{\bf p\cdot k}
\pm k_0^2\approx 2p\cdot k(1\pm k_0/2p_0)+{\cal O}(k_0^3,{\bf k}^3),
\nonumber\\
p_0&\rightarrow& p_0\mp {{\bf p\cdot k}\over 2p_0}.
\eea
Performing the expansion of the denominators and of the arguments of
$\tilde n (p_0)/p_0$ to linear order in ${\bf k}, k_0$ one finds the
following approximate expression for $J_1$:
\bea 
J_1(k)&=&-{\lambda^2\bar\phi^2\over 4}\int {d^3p\over (2\pi )^3}
\left\{\left[
{1\over p_0^2}{d\tilde n(p_0)\over dp_0} {k_0p_0\over p\cdot k}
+{1\over p_0} \left({1\over 2}+\tilde n(p_0)-p_0{d\tilde n(p_0)\over dp_0}
\right)\right]\right.\Bigg|
_{p_0=({\bf p}^2+M_1^2)^{1/2}}\nonumber\\
&&\left.
+{N-1\over 9}{1\over p_0^2}\left[{d\tilde n(p_0)\over dp_0}
{k_0p_0\over p\cdot k}
+{1\over p_0}
\left({1\over 2}+\tilde n(p_0)-p_0{d\tilde n(p_0)\over dp_0}\right)\right]
\Bigg|_{p_0=|{\bf p}|}\right\}
\phi_1(k).
\eea
The IR-cutoff $\Lambda \gg|\k|$ is important to prevent the singularity
of the contribution from the Goldstone-modes.
\footnote{ The proper solution of the effective dynamics of the heavy
field surrounded by massless Goldstone-quanta will require the
application of methods analogous to the Bloch-Nordsieck resummation at
$T=0$ \cite{boyanovsky98c}.  We thank M. Simionato for an enlightening
discussion on this point.}  The contributions from the second terms in each line
are independent of $k$, therefore they are simply understood as a finite
shift in the squared mass of $\phi_1$.  With these steps one arrives at the
final expression for the nonlocal part of the induced ``Higgs'' current:
\be
\hspace*{-0.25cm}
J_1(k)= -{\lambda^2\bar\phi^2\over 4}\int{d^3p\over (2\pi )^3}
\left[{1\over p_0^2}{d\tilde n(p_0)\over dp_0}\Big|_{p_0=({\bf p}^2+
M_1^2)^{1/2}}
+{N-1\over 9}{1\over p_0^2}
{d\tilde n(p_0)\over dp_0}\Big|_{p_0=|{\bf p}|}\right]{1\over v\cdot k}
k_0\phi_1(k).
\label{j2_ind}
\ee 
Here $v^\mu =(1,{\bf p}/p_0)$.  In this case it is sufficient to
introduce only two real auxiliary fields $W^1(x,{\bf v})$ and
$W^i(x,{\bf v})$ in order to make the dynamical equations local. They
are defined through the equations:
\be
v\cdot kW^a(k,{\bf v})=k_0\phi_1(k), \qquad a=1,i.
\label{aux1eq}
\ee
The weighting factors in the integrals over the auxiliary field can be
interpreted as the deviations from equilibrium of the Higgs and
Goldstone particle distributions, $\delta n_1({\bf p})$ and $\delta
n_i({\bf p})$, due the scattering of the high frequency particles off
the $\phi_1$ condensate. Then one writes
\be
J_1(x)=-{\lambda^2\bar\phi^2\over 4}\Bigl (\delta\langle W^1(x,{\bf v})
\rangle +{N-1\over 9}\delta\langle W^i(x,{\bf v})\rangle\Bigl ),
\ee
where $\delta\langle ... \rangle$ means a phase space integral
weighted with the non-equilibrium part of the relevant distributions.

For this case it is easy to construct the energy-momentum vector
corresponding to the nonlocal piece of the induced current. One can follow 
the procedure
proposed by Blaizot and Iancu \cite{iancu94} and investigate the
divergence of the $( \mu ,0)$ component of the energy-momentum tensor:
\be
\partial_\mu T^{\mu 0}_{induced}=-J_1\partial^0\phi_1,
\ee
The task is to transform the right hand side into the form of a
divergence.  Using Eqs.(\ref{j2_ind}) and (\ref{aux1eq}) one finds the
following expression:
\bea
-J_1\partial^0\phi_1&=&{\lambda^2\bar\phi^2\over 4}\int{d^3p\over (2\pi )^3}
\left(
W^1(x,{\bf v}){1\over p_0^2}{d\tilde n(p_0)\over dp_0}
(v\cdot \partial_x)W^1(x,{\bf v})\Big|_{p_0=({\bf p}^2+M_1^2)^{1/2}}
\right.
\nonumber\\
&&\left.+{N-1\over 9}W^i(x,{\bf v}){1\over p_0^2}{d\tilde n(p_0)
\over dp_0}(v\cdot \partial_x)W^i(x,{\bf v})\Big|_{p_0=|{\bf p}|}\right).
\eea
From here one can read off the  induced energy-momentum function of
the approximate Higgs dynamics:
\bea
T^{\mu 0}_{induced} & = & {\lambda^2\bar\phi^2\over 8}
\int{d^3p\over (2\pi )^3} v^\mu\left[
{1\over p_0^2}{d\tilde n(p_0)\over dp_0}
W^1(x,{\bf v})^2\Big|_{p_0=({\bf p}^2+M_1^2)^{1/2}}\right. \nonumber\\
&& \left.+ {N-1\over 9}{1\over p_0^2}{d\tilde n(p_0)\over dp_0}
W^i(x,{\bf v})^2\Big|_{p_0=|{\bf p}|}\right].
\eea

For the exact (essentially non-local) dynamics we did not attempt the
construction of an energy functional, but its existence for the above
limiting case hints for the Hamiltonian nature also of the full one-loop
dynamics as expressed in terms of the auxiliary variables.

The analysis of the induced Goldstone source, appearing in
Eq.(\ref{expl_source}) goes in fully analogous steps. The temperature
dependent part to which a statistical interpretation can be linked is
rewritten with help of two complex auxiliary fields $V^a(x,{\bf p}),
a=1,i$ as
\be 
J_i(x)={\lambda\over 3} \int{d^3p\over (2\pi )^3}
\Bigl [{1\over p_0}\tilde n(p_0=({\bf p}^2+M_1^2)^{1/2})
\textrm{Re}\, V^1(x,{\bf p})-{1\over p_0}\tilde n(p_0=|{\bf p}|) 
\textrm{Re}\,V^i(x,{\bf p})\Bigr ],
\label{j2_i}
\ee
where the auxiliary fields fulfil the equations
\bea
(2p\cdot k+k^2+M_1^2)V^1(k,{\bf p})&=&(2p\cdot k+k^2)\phi_i(k),\nonumber\\
(2p\cdot k-k^2+M_1^2)V^i(k,{\bf p})&=&(2p\cdot k-k^2)\phi_i(k).
\label{Veq}
\eea
In this case even for very small values of $k_0$ we were not able to
derive any limiting case in which one could treat the time evolution
of the low frequency Goldstone modes as a truly Boltzmannian kinetic
evolution.

The solution of the system of equations (\ref{effeqs}),
(\ref{j1_ind}), (\ref{auxsymeq}), (\ref{j2_i}), (\ref{Veq}) requires
the specification of initial conditions also for the auxiliary fields
$W^a+W^{a*}, V^a$. If one uses the ``past history'' condition for the
physical fields $\phi_1(x)={\rm const}, \phi_i(x)= {\rm const}$ for
$t<0$, then the linear integral equation form of the equations for the
auxiliary fields and their retarded nature imply vanishing $W^a, V^a$
for $t=0$.

Before concluding this Section we show that the method of introducing
auxiliary fields is equivalent with the method of mode-function expansion.
We choose $N=1$ for simplicity.

We start from the linearised equation for the quantum fluctuation $\varphi$ 
that is also linear in the background $\phi$:
\be
\left(\partial^2+m^2+\frac{\lambda}{2}\bar\phi^2\right)\varphi+
\frac{\lambda}{2}\bar\phi\phi\varphi=0.
\ee
If the amplitude of the background fluctuations is small, we can try to
expand with respect to it. As a first step we can stop after the linear
approximation. The different orders read
\bea
\textrm{0th order:}&\quad& (\partial^2+m_H^2)\varphi_0=0,\nn
\textrm{1st. order:}&\quad& (\partial^2+m_H^2)\varphi_1+
\lambda\bar\phi\phi\varphi_0=0,
\eea
where $m_H^2=m^2+\frac{\lambda}{2}\bar\phi^2$.

The 0th order solution is a harmonic function
\be
\varphi_0(x)=\sum_{\bf k}\frac{1}{2\omega_{\bf k}}
\left( a_{\bf k}e^{-ikx}+a^+_{\bf k}e^{ikx}\right),
\label{Eq:0th}
\ee
where $k_0=\omega_\k=\sqrt{\k^2+m_H^2}$. At 1st order we perform a Fourier
transformation obtaining in $\k$-space
\be
\varphi_1(k)=-\lambda\sum_{\bf q}\frac{1}{2\omega_{\bf k}}
\frac{1}{k^2-m_H^2}\left( a_{\bf q}\phi(k-q)+a_{\bf q}^+\phi(k+q)\right).
\ee
When transforming back in $x$-space we introduce the auxiliary field
$W(x,q)$ defined by
\bea
\nonumber
-W(x,q)\bar\phi&=&\lambda\int \frac{d k_0}{2\pi}\sum_{\bf k}
e^{-i(k-q)x}\frac{\phi(k-q)}{k^2-m_H^2}\\
&=&
\lambda\int\frac{dk_0}{2\pi}\sum_{\bf k}\frac{e^{-ikx}\phi(x)}{(k+q)^2-m_H^2}=
\lambda\int\frac{dk_0}{2\pi}\sum_{\bf k}\frac{e^{-ikx}\phi(x)}{k^2+2kq}
\label{W_bevezetes}
\eea
and we obtain
\be
\varphi_1(x)=-\sum_{\bf q}\left(
a_{\bf q}e^{-iqx}W(x,q)+a_{\bf q}^+e^{iqx}W^*(x,q)\right).
\label{Eq:1st}
\ee
from this form it is evident that $W$ satisfies 
\be
(\partial_x^2-2iq\partial_x)W(x,q)=\lambda\phi(x),
\ee
that corresponds to Eq. (\ref{auxeq}).
\rt{ if we replace 
$\phi\rightarrow\phi+\bar\phi$ in $s(x)$ and keep only the linear term in 
$\phi$. The $W$ introduced here is $\bar\phi$ times the auxiliary field
introduced in  Eq. (\ref{auxeq}).}
 
The complete solution up to linear order in $\phi$ reads cf. Eq.(\ref{Eq:0th})
and Eq. (\ref{Eq:1st})
\be
\varphi=\sum_{\bf k}\frac{1}{2\omega_{\bf k}}
\left(
a_{\bf k}e^{-ikx}(1-W(x,k))+a_{\bf k}^+e^{ikx}(1-W(x,k))^*
\right).
\ee
This expression makes the connection with the mode-expansion since by
looking at Eq.(\ref{Eq:modeexpantion}) we can identify
\be
\psi_{\bf k}(x)=e^{-ikx}(1-W(x,k)).
\ee

\subsection{\label{ss:On_concl}Conclusions}

In this Section we have performed a complete 1-loop analysis of the time
evolution of low frequency ($k_0\ll M_1$) field configurations in the broken
phase of the $O(N)$ symmetric scalar field theory.  

\rt{We have shown explicitly
the full formal equivalence of the leading order iterative solution of the
Dyson-Schwinger equations and the perturbation theory computation of the
two-point function.}

We have analysed explicitly the case when $T>M_1$, which occurs, for
instance, in the vicinity of the second order phase transition of the model.
Here we could compare our results with the results arising from the
dynamical equations of the classical $O(N)$ field theory with appropriately
chosen parameters. For the ``Higgs'' particle we have found complete agreement
for the $|\k|\ll M_1<T$ modes. Related to this is the fact that we were able
to show that the exact one-loop equations derived for the auxiliary fields
$W^a$ have a simple Boltzmannian collisionless kinetic form for small $|\k|$.

However, the analysis gave different conclusions for the linear response of
the Goldstone-modes. The agreement with the classical theory is restricted
to the interval $M_1^2/4T\ll |\k|\ll M_1$. Below the new characteristic scale
$\displaystyle M_2\equiv\frac{M_1^2}{4T}
\label{newscale}$
the damping of the Goldstone modes becomes exponentially small and it
vanishes non-analyti\-cally for $|\k|\rightarrow 0$. This is a very natural
manifestation of the Goldstone-theorem in a dynamical situation: no
homogeneous ground state will relax to any ``rotated'' nearby configuration. 
In the light of this suggestive picture it is not surprising that we could
not find a classical kinetic interpretation of the exact one-loop dynamics
of the Goldstone modes even for very small values of $|\k|$.
Introducing a small explicit symmetry breaking term $\sim h$ we have checked
that the damping  of the Goldstone modes with wave number 
$|{\bf k}|\rightarrow 0$ vanishes non-analytically as $h\rightarrow 0$.
This is a convincing argument demonstrating that the behaviour 
outlined above is a manifestation of the O(N) symmetry.

Besides the exponential damping analysed above, $(1-Z)$ fraction of the
initial configuration follows a different time evolution determined by the
particle production and Landau-damping cuts. In case of the lowest wave
number Higgs fluctuations this leads to the result that the exponential
regime will be followed by a power-decay for times $t>M_1^{-1}$. In case of 
the Goldstone modes the cut contribution is $\sim\exp{(-4\sqrt{M_2t})}$, and
for modes above the scale $M_2$ this will dominate for large times. Below the 
scale $M_2$ the cut contribution will be observable for intermediate times, 
while the pole dominated damping, because of the exponentially small damping 
rate,  becomes relevant only for very long times.

It will be interesting to see through numerical investigations, if the
onset of the nonlinear regime will influence the damping scenario of
the Goldstone fluctuations. The effect of interaction among the
high-$k$ modes (in addition to their scattering off the low-$k$
background) on the effective theory merits also further study.

\rt{Finally, we work on the extension of our analysis to the Gauge+Higgs
models in the broken phase, relevant to the physics of the standard
model below the electroweak phase transition.}

\newpage

%% file: scalar2nl.tex
\section{\label{s:MW}
Nonlinear relaxation of classical O(2) symmetric fields in $2+1$ dimensions}

In this Section we wish to gain insight into the validity of the classical
linear response theory, by comparing its results to exact solutions. A
similar attempt was already done in \cite{aarts00,aartsPRD63}. There the early, 
far from equilibrium time evolution was confronted with the linear response
results, and a relaxation slower than expected has been observed.  We will
show later (see~Fig.~\ref{goldstonedecay}) that the decay in the linear
regime realizes the fastest relaxation rate of this quantity during the
whole time evolution.  Exploration of the range of validity of the linear
regime is possible only by following the evolution very long, and averaging
over initial conditions in order to raise the signal over the noise level.

\subsection{The model and its numerical solution}

We concentrate on the $O(2)$ symmetric classical scalar theory with
a small explicit breaking term in its Lagrangian:
\be
{\cal L}=\frac12(\partial_\mu\Phi_1)^2+
\frac12(\partial_\mu\Phi_2)^2
-\frac12m^2\Phi_1^2-\frac12m^2\Phi_2^2
-\frac{\lambda}{24}\left(\Phi_1^2+\Phi_2^2\right)^2
+h\Phi_1,
\label{lagrange}
\ee
where $h$ controls the explicit symmetry breaking, and $m^2<0$. We have 
studied the time evolution of a system, discretized on lattice in two space 
dimensions at so low energy density (see below) that according to the mean 
field analysis this would correspond to the broken symmetry phase. 

{The dynamics eventually drives the system towards thermal equilibrium. For
$h\neq 0$, the equilibrium state has large magnetisation, nearly corresponding 
to the minimum of the classical potential. Fluctuations around this state
are naturally divided into Goldstone (light) and Higgs (heavy) excitations.
These excitations experience an effective potential, which agrees very well
with the result of finite volume perturbation theory. The relaxation into this 
state is the main subject of this investigation. We shall present a detailed 
comparison of the exact time evolution with the linear response theory.

A second relaxation process can be initiated from the equilibrium with 
$h\neq0$, if the magnetic field is switched off. The final $h=0$ equilibrium,
however, obeys in the thermodynamical limit the Mermin-Wagner-Hohenberg
theorem \cite{mermin}, which states the absence of spontaneous magnetisation
in two dimensional systems with continuous symmetry.  For equilibrium two
dimensional {\it finite volume} systems even for vanishing external source,
at very low temperature there is a non-zero magnetisation with a well
defined direction as shown in \cite{archambault}. The finite volume
magnetisation selects the direction with respect to which we can define a
parallel and a perpendicular mode. This natural choice of base suggests the
use of the Goldstone boson terminology in the $h=0$ case too. The way the
finite magnetisation disappears as the volume goes to infinity will be
touched upon shortly at the end of our discussion.
}

The exact time evolution of the lattice system was studied by introducing
dimensionless field variables $\Phi_i\rightarrow\sqrt{a}\Phi_i$
and rescaling further these fields like 
$\Phi^{\textmini{rescaled}}_{1,2}=\Phi_{1,2}\sqrt{\lambda/6}$
and $h^{\textmini{rescaled}}=h\sqrt{\lambda/6}$.
This sets the bare coupling to $\lambda^{\textmini{rescaled}}=6$.
We have chosen $m^2a^2=-1$, therefore all dimensionful quantities are 
actually expressed in units of $|m|$. Squared lattices of size
ranging from $64\times64$ to $512\times512$ were used. 

The explicit symmetry breaking parameter $h$ is chosen in the range
$0\dots0.0025/\sqrt{6}$. 
{Initially the system is at rest near the origin
of the $\Phi$-space. }
A small amplitude white noise determines the field in
every lattice point ($\Phi_1(\xvon,t=0)=\eta_1(\xvon),
\Phi_2(\xvon,t=0)=\eta_2(\xvon),
\dot\Phi_1(\xvon,t=0)\equiv\dot\Phi_2(\xvon,t=0)\equiv0,
\langle\eta_i(\xvon)\eta_j({\bf y})\rangle\sim\delta_{{\bf x},{\bf y}}
\delta_{i,j}$).
This means, that the space average of $\Phi_i$, the two-component order 
parameter (OP) $(\overline{\Phi_1}^V,\overline{\Phi_2}^V)$ is very close 
initially to zero, which is (approximately) a local maximum of the effective
potential. {The roll-down towards the minimum is oriented by the external
field $h$.} It is the potential energy difference of the initial and the
final states, which is redistributed between the modes. The magnitude of the
noise was so small, that the final energy density of the system is
exclusively determined by this difference in the potential energy.

{
Numerical diagonalization of the measured $2\times2$ matrix of equal
time fluctuations 
($C_{i,j}(t)=\overline{\Phi_i(\xvon ,t)\Phi_j(\xvon ,t)}^V-
\overline{\Phi_i(\xvon ,t)}^V
\overline{\Phi_j(\xvon ,t)}^V$) gave as uncorrelated degrees of freedom the
radial (Higgs) and angular (Goldstone) components of $(\Phi_1,\Phi_2)$:
\be
\Phi_H(\xvon,t)=\sqrt{\Phi_1^2(\xvon,t)+\Phi_2^2(\xvon,t)},\qquad
\Phi_G(\xvon,t)=\Phi_{H,0}\arctan\frac{\Phi_2(\xvon,t)}{\Phi_1(\xvon,t)},
\label{eq:felbontas}
\ee
$\Phi_{H,0}=|m|\sqrt{6/\lambda}+h/2|m|^2+{\cal O}(h^2)$ being the
minimum of the bare potential valley. This statement means
that $\overline{\Phi_H(t)\Phi_G(t)}^V\approx
\overline{\Phi_H(t)}^V\overline{\Phi_G(t)}^V$ for all $t$. 
The boundedness of $\Phi_2$ fluctuations follows
from the IR-regulating effect of the explicit symmetry
breaking term.
The position of the tree level minimum scaled by $|m|$ is 
$\Phi_{H,0}=1+h/2+{\cal
O}(h^2)$, and the scaled squared masses read as 
$m_H^2=2+h/2+{\cal O}(h^2)$ for the radial mode and $m_G^2=h+{\cal O}(h^2)$ 
for the angular one.
}

\subsection{Relaxation to the $h\not=0$ equilibrium}

During and immediately after the roll-down the main mechanism of the fast
excitation of spatial fluctuations is the parametric resonance, as observed
and studied by many authors, e.g. \cite{khlebnikov,greene97}. Also the
dephasing of different oscillation modes can be observed as reported in
\cite{Wetterich99,mottola97}. The time scales of this stage are
characterised by the cutoff and by the masses of the radial and angular
modes. The measurement of these latter is discussed later.

\begin{figure}[htbp]
\begin{center}
\includegraphics[width=8cm]{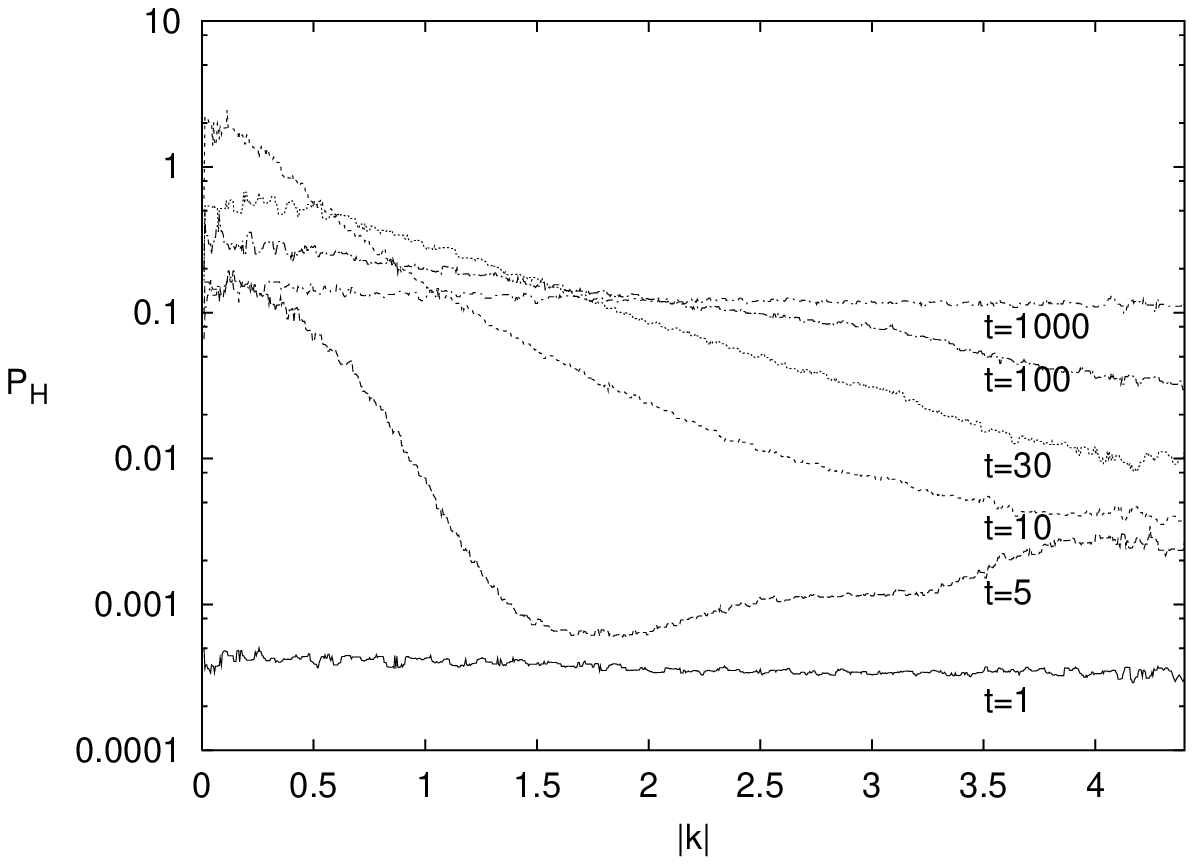}
\includegraphics[width=8cm]{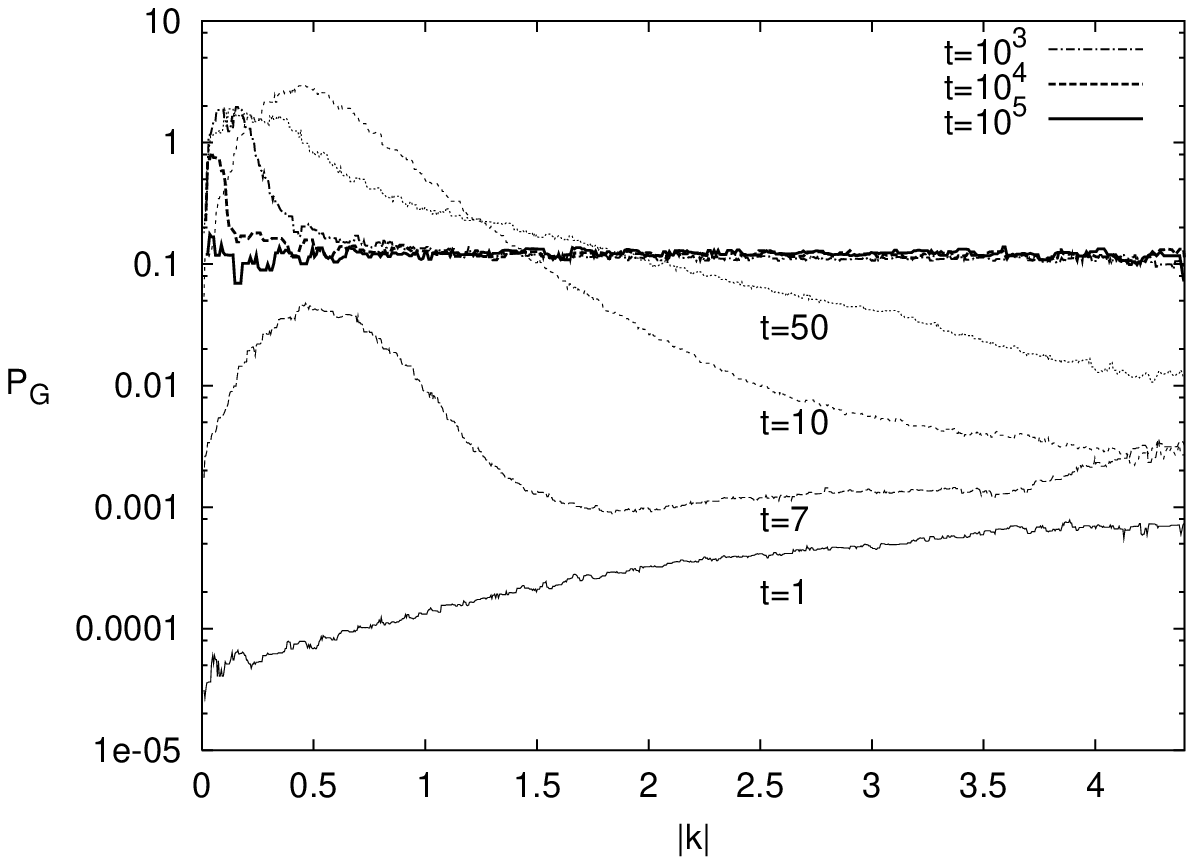}
\end{center}
\vspace*{-0.8cm}
\caption{Time evolution of $\kabs$ power spectra of Higgs (left) and
Goldstone
(right) field components ($512\times512$ lattice,\, $h=0.0025/\sqrt{6}$ )}
\label{powerspectra}
\end{figure}

The early time evolution can be characterised by the variation of the
kinetic power spectra of the Higgs and Goldstone fields (see Fig.
\ref{powerspectra}). After dephasing the potential and kinetic energy of
each mode is balanced in itself, therefore the energy content of different
modes may be characterised by the kinetic power spectrum
\be
P_{H,G}(\kabs)=\overline{{\dot\Phi_{H,G}}^2}^{\Omega_{\bf k}},
\ee
where the averaging should be taken over the polar angle in momentum
space. 
In the final equilibrium state 
this quantity does not depend neither on $\kabs$,
{nor on which field it refers to}. Its value
corresponds to the temperature, which
{was measured} in our case to be
$P_{H,G}(\kabs)\equiv T^{kin}=0.125\pm0.0005$. 
This kinetic temperature can obviously be defined also
out-of-equilibrium
{as $T^{kin}=\overline{{\dot\Phi_{H,G}}^2}^{{\Omega_{\bf k}}}$}.

The initial parametric peak in the power spectrum of the Higgs field
disappears rapidly. It proceeds mainly through decays into pairs of the
lighter Goldstone modes. This damping is reflected by an energy transfer
from the Higgs to the Goldstone modes. In addition, the excitation of
spinodally unstable Higgs modes also contributes to the relaxation, as it
was shown in Subsection \ref{ss:time-history}.

The parametric low $\kabs$ Goldstone peak, however, survives as long as
$t=10^4\dots 10^5$. During its long relaxation the energy is slowly
transferred back to the Higgs modes.  This process qualitatively corresponds
to off-shell emission of Higgs waves. The final equilibrium is reached
in a slow, non-exponential relaxation of $T^{kin}_G$ to $T^{kin}_H$.

The mechanism sketched above is suggested by the structure of the analytic
formulae derived in the perturbative relaxation analysis, which should be
relevant at least to the large time asymptotics of the evolution. The
forward time evolution of an initial configuration is determined by the
classical self-energy function
$\Pi (\xvon,t)$:
\be
\Phi_{H,G}(t,\k)=\int\limits^{\infty}_{-\infty}\frac{dk_0}{2\pi}
\frac{z(\k)}{k^2-m_{H,G}^2-\Pi_{H,G}(k)}e^{-ik_0t},
\label{forward}
\ee
where $z(k)=ik_0F(\k)-P(\k)$ is determined by the corresponding
initial configuration
$F(\k)=\,\,\Phi(t=~0,\k)$, $P(\k)=\partial_t\Phi(t=0,\k)$.

For the calculation of the Fourier transform of $\Pi (\xvon ,t)$ 
the procedure described in \cite{Aarts97,jako97} was used. 
(For the quantum treatments, which lead for
large occupation numbers to the same results, see
\cite{boyanovsky,mrow90}). 

First one makes the shift
$\Phi_1\to\Phi_1+\bar\Phi$ in order to describe the constant equilibrium
background, where $\bar\Phi$ is the classical ensemble average of
$\Phi_H(\xvon)$. Its direction is actually selected by the explicit 
symmetry breaking. Then in the linear approximation one finds
$\Phi_H(\xvon)\approx\Phi_1(\xvon)$ 
{and 
$\Phi_G(\xvon)\approx\Phi_2(\xvon)\Phi_{H,0}/{\bar\Phi}$.
We refer to Appendix \ref{calss} for the detailed derivation of the self 
energies, which read now for the Higgs and Goldstone fields, as follows:
\bea
\Pi_{H}(k)
&=&-\left(\frac{\lambda\bar\Phi}{3}\right)^2T\int\frac{d^2q}{(2\pi)^2}
\int\frac{d^2p}{(2\pi)^2}\int d^2xdte^{i(k_0 t-{\bf k}{\bf x})}
e^{i({\bf q}-{\bf p}){\bf x}}\Theta(t)\times\nonumber\\
&&\times\bb
\frac{9\sin \omega_{H,{\bf q}}t}{\omega_{H,{\bf q}}}
\frac{\cos \omega_{H,{\bf p}}t}{\omega_{H,{\bf p}}^2}+
\frac{\sin \omega_{G,{\bf q}}t}{\omega_{G,{\bf q}}}
\frac{\cos \omega_{G,{\bf p}}t}{\omega_{G,{\bf p}}^2}
\eb,\\
\Pi_{G}(k)
&=&-\left(\frac{\lambda\bar\Phi}{3}\right)^2T\int\frac{d^2q}{(2\pi)^2}
\int\frac{d^2p}{(2\pi)^2}\int d^2xdte^{i(k_0 t-{\bf k}{\bf x})}
e^{i({\bf q}-{\bf p}){\bf x}}\Theta(t)\times\nonumber\\
&&\times\bb
\frac{\sin \omega_{G,{\bf q}}t}{\omega_{G,{\bf q}}}
\frac{\cos \omega_{H,{\bf p}}t}{\omega_{H,{\bf p}}^2}+
\frac{\sin \omega_{H,{\bf q}}t}{\omega_{H,{\bf q}}}
\frac{\cos \omega_{G,{\bf p}}t}{\omega_{G,{\bf p}}^2}
\eb,
\eea
where $\omega_{H/G,{\bf k}}^2=m_{H/G}^2+|{\bf k}|^2$. 
{The relevant mass values were calculated numerically from the relaxed field 
configurations with the method developed in Subsection \ref{ss:metastable}.} 
At the parameter values specified in our simulations the typical masses are 
$a m_H\approx 1.3$ and $a m_G\lesssim 0.033$. 

The imaginary part of the self energy, which accounts for the damping
phenomena, may be extracted using the principal value theorem. There is a
technically important difference between our present formulae and the ones
derived in Subsection \ref{ss:dispersion}. It follows from the fact that the
system investigated in this section is defined in two space dimensions. This 
circumstance modifies the volume element in the momentum space and an integrand 
of more complicated analytical structure appears when calculating the imaginary
part of the classical self energy function. The structure of the cuts in the  
Higgs and Goldstone self-energy functions is presented in Fig.
\ref{Fig:cuts}~.

\begin{figure}[htbp]
\begin{center}
\includegraphics[width=7.5cm]{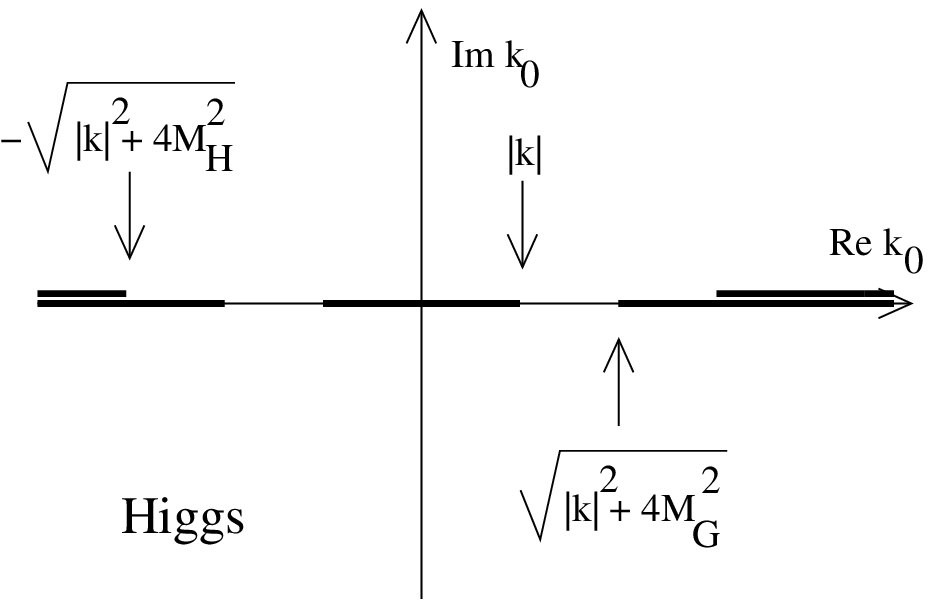}
\hspace*{1cm}
\includegraphics[width=7.5cm]{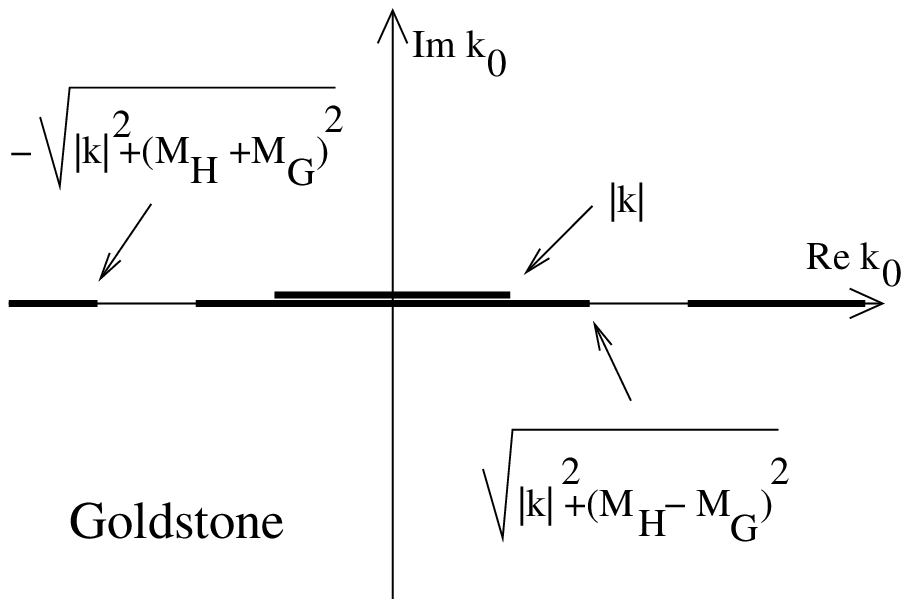}
\end{center}
\vspace*{-0.8cm}
\caption{The analytical structure of the classical Higgs and Goldstone
self-energy functions.}
\label{Fig:cuts}
\end{figure}

The explicit expressions of the imaginary parts are the following:
\bea
\nonumber
\Im\Pi_H
&=&-\left(\frac{\lambda\bar\Phi}{2}\right)^2\Theta(-k^2)\frac{2T}{\pi}
\frac{1}{R_H}
\left(\frac{\pi}{2}
-\arctg\frac{R_H}{|k_0|\sqrt{-k^2}}\right)\nn
&&-\left(\frac{\lambda\bar\Phi}{6}\right)^2\Theta(-k^2)\frac{2T}{\pi}
\frac{1}{R_G}
\left(\frac{\pi}{2}
-\arctg\frac{R_G}{|k_0|\sqrt{-k^2}}\right)\\
&&-\left(\frac{\lambda\bar\Phi}{2}\right)^2\Theta(k^2-4m_H^2)
\frac{T}{R_H}-\left(\frac{\lambda\bar\Phi}{6}\right)^2\Theta(k^2-4m_G^2)
\frac{T}{R_G},
\label{kiintegraltpihiggs}\nn
\Im\Pi_G&=&
-\left(\frac{\lambda\bar\Phi}{6}\right)^2 T
\left(P_G+P_H\right)\Theta(k^2)\Theta(k^2-(m_H+m_G)^2)\nn
&&-\left(\frac{\lambda\bar\Phi}{6}\right)^2 T
\left(P_G-P_H\right)
\Theta(k^2)\Theta((m_H-m_G)^2-k^2)\nn
&&-\left(\frac{\lambda\bar\Phi}{6}\right)^2
\frac{2T}{\pi} \Theta(-k^2)\left[ P_H
\arcsin\frac{k_0(m_G^2-m_H^2-k^2)}{\kabs\sqrt{\Delta}}\right.\\
&&\left.\qquad\qquad+P_G
\arcsin\frac{k_0(m_H^2-m_G^2-k^2)}{\kabs\sqrt{\Delta}}\right],
\label{kiintegraltpigoldstone}
\eea
with
\bea
&\Delta=(k^2-m_H^2+m_G^2)^2-4m_G^2k^2,&\nn
&R_{H,G}=\sqrt{k^4+4\kabs^2m_{H,G}^2},\qquad
P_{H,G}=1/\sqrt{\Delta+4m_{H,G}^2k_0^2}.&
\eea

It is the off-shell damping coming from the contributions of the cuts to
(\ref{forward}) which has a direct impact on the {late} time evolution of
the OP we are analysing. We followed the method, developed in
\cite{boyanovsky96b,boyanovsky}, which determines the leading power law
tail of the relaxation.

{Contrary to the $3+1$ dimensional case in $2+1$ dimensions the integrand of
Eq. (\ref{forward}) has extra cuts. These can, however, be directed in a way
that either they do not contribute (Goldstone case) or their contribution is
suppressed by a factor of $e^{-m_Ht}$ (Higgs case). The large time asymptotics 
of the two field components is the following: }
\bea
\nonumber
\Phi_H(t,\k)&=&
\frac{T}{\pi t}
\left(\frac{\lambda\bar\Phi}{6}\right)^2
\left[
\frac{1}{m_H^5}
\left(
\frac{P(\k)\cos(t\Omega_{H,{\bf k}})}{\Omega_{H,{\bf k}}}
-F(\k)\sin(t\Omega_{H,{\bf k}})
\right)\right.\nn
&&+\left.
\frac{1}{m_G(4m_G^2-m_H^2)^2}
\left(
\frac{P(\k)\cos(t\Omega_{G,{\bf k}})}{\Omega_{G,{\bf k}}}
-F(\k) \sin(t\Omega_{G,{\bf k}})
\right)\right]\nn
&&-
\frac{T}{(2\pi t)^{3/2}}
\left(\frac{\lambda\bar\Phi}{6}\right)^2
\frac{2\sqrt{\kabs}}{m_H^4}
\left(\frac{9}{m_H^2}+\frac{1}{m_G^2}\right)\times\\
&&\quad\times\left(\frac{P(\k)}{\kabs}\cos(\kabs t-\frac{\pi}{4})
-F\sin(\kabs t- \frac{\pi}{4})\right),\nn
{\bar\Phi\over {\Phi_{H,0}}}\Phi_G(t,\k)&=&
-\left(\frac{\lambda\bar\Phi}{6}
\right)^2
\frac{T}{\pi t}
\left[
\frac{1}{m_H^2(m_H-2m_G)^2}
\left(\frac{1}{m_G}-\frac{1}{m_H}\right)\right.\nn
&&\times
\left(
\frac{P(\k)\cos \Omega_{H-G,\k}t}{\Omega_{H-G,\k}}
- F(\k) \sin t\Omega_{H-G,\k}
\right)\nn
&&+\frac{1}{m_H^2(m_H+2m_G)^2}
\left(
\frac{1}{m_G}+\frac{1}{m_H}
\right)\\
&&\left.\times\left(
\frac{P(\k)\cos \Omega_{H+G,\k}t}{\Omega_{H+G,\k}}
- F(\k) \sin t\Omega_{H+G,\k}
\right)\right],
\label{goldtimedep}
\eea
where the threshold frequency values, appearing above are given by
$\Omega_{H/G,{\bf k}}^2={\bf k}^2+m^2_{H/G}$ and
$\Omega_{H\pm G,{\bf k}}^2={\bf k}^2+(m_H\pm m_G)^2$.

In order to obtain numerical evidence for the presence of the predicted
power law tail, we measured the quadratic spatial fluctuation moments
of both fields as a function of time. They are defined as
$Fluct(t)=\overline{\Phi(\xvon,t)^2}^V-\left(\overline{\Phi(\xvon,t)}^V\right)^2$.
Substituting (\ref{goldtimedep}) into this definition we get $\sim t^{-2}$ 
like relaxation for late times. This behaviour is indeed observed as the
fitted {power} shows in Fig. \ref{goldstonedecay} for the example of the
Goldstone mode.

\begin{figure}[htbp]
\begin{center}
\includegraphics[width=8cm]{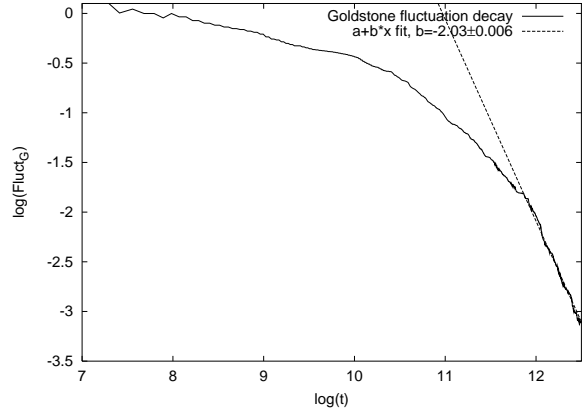}
\end{center}
\vspace*{-0.8cm}
\caption{Relaxation of Goldstone spatial fluctuations with time
on ($e$-based) log-log scale plot.
($256\times 256$ lattice, $h=0.0001/\sqrt{6}$, $m_G=(62\pm5)\cdot10^{-4}$,
$m_H=1.329\pm0.001\cdot10^{-3}$, the average over 69 runs is shown.)}
\label{goldstonedecay}
\end{figure}
The formulation of the theory upon which the above formulae were obtained
actually uses ensemble averages over the initial data of the classical
evolution. We found, that although every single run seems to relax even
quantitatively in the same way, this late time evolution can be extracted
from the noisy data only, if this averaging is performed indeed (116 runs
were involved). We could recognise a $\sim1/t^2$ decay for Higgs modes too,
but because of the low signal/noise rate we could not do quantitative
analysis. In particular the analysis is made difficult by the fact, that
Higgs modes decay rapidly and vanish in the noise before the power law tail
would be reached.

Relaxation behaviour of OP ($\Phi_H(${\bf k}=0$)$) was also investigated, and
--- in accordance with the expectation above --- an oscillation damped
by $\sim t^{-1\pm0.02}$ was found. Moreover, this oscillation, is observed
around a value, monotonically approaching its equilibrium value as
$\sim t^{-2\pm0.01}$. This latter behaviour is explained by the fact,
that the exact equation of motion for OP contains $Fluct(t)$ as a time
dependent parameter \cite{khlebnikov,maxwell}.
Its $\sim t^{-2}$ like behaviour is inherited by
the slowly varying part of OP. (The errors of the exponents come from
averaging over 92 runs on a $256\times256$ lattice.)

\subsection{The $h\not=0$ equilibrium}

When one follows the evolution of the system for long times ($t=10^6$),
there seems no further relaxation to take place, i.e. the initial signal has
been lost in the thermal noise. In order to convince ourselves of having
arrived to thermal equilibrium we compare the measured values of
$\overline{\Phi_H}^V$, and the masses $m_G$ and $m_H$ to the analytical
estimates of these quantities coming from the one-loop lattice effective
potential evaluated for the same size lattice as used in the simulation. In
the analytic expression we used the measured kinetic temperature. The
equilibrium masses $m_G$ and $m_H$ were determined in the simulation both
from the corresponding correlation functions and by fitting the oscillatory
motion around the equilibrium of the corresponding OP-components as
described in Subsection \ref{ss:metastable} and \cite{maxwell}.

We have checked that the system with the present initial conditions is
deeply in the Coulomb phase i.e. the temperature was about four times
smaller than the Kosterlitz-Thouless critical temperature $T_{KT}$. In the
absence of explicit symmetry breaking on the critical line between $T=0$ and
$T_{KT}$ the Goldstone correlation length is expected to diverge with the
lattice size. The explicit symmetry breaking parameter $h$ acts as an
infrared regulator. Indeed, the measured Goldstone correlation length is
found to be proportional to $L$, but for the largest sizes ($L=256$), where
the IR cutoff $h$ begins to dominate. On the other hand, the inverse
correlation length in the Higgs direction is finite, its value being $3\%$
smaller than the two-loop mass and $6\%$ smaller than the one-loop value.

A very good agreement of $\overline{\Phi_H}^V$ with $\bar\Phi$ is found where
$\bar\Phi$ is given by the minimum of the effective potential in the radial 
direction, the relative deviation being ${\cal O}(10^{-4})$. The discrepancy 
between the measured (with the method described in Subsection 
\ref{ss:metastable} or \cite{maxwell}) and 
calculated masses was less than $1\%$ and $5\%$ for the Higgs and Goldstone 
modes, respectively, the perturbative values being systematically smaller. Our 
measurements of the Goldstone mass became very noisy for small values of 
$h\lesssim0.00001/\sqrt{6}$.

\begin{figure}
\begin{center}
\includegraphics[width=8cm]{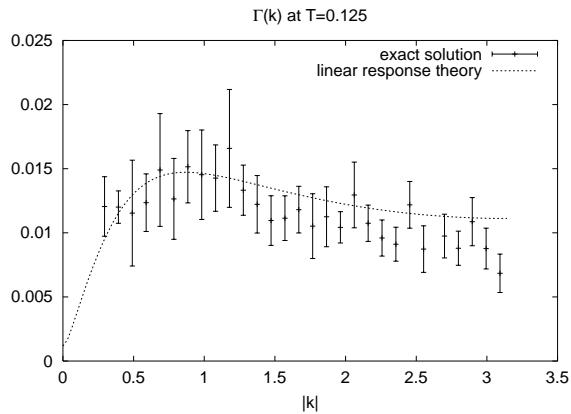}
\end{center}
\caption{Numerical and analytical values for Goldstone on-shell decay
rate as a function of $\kabs$
($128\times128$ lattice, $h=0.0025/\sqrt{6}$, $m_G=0.0319$,
$m_H=1.33$, $\bar\Phi=0.96$, $T=0.125$)}
\label{gammacomparison}
\end{figure}

In the equilibrium system one can proceed to ``experiments'', which check
the correctness of the on-shell decay rates computed in the linear response
theory. These are the simple zeros of the denominator of Eq.(\ref{forward})
which determine the exponential damping of on-shell excitations.  The
damping rates are obtained by substituting $\kabs=k_0$ into Eqs.
(\ref{kiintegraltpihiggs}) and (\ref{kiintegraltpigoldstone}). To leading
order in $(\lambda\bar\Phi/6)^2T$ the rates read (assuming $m_H>2m_G$) as
\bea
&&\Gamma_H=-\frac{\Im\Pi_H(k^2=m_H^2)}{2\omega_{H,\bf k}}=
\left(\frac{\lambda\bar\Phi}{6}\right)^2\frac{T}{2\omega_{H,\bf k}}
\frac{1}{\sqrt{m_H^4+4\kabs^2m_G^2}}\\
&&\Gamma_G=\left(\frac{\lambda\bar\Phi}{6}\right)^2
\frac{T}{2\omega_{G,{\bf k}}}
\left[
\frac{1}{\sqrt{(m_H^2-2m_G^2)^2+4m_G^2\kabs^2}}
-\frac{1}{m_H\sqrt{m_H^2+4\kabs^2}}
\right].
\eea

On equilibrium configurations single $\bf k$-modes have been superimposed
with an amplitude between $0.01\dots0.2$. Time evolution of these excited
modes showed perfect exponential decay. The exponents did, however, depend
on the amplitude. For small amplitudes the fit was unstable, the mode was
quickly lost in the noise. For bigger values nonlinear effects did show up.
Therefore we were looking for a plateau in the amplitude dependence of the
decay rates between these two extremes. Then its value has been compared to
the analytical estimates (see Fig.\ref{gammacomparison}).  The error bars
show how well-defined plateau could be found. This numerical experiment was
performed both for Higgs and Goldstone waves, but the errors for Higgs decay
were too large again. The value of $\Gamma_H$ computed from the linear
response theory was reproduced within an accuracy of $\pm50\%$, but we can
not say anything about its predicted $\k$-dependence. The reason for this
seems to be, that the evolution becomes very soon nonlinear as we try to
increase the excitation amplitude higher than the noise level.

As was indicated in \cite{hasenfratz} finite size effects may be highly
important in systems with Goldstone-like excitations. This circumstance made
necessary the use of finite volume perturbation theory in the analysis
above. Neither the lattice summed perturbative effective potential, nor the
numerically found mass values showed any $L$ dependence if $L>64$. We have
checked for the $L$ independence in all cases where the exact evolution and
linear analysis were confronted (e.g. the values of damping rates).

\subsection{Diffusion to the $h=0$ equilibrium}

The most care must be taken when the macroscopic magnetisation $M(L)$ is
considered in a system with $h=0.$ For finite volume, the ensemble average
computed from the low temperature spin-wave approximation is non-zero
\cite{archambault} even in the absence of explicit symmetry breaking:
\be
M^2(L)\equiv\left(\overline{\Phi_1}^V\right)^2+
\left(\overline{\Phi_2}^V\right)^2 
\approx (2L)^{-T/(2\pi\bar\Phi^2)}.
\ee
For each realization of the canonical ensemble observed in Monte Carlo
simulations the magnetisation vector has a well-defined direction
$\theta =\arctan\left(\overline{\Phi_2}^V/\overline{\Phi_1}^V\right)$.

It is an interesting question, by what mechanism the MWH-theorem is realized
as $L\to\infty$.  It was shown in Monte Carlo simulations that the erasure
of the order is realized by the diffusion like displacement of the direction
of $\theta$ \cite{archambault}.

In our real time study, for the investigation of the onset of the finite
volume version of MWH-theorem a number of runs were continued from the
equilibrium state reached with $h\neq 0$, after the magnetic field was
switched off. In each individual system OP begins to circle around the
origin with an average radius $M(L)$. This motion is probably due to a
random nonzero angular momentum of the initial configuration. Subtracting
the angular momentum of OP, the relevant diffusion-like motion remains.  MC
studies employing first order Monte Carlo ``time''-evolution of
non-equilibrium one dimensional systems have shown that the sign for SSB is
the exponential decrease of the diffusion constant with $L$
\cite{mukamel}. In agreement with the expectations based on the 
MWH-theorem we find in the present case that the diffusion constant
decreases with increasing $L$ following  a power law with the exponent 
$-1.16\pm0.1$. 

\subsection{Conclusions}
In conclusion we can state that the numerical study of the Hamiltonian
dynamics of the classical O(2) symmetric scalar model in 2+1 dimensions
provides a non-trivial check of our understanding of real time relaxation
phenomena. Numerical results both for the late-time OP-asymptotics and for
the decay rate of on-shell waves (with well-defined wave vector $\bf k$)
were found to be in agreement with the linear response theory. In the $h\neq
0$ equilibrium the agreement of the masses coming from the perturbatively
calculated effective potential with the numerically established excitation
masses was also verified. Finally, we have demonstrated the real time
manifestation of the finite volume MWH-theorem by measuring the large $L$
asymptotics of the angular diffusion rate of the macroscopic order
parameter.

%% file: higgs.tex
\chapter{Real time dynamics of the Higgs effect}

The real-time transformation of light gauge fields into massive intermediate
vector bosons is a subject of increasing cosmological interest.

Recently, Garcia-Bellido {\it et al.} \cite{Garcia99} (see also
\cite{Trodden}) assumed that the coupled inflaton+Higgs system was far out
of equilibrium when the energy density of the expanding Universe has passed
the point corresponding to the thermal Higgs transition. With the
supplementary condition that the reheating temperature after the exit from
the inflationary period did not exceed the electroweak critical temperature,
they demonstrated in a (1+1)-dimensional toy model, that large enough
matter--antimatter asymmetry could have survived till today. The existence
of non-equilibrium Higgs transitions in 3+1 dimensions has been demonstrated
by Rajantie {\it et al.} \cite{Rajantie00}.  In these investigations the
classical mean field equations were used with renormalised and
temperature-dependent couplings.

Traditionally the evolution of the baryon asymmetry through the electroweak
phase transition, accompanying the onset of the Higgs effect has been
thoroughly discussed, under the assumption that the system is in thermal
equilibrium, with specific emphasis on the high-temperature sphaleron rate
\cite{Moore01,Moore00,Ambjorn95}. Near equilibrium, the gauge-invariant HTL
action dominates the influence of the thermalized quantum fluctuations on
the motion of the mean fields with  $k\simeq{\cal O}(gT)$ 
\cite{Braaten90,Frenkel90,Blaizot94}. This term was taken into account in 
real-time simulations of the sphaleron rate \cite{Moore00} 
(see also \cite{Hindmarsh99}).

The framework for the theoretical study of the baryon asymmetry has been
somewhat modified with the realization that, within the Standard Model, the
Higgs effect sets in via smooth phase transformation, characterised by an
analytic variation of the order parameter
\cite{Kajantie97,Karsch97,Guertler97,Csikor99}.  Under such
circumstances the expectation value of the Higgs field is different from
zero also above the electroweak energy scale.  In this context it is
remarkable that the leading HTL correction is insensitive to the presence of
the scalar condensate \cite{Rebhan95, Moore00}.

In this Chapter we go one step beyond the HTL correction.  Generalising the
discussion of the damping rates of scalar fields in the previous chapters,
we complete the kinetic theory of the pure gauge fields by appropriate terms
reflecting the presence of scalar fields. Our aim is to find the leading
effect due to the presence of an arbitrary constant background field:
$\bar\Phi$. In equilibrium, such a background is sufficient for the
calculation of the effective potential (see for example \cite{Kapusta89}),
but not the full effective action. Similarly here, we restrict somewhat the
generality of our correction terms to the equations of motion, namely the
corrections to the $\Phi^2$--$A^2$ vertex will be determined neglecting the
non-local effects in the scalar field.

The high-temperature limit of the resulting expression of the induced
current shows that the next-to-leading ``mass and vertex'' corrections can
be uniquely decomposed into gauge-fixing invariant and gauge-fixing
dependent modes. In this way we can propose an expression of wider
applicability for the Landau damping effect also including the effects of a
scalar condensate. We propose a physical gauge for the
numerical solution of these equations where the gauge-fixing dependent mode
does not propagate.

Baacke and his collaborators \cite{Baacke97,Baacke99,Heitmann01} have
applied, in a series of papers, a complementary approach to the Higgs
effective action without assuming the presence of a general gauge
background.

A non-trivial mean gauge field configuration reacts back on the scalar
condensate. Our final goal is to derive a coupled set of equations for the
scalar condensate and the mean gauge field, which is appropriate for
studying the real-time onset of the Higgs effect and the variation of the
sphaleron generation and decay rate.

\rt{The generalisation of the derivation to non-Abelian systems presents only
technical complications.  The Abelian calculation has been performed both in
the standard real-time generator functional formulation \cite{LeBellac}
and with an iterated solution of the linearised Heisenberg equations of the
quantum fluctuations \cite{Boyanovsky98,Heinz86,mrow90}. We shall use 
below the latter approach, where the adequate two-point functions enter more
naturally. }

\section{Mean Field Equations in the Abelian Higgs model}

The Lagrangian of the model in the $O(2)$ notation is the following:
\begin{equation}
\cL = -\frac14 \hat F_{\mu\nu}\hat F^{\mu\nu} +\frac12 
(\d_\mu\hat\Phi)^2 - \frac12 m^2\hat \Phi^2 + 
e\left(\hat\Phi_2\d_\mu\hat\Phi_1-\hat\Phi_1\d_\mu\hat\Phi_2\right)\hat A^\mu  
 +\frac{e^2}2 \hat A^2\hat \Phi^2 -\frac\lambda{24}(\hat \Phi^2)^2,
\label{lagrangian}
\end{equation}
where $\hat \Phi=(\hat \Phi_1,\hat \Phi_2)$.
We split the
fields into a mean field ($A_\mu,\Phi$) and a fluctuation contribution
\begin{equation}
  \hat A_\mu =A_\mu+a_\mu,\quad \hat \Phi=\Phi+\ph,\quad
\avr{a_\mu}=\avr{\ph}=0
\end{equation}
at any time (averaging is understood with respect to the initial
density matrix). We assume that the scalar background is constant and
that it points to the $\Phi_1$ direction, its value being $\bP$. We
use in the sequel the notations $m_W=e\bP,\, m_H^2=m^2+\lambda
\bP^2/2,\, m_G^2=m^2+\lambda\bP^2/6$, although these are not the
vacuum values.

We fix a 't Hooft $R_\xi$ gauge by changing the Lagrangian as
\begin{equation}
  \cL\to\cL -\frac1{2\xi}(\d_\mu \hat A^\mu + \xi m_W\ph_2)^2 -\bar
  c(\d^2+\xi m_W^2 +e\xi m_W\ph_1)c,
\label{gaugeEOM}
\end{equation}
where $c$ is the ghost field. In this way the field $\ph_2$ receives
an additional mass contribution $\xi m_W^2$.

The operatorial equation of motion (EOM) separately induces EOMs for
the mean fields and the fluctuations. The average of the operatorial
EOM for the gauge field reads as
\begin{equation}
  \left[(\d^2+m_W^2)g_{\mu\nu} - \left(1-\frac1\xi\right) \d_\mu\d_\nu
  \right] A^\nu(x) + j^{ind}_\mu(x) =0,
\label{gauge_eq}
\end{equation}
with the induced current
\be
\label{ind_curr}
j^{ind}_\mu(x) = e\avr{j_\mu(x)} + 
2e^2\bP\avr{a_\mu(x)\ph_1(x)} + e^2\avr{\ph^2(x)}A_\mu(x)
+ e^2\avr{a_\mu(x) \ph^2(x)},
\ee
where $j_\mu(x)=\ph_2(x)\d_\mu\ph_1(x)-\ph_1(x)\d_\mu\ph_2(x)$.
We will perform the
calculations at the one-loop level, when the last term does not
contribute. The subtraction of (\ref{gauge_eq}) from the full equation
yields the EOM for the fluctuations. At one loop it is sufficient to
consider only the equations linearised in the fluctuations.  On the
other hand, since we want to calculate $j^{ind}_\mu$ in the linear
response approximation, we can neglect all terms non-linear in $A$.
These assumptions make the equations very simple:
\begin{eqnarray}
\nonumber
&&\left[-(\d^2+m_W^2)g_{\mu\nu} +\left(1-\frac1\xi\right)
 \d_\mu\d_\nu \right] a^\nu(x) - 2e^2\bP \ph_1(x) A_\mu(x) = 0,\nn
&& \left(\d^2+m_H^2\right) \ph_1(x) + 2eA(x)\cdot\d\ph_2(x) +e\ph_2(x)\d A(x) 
-2e^2 \bP A(x)\cdot a(x) =0,\\
&& \left(\d^2+m_G^2+\xi m_W^2\right) \ph_2(x)-2eA(x)\cdot\d\ph_1(x) 
-e\ph_1(x)\d A(x)=0.
\label{linopeom}
\end{eqnarray}
The ghost fields follow a free EOM, so that they do not influence the
present calculation. These equations are solved to linear order in the
$A$ background:
\begin{eqnarray}
\nonumber
  a_\mu(x) & =& a_\mu^{(0)}(x) - 2e^2\bP \int\! d^4z\,
  G^R_{\mu\nu}(x-z) A^\nu(z) \ph_1^{(0)}(z),\nn
  \ph_1(x) & =& \ph_1^{(0)}(x) + e\int\! d^4z\, G^R_1(x-z) \left[ 2
  A^\mu(z) \d_\mu\ph_2^{(0)}(z) + (\d A)(z)\ph_2^{(0)}(z)  - 2e\bP
  A^\mu(z)a_\mu^{(0)}(z)\right],\\
  \ph_2(x) & =& \ph_2^{(0)}(x) -e \int\! d^4z\, G^R_2(x-z) \left[
  2A^\mu(z) \d_\mu \ph_1^{(0)}(z) + (\d A)(z)\ph_1^{(0)}(z) \right],
\end{eqnarray}
where the superscript zero denotes the solutions of the free EOM, and
$G^R$'s are the free retarded Green functions\footnote{Here the
  definition $KG^R=-\delta$ is used, where $K$ denotes the free
  kernel.}.

For the induced current (\ref{ind_curr}) we need the expectation
values of certain local products of the fluctuating fields. For
example, it directly follows from the above equations (with
$\avr{AB}_0\equiv\langle A^{(0)}B^{(0)}\rangle$), that
\begin{eqnarray}
\nonumber
  \avr{\ph_2(x)\d_\mu\ph_1(x)} =&& e\!\int\! d^4z \,\biggl\{
  \d_\mu^xG^R_1(x\!-\!z) \biggl[ 2 A^\nu(z)
  \avr{\ph_2(x)\d_\nu^z\ph_2(z)}_0 + (\d A)(z) \avr{
  \ph_2(x)\ph_2(z)}_0\biggr]\\ 
  - && G^R_2(x\!-\!z)\biggl[ 2 A^\nu(z)\avr{\d_\mu\ph_1(x)
  \d_\nu\ph_1(z)}_0 + (\d A)(z) \avr{\d_\mu\ph_1(x) \ph_1(z)}_0
  \biggr]\biggr\}.
\end{eqnarray}
We define the local products in a symmetric way (i.e.
$\frac12\avr{\ph_2(x)\d_\mu\ph_1(x) + \d_\mu\ph_1(x)\ph_2(x)}$) and
introduce
\begin{equation}
  \Delta(x-z) = \frac12\avr{\ph(x)\ph(z) + \ph(z)\ph(x)}_0.
\end{equation}
Then the above expression can be written in Fourier space as
\begin{equation}
  \avr{\ph_2\d_\mu\ph_1}(Q) = -e A^\nu(Q) \pint4p p_\mu
  (2p-Q)_\nu\left[ G^R_1(p) \Delta_2(Q-p) +
  G^R_2(Q-p)\Delta_1(p)\right].
\end{equation}
The evaluation of other expectation values goes along the same
line, finally giving
\begin{eqnarray}
\nonumber
  && e\avr{j_\mu}(Q) = -e^2 A^\nu(Q) \pint4p (2p-Q)_\mu (2p-Q)_\nu\left[
  G^R_1(p) \Delta_2(Q-p) + G^R_2(Q-p)\Delta_1(p)\right],\nn
  && 2e^2\bP \avr{a_\mu\ph_1}(Q)= -4e^2 m_W^2 A^\nu(Q) \pint4p \left[
  G^R_{\mu\nu}(p) \Delta_1(Q-p) + G^R_1(Q-p)\Delta_{\mu\nu}(p)\right],\\
  && e^2 A_\mu\avr{\ph^2} = e^2 A_\mu(Q) \pint4p
  \left[\Delta_1(p)+\Delta_2(p) \right].
\label{two_products}
\end{eqnarray}
Assuming that the free fluctuations are in thermal equilibrium,
the propagators can be related to the corresponding spectral functions
\cite{Landsmann}, which are the discontinuities of the free kernels of
eq.~\eqref{linopeom}. Introducing
\begin{equation}
  \Delta_{m^2}(p)= 2\pi\left(\frac12 + n(|p_0|)\right)
  \delta(p^2-m^2), \quad \textrm{and}\quad G^R_{m^2}(p) =
  \frac1{p^2-m^2 + i\ep p_0},
\end{equation}
where $n$ is the Bose--Einstein distribution, we have
\begin{equation}
  \Delta_1= \Delta_{m_H^2},\quad  \Delta_2 = \Delta_{m_G^2+\xi
  m_W^2},\quad \Delta_{\mu\nu} = -g_{\mu\nu} \Delta_{m_W^2} +
  \frac{p_\mu p_\nu}{m_W^2} \left( \Delta_{m_W^2} - \Delta_{\xi m_W^2}
  \right),
\end{equation}
and analogously for the corresponding $G^R$'s.

\section{High-temperature expansion beyond the HTL approximation}

The leading HTL term of $j^{ind}_\mu$ in the high-temperature
expansion comes from $e\avr{j_\mu}$ by neglecting all the masses
\cite{Rebhan95}.  Our aim is to calculate the first subleading term
proportional to $\bP^2$ instead of $T^2$ in the high-temperature
expansion.

We emphasise that $j^{ind}(Q)= j^{ind}(-Q)^*$, therefore all partial
contributions that are odd under hermitian $Q$-reflection can be
freely omitted.

A useful formula, which simplifies the calculations, is the following
\cite{JPPSz}:
\begin{eqnarray}
\nonumber
  I_F(Q,m^2,M^2) &=& \pint4p F(Q,p) \left(G^R_{m^2}(p)\Delta_{M^2}(Q-p) +
  G^R_{M^2}(Q-p)\Delta_{m^2}(p)\right) \\ &=& \pint4p F(Q,\frac
  Q2-p)\frac{\Delta_{m^2}(p-\frac Q2) - \Delta_{M^2}(p+\frac Q2)} {2pQ
  +m^2-M^2}.
\label{fund_eq}
\end{eqnarray}
The tensorial structures appearing in (\ref{two_products}) when cast
into the form (\ref{fund_eq}) imply the appearance in $F$ of the
$p$-dependent terms $p_\mu p_\nu, p_\mu Q_\nu +p_\nu Q_\mu$ and also
of terms independent of $p$.

\subsection{Even terms}

We start the discussion with the terms even under $p$-reflection and
introduce the notation $f(p)=(F(Q,Q/2-p)+F(Q,Q/2+p))/2$. Here the
$m^2-M^2$ term in the denominator can be neglected since in the mass
expansion 
\begin{equation}
  \pint4p f(p) (M^2-m^2) \frac{\Delta_0(p-\frac Q2) - \Delta_0(p+\frac
  Q2)}{(2pQ)^2} =0
\end{equation}
because of the odd $p\to-p$ behaviour of the integrand. Then we only
need to expand the difference of $\Delta$'s of the numerator.  The
numerator can be expanded with respect to $Q$, when low-momentum mean
fields are considered. The leading term of the gradient expansion
gives zero, again because of the $p$-odd integrand.

The first non-zero contribution is therefore
\begin{equation}
  I_F(m^2,M^2) = -\frac{Q_\rho}2 \pint4p \frac{f(p)}{2pQ} \frac\d{\d
  p_\rho}(\Delta_{m^2}(p) + \Delta_{M^2}(p)).
\end{equation}

Now, we treat separately the two actual cases: $f=p_\mu p_\nu$ and
$f=\,$constant. The $f=p_\mu p_\nu$ case contributes to the full
expression of $j^{ind}_\mu (Q)$
\begin{equation}
  2e^2 A^\nu(Q) Q^\rho \pint4p \frac{p_\mu p_\nu}{2pQ}\frac\d{\d
  p^\rho}(\Delta_1+\Delta_2 +\Delta_{m_W^2} -\Delta_{\xi m_W^2}).
\label{pp_contr}
\end{equation}
We introduce the field-strength tensor by $A^\nu Q^\rho=-i F^{\nu\rho}
+ A^\rho Q^\nu$. The local term proportional to $A$ reads as
\begin{equation}
  e^2 A^\rho(Q) \pint4p p_\mu \frac\d{\d
  p^\rho}(\Delta_1+\Delta_2 +\Delta_{m_W^2} -\Delta_{\xi m_W^2}).
\end{equation}
After partial integration the first two terms cancel with $e^2
A_\mu\avr{\ph^2}$ in the induced current. What remains is a
contribution vanishing at zero $m_W^2$. In the mass expansion they
give
\begin{equation}
  j_\mu^{local,1} = -e^2 (1-\xi) m_W^2 A_\mu(Q) \pint4p
  \frac{\d\Delta_0}{\d\p^2}.
\label{coupling1}
\end{equation}
For each $\Delta_{m^2}$ the field strength contribution of
(\ref{pp_contr}) is rewritten with the help of the relation
\begin{equation}
  s^\rho\frac{\d\Delta}{\d p^\rho} = -2ps \frac{\d\Delta}{\d\p^2} +
  s_0 \frac{dn(|p_0|)}{dp^0} 2\pi \delta(p^2-m^2)
\label{aux_rel}
\end{equation}
in the form
\begin{equation}
  -2ie^2 \pint4p \left( -\frac{p_\mu}{pQ} p_\nu p_\rho F^{\nu\rho}(Q)
   \frac{\d\Delta_{m^2}}{\d\p^2} + F^{\nu 0}(Q) \frac{p_\mu p_\nu}{2pQ}
   \frac{dn(|p_0|)}{dp_0} 2\pi\delta(p^2-m^2)\right).
\end{equation}
The first term drops out because of the antisymmetry of $F$. In the
second term we perform the mass expansion. After adding the different
contributions we find
\begin{equation}
\hspace*{-0.25cm}
  j_\mu^{(1)} =-4 i e^2 F^{\nu 0} \pint4p \frac{p_\mu p_\nu}{2pQ}
   \frac{dn(|p_0|)}{dp_0} \left[ 2\pi\delta(p^2) - \frac{m_H^2+m_G^2 +
   m_W^2}2 2\pi\delta'(p^2)\right]\equiv \Pi_{\mu\nu}A^\nu.
\label{jmu1}
\end{equation}
The first term of this expression is the usual HTL contribution, the
second is a mass correction to it. In the mass correction, originally,
there was $m_1^2+m_2^2+m_W^2-\xi m_W^2$, but because of
$m_2^2=m_G^2+\xi m_W^2$ the $\xi$ dependence drops out.

The $f=\,$constant contribution to the induced current appears as
\begin{equation}
  -4e^2m_W^2A_\mu(Q) Q^\rho\pint4p \frac1{2pQ} \frac{\d\Delta_0}{\d
   p^\rho}.
\end{equation}
Since it is already proportional to $m_W^2$ we have dropped the mass
dependence of the integral. Using once more the relation
(\ref{aux_rel}) we write for it
\begin{equation}
  -4e^2 m_W^2 A_\mu(Q) \pint4p \frac1{2pQ} \left(
   -2pQ\frac{\d\Delta_0}{\d \p^2} + Q_0 \frac{dn(|p_0|)}{dp_0}
   2\pi\delta(p^2)\right).
\end{equation}
The first term is again local:
\begin{equation}
  j_\mu^{local,2} = 4e^2 m_W^2 A_\mu(Q) \pint4p \frac{\d\Delta_0}{\d \p^2}.
\label{coupling2}
\end{equation}
The second term reads as
\begin{equation}
  j_\mu^{(2)} = -4e^2 m_W^2 Q_0 A_\mu \pint4p \frac1{2pQ}
  \frac{dn(|p_0|)}{dp_0} 2\pi\delta(p^2).
\label{jmu2}
\end{equation}

\subsection{Odd contributions}

Odd contributions come exclusively from $\avr{a_\mu\ph_1}$. 
Using \eqref{fund_eq} and denoting the current by $j_\mu^{(3)}$ we find
\begin{equation}
  j_\mu^{(3)} = 2e^2A^\nu \pint4p (p_\mu Q_\nu + p_\mu Q_\nu)
  \frac{\Delta_{m_W^2}(p-\frac Q2) - \Delta_1(p+\frac Q2)} {2pQ
  +m_W^2-m_1^2} - \{m_W^2 \to \xi m_W^2\}.
\end{equation}
Finally, with the help of \eqref{aux_rel}, performing the mass and
external momentum expansion according to the method followed in the
previous subsection we arrive at
\begin{equation}
  j_\mu^{(3)} = 2e^2(1-\xi)m_W^2 Q_0 A^\nu \pint4p \frac{p_\mu Q_\nu +
  p_\mu Q_\nu} {(2pQ)^2} \frac{dn(|p_0|)}{dp_0} 2\pi\delta(p^2)\equiv
  Z_{\mu\nu} A^\nu.
\label{jmu3}
\end{equation}

\subsection{Linear response and physical modes}

The full induced current is the sum of the different parts coming from
\eqref{coupling1}, \eqref{jmu1}, \eqref{coupling2}, \eqref{jmu2} and
\eqref{jmu3}:
\begin{equation}
  j_\mu^{ind} = j_\mu^{local} + j_\mu^{(1)} + j_\mu^{(2)} + j_\mu^{(3)},
\end{equation}
where $j_\mu^{local}=j_\mu^{local,1}+j_\mu^{local,2}$.

Here $j_\mu^{(1)}$ and $j_\mu^{(2)}$ are $\xi$-independent, while
$j_\mu^{local}$ and $j_\mu^{(3)}$ depend on the gauge fixing.  For the
physical characterisation of the system (for instance, damping rates)
we have to find the independently evolving modes. In order to do this
we decompose the polarisation tensor in the tensor basis, appropriate
for finite-temperature studies \cite{Buchmuller94}:
\bea
\nonumber
&&  P^T_{\mu\nu} = - g_{\mu i}(\delta_{ij} - \hat Q_i\hat Q_j) g_{\nu j},
  \quad P^L_{\mu\nu} = - \frac{Q^2}{q^2} u_\mu^T u_\nu^T,\\
&&  P^G_{\mu\nu} = \frac{Q_\mu Q_\nu}{Q^2},\quad S_{\mu\nu} =
  \frac1q\left(Q_\mu u_\nu^T + u_\mu^T Q_\nu\right),
\label{projectors}
\eea
where $Q=(q_0,\q)$, $\q^2=q^2$ and $u_\mu^T =g_{\mu0} - Q_\mu q_0
/Q^2$ was used.

Since $\Pi_{\mu\nu}$ of (\ref{jmu1}) is transverse
($Q^\mu\Pi_{\mu\nu}=0$), it is the combination of $P^T$ and $P^L$.
Introducing $\Pi_L=\Pi_{00}$ and $\Pi_T = 1/2P^T_{\mu\nu}\Pi^{\mu\nu}$
we find
\begin{equation}
  \Pi = -\frac {Q^2}{q^2} \Pi_LP^L + \Pi_T P^T.
\end{equation}
The $Z_{\mu\nu}$ term in (\ref{jmu3}) can be written as a combination
of $P^G$ and $S$. Introducing $Z_G=Z_\mu^\mu$ and $Z_S=(q_0^2 Z_G -
Q^2 Z_{00})/(qq_0)$ we find
\begin{equation}
  Z = Z_GP^G + Z_S S.
\end{equation}
Using the completeness relation $P^L+P^T+P^G=g$ the Fourier transform
of the EOM (\ref{gauge_eq}) has the following form:
\begin{equation}
  \left[(Q^2-R-\Pi_T)P^T + \left(Q^2-R+\frac{Q^2}{q^2} \Pi_L\right)
  P^L + \left(\frac1\xi Q^2 -R-Z_G\right)P^G - Z_SS\right]A =0,
\label{Eq:Abd}
\end{equation}
with $R$ defined from $ m_W^2 A_\mu +j_\mu^{local} + j_\mu^{(2)}
\equiv R A_\mu .$

Since $S$ is not a projector, but mixes the subspaces belonging to
$P^L$ and $P^G$, in this two-dimensional subspace the inverse
propagator matrix still has to be diagonalized:
\begin{equation}
  \left(\begin{array}[c]{cc} Q^2-R+\frac{Q^2}{q^2} \Pi_L &
  -Z_S\cr Z_S &\frac1\xi Q^2 -R-Z_G\cr \end{array}\right).
\end{equation}
The eigenvalues should be found with one-loop accuracy, which means
that terms proportional to the square of one-loop corrections (i.e.
$\Pi^2_L$, $Z_G^2$ and $Z_S^2$) should be neglected. This, however,
implies that the eigenvalues are the diagonal entries. With
appropriately rotated projectors in the $(P^G,P^L)$ plane, we can
therefore write (\ref{Eq:Abd})
\begin{equation}
  \left[(Q^2-R-\Pi_T)P^T + \left(Q^2-R+\frac{Q^2}{q^2} \Pi_L\right)
  \tilde P^L + \left(\frac1\xi Q^2 -R-Z_G\right)\tilde P^G\right]A =0,
\label{gauge_phys_eq}
\end{equation}
As expected, the transverse modes plus the $\tilde P_L$ mode, which
might be called longitudinal, can be made independent of the
gauge-fixing parameter (see the discussion on the renormalization of
$R$ below).

\subsection{The infrared separation scale}

The $p$-integrals in all terms of $j_\mu^{ind}$ are factorized into a
radial and an angular integral. It is well known that the HTL term
describes the dynamical screening of the electric fluctuations below
the Debye scale, an infrared separation scale (``IR cut-off'') has to
be therefore introduced into the radial integration at $p=M=C_M\times
eT$. The effective equations one arrives at in this way are to be used
for the modes $p\le M$.

Applying this cut-off to the integrals appearing in the expressions of
$R(Q)$ and $\Pi_{\mu\nu}$, one finds
\begin{eqnarray}
\nonumber
  && R = m_W^2 \left[ 1 + \frac{3+\xi}{8\pi^2}e^2 \ln\frac{\kappa
  T}\Lambda + \frac{e^2T}{2\pi^2 M} \frac{q_0}q \ln\frac{q_0+q}{q_0-q}
  \right],\nn
  &&\Pi_L = m_D^2 \left(1-\frac{q_0}{2q}\ln\frac{q_0+q}{q_0-q}\right)
  + \frac{q^2}{Q^2} \frac{e^2 T}{4\pi^2 M}(m_W^2+m_G^2+m_H^2), \\
  &&\Pi_T = m_D^2 \frac{q_0}{2q}\left[\frac{q_0}q-\frac{Q^2}{2q^2}
    \ln\frac{q_0+q}{q_0-q}\right] + \frac{e^2 T}{4\pi^2M}
  (m_W^2+m_G^2+m_H^2) \frac{q_0}q\ln\frac{q_0+q}{q_0-q},
\label{projections}
\end{eqnarray}
with $\kappa=2\pi \exp (-\gamma_E)$ and $m_D^2=e^2T^2/3 -
e^2MT/\pi^2$.  The logarithmical UV divergence in $R$ can be absorbed
into the $e^2$ renormalization. With appropriate renormalization scale
$\mu=\kappa T$ the $\xi$ dependence can be made to vanish.

The intermediate IR scale $M$ contributes a term to the Debye mass,
which depends linearly on $M$. It will be cancelled by the linear
$M$-dependence of the self-energies, which shows up in the (one-loop)
solution of the classical effective EOM (\ref{gauge_eq})
\cite{bodeker95,ASY95,Hindmarsh99,AaNauta99}.

The mass corrections in (\ref{gauge_phys_eq}), proportional to $\bP^2
A^\nu$, should be interpreted as coming from an induced (non-local)
$\Phi^2 A^2$ vertex correction. In the subsequent classical time
evolution it contributes to the self-energies of both fields.  When
(perturbatively) combined with the $\sim MT$ terms coming from the
tadpoles, it results in a finite contribution of order $T^2$,
independent of the exact choice of the coefficient $C_M$ in the
expression of $M$. Therefore, these terms can by no means be
neglected.

\section{Discussion}

In this Chapter we have investigated the induced current in the
effective EOM of the gauge field in the Abelian Higgs model.
The main result is the determination of the gauge
polarisation tensor beyond the well-known HTL expression in presence
of a constant scalar background. The expressions appear in
\eqref{gauge_phys_eq} and \eqref{projections}.
 
The most important property of the corrections is
that, just as the leading HTL term, in the high-temperature
expansion for the physical degrees of freedom they are independent of the
gauge-fixing procedure (after appropriate renormalization).

The result is compatible with gauge invariance if the pieces which
are proportional to $\bar\Phi^2$ are considered as the simplest 
manifestation of the 
nonlocal corrections to the $\frac{e^2}{2}(A^2(x)\Phi^*(x)\Phi (x))$ vertex 
appearing in \eqref{lagrangian}. Though for constant scalar background they
break the gauge symmetry, it is 
rather easy to embed them into a nonlocal gauge invariant structure.
For example, $j_\mu^{(2)}(x)$ of \eqref{jmu2} follows from the
functional derivation
with respect to the vector potential of the expression
\begin{equation}
-2e^2\int d^4x(D_\mu\Phi(x))^*{\hat{\cal D}}(D)D^\mu\Phi(x),
\label{vertex_corr}
\end{equation}
where $D_\mu$ is the operator of the covariant derivation and
the operator $\hat{\cal D}(D)$ is the obvious covariant generalisation of
\begin{equation}
\int{d^4p\over (2\pi )^4}2\pi\delta (p^2)
{dn(|p_0|)\over dp^\rho}{\partial^\rho\over 2p\cdot\partial}.
\end{equation}
Up to partial integrations Eq.\eqref{vertex_corr} is a unique gauge
invariant completion of \eqref{jmu2} and similar type of completion
can be written down for $j_\mu^{(3)}$. For the determination of the
finite temperature effective theory in imaginary time
the same strategy was used by two of us successfully
\cite{jakovac94}.
 
For us the only
important point is that the corrections to the HTL-current
are compatible with the gauge invariance of the full theory. The piece
we explicitly picked up in the linear response approximation is certainly
dominant for slowly varying scalar background. 

For consistency we have applied an IR cut-off $M=C_M\times eT$ to the
fluctuations, and the effective EOM is valid below this scale. The
subsequent 3D time evolution will be insensitive to the accurate
choice of $M$.  Partly it is cancelled by the 3D
(Rayleigh--Jeans-type) divergences, partly it yields also non-zero,
$M$-independent contributions, when the gauge--scalar vertex
corrections are combined with the 3D would-be divergences.
The order of magnitude of the tadpole contribution, remaining after
the 3D divergences are canceled is 
$(e^2T/M)\times (e^2MT)\sim {\cal O}(e^4T^2)$, if the mean field masses
are negligible (for example near the phase transition), relative to
the effective cut-off. If the temperature decreases, the mean field
masses might dominate over the momentum cut-off, and one has the
estimate $e^4T\bar\Phi^2/M\sim e^3\bar\Phi^2$. That is, deeper in the broken
phase the terms proposed in this note become more important. For
$e\bar\Phi^2/T^2\sim 1$ the very idea of this scheme breaks down.
 
We shortly discuss here a possible strategy for the numerical
implementation of the effective
gauge field equation. The most convenient gauge fixing seems to be the
Landau-gauge, where the $\xi$-dependent mode does not propagate. Then
$j_\mu^{(3)}$ can be left out of the discussion and the effective
equations of motion are to be solved under the constraint
$\partial_\mu A^\mu =0$. It can be implemented, for example, by
solving the equations for $A_i$ and computing $A_0$ from the
constraint.

The non-local induced currents are written in local form with the help
of auxiliary fields. With the well-known form of the leading HTL
current, we can write in this form also the corrections due to the
non-zero scalar background:
\begin{eqnarray}
&& j_i^{(1)}(x)=m_D^2\int\frac{d\Omega_v}{4\pi}\left(v_iv_j -
  \frac{e^2T}{2\pi^2M}\frac{m_W^2+m_G^2+m_H^2}{m_D^2}\delta_{ij}\right)
  W_j^{(1)}(x,{\bf v}),\nn
&& j_i^{(2)}(x)=\frac{e^2m_W^2T}{\pi^2 M}\int \frac{d\Omega_v}{4\pi}
W_i^{(2)}(x,{\bf v}),
\end{eqnarray}
where the auxiliary fields satisfy
\begin{equation}
(\d_0-\mbox{\boldmath $v\d$}) W_i^{(1)} = F_{0i},\quad
  (\d_0-\mbox{\boldmath $v\d$}) W_i^{(2)} = \d_0 A_i.
\end{equation}
In the derivation of the expression of $j_\mu^{(1)}$ we have performed
a partial $p$-integration in the part of (\ref{jmu1}) proportional to
$\delta'(p^2)$, and used the fact that, to the accuracy of our
calculation, $Q_\nu F^{\nu\mu}(Q)=0$ can be exploited in the
expression of the induced current.The auxiliary fields account
for the energy flow from the low frequency electromagnetic field
towards the high frequency $(k>M)$ modes, due to Landau damping. They
represent generalisation of the well-understood effect in
electromagnetic plasma \cite{LeBellac} to the case of Higgs-effect.

The equations for the scalar field receive mainly local corrections
(renormalization and $T$-depen\-dence of the couplings in the classical
equations). However, for consistency also the $\Phi$-derivative of the
$\Phi^2$--$A^2$ vertex correction \eqref{vertex_corr} should be introduced.

The logics of the derivation followed in the Abelian Higgs model seem
to be robust enough for us to attempt its generalisation to the
non-Abelian case, which will be the subject of a future study.  A
numerical study of the corrected EOMs can lead us to a deeper
understanding of the reliability of our present views on the
high-temperature Higgs models, especially where the sphaleron rate is
concerned. We will also learn in more detail of the non-equilibrium
aspects of the cosmological onset of the Higgs regime.

\clearemptydoublepage

%% file: acknowledgments.tex
\section*{Acknowledgements}
First of all I would like to thank my advisor Prof. Andr\'as Patk{\'o}s for
giving many useful advice during the years, for his constant support and
encouragement. I would also like to thank his patience and help 
during the elaboration of this thesis. I express my gratitude to all my
collaborators Szabolcs Bors\'anyi, Antal Jakov\'ac, P{\'e}ter Petreczky and
J\'anos Polonyi for many interesting and illuminating discussions. Special
acknowledge to Szabolcs Bors\'anyi since without his skills the numerical
work would hardly have been accomplished. We thank Zolt\'an Fodor for
generously providing computing resources at the Inst. for Theoretical
Physics of the E{\"o}tv{\"o}s University. Parts of the work presented
profited from valuable discussions with Prof. Zolt\'an R\'acz. 
Travel grants from the doctoral program of E{\"o}tv{\"o}s
University are gratefully acknowledged.

%% file: appendix.tex
\pagestyle{plain}
\chapter{\normalsize The introduction of the one-particle distribution function
and the calculation of the damping rate}
\label{app:partition}

In the relativistic kinetic theory (see for example \cite{groot1,groot2})
the number of world lines $\Delta N(x,p)$ representing particles with 
four-momenta $p_\mu=m_{eff}dx_\mu(\tau)/d\tau$ which intersects at point $x$
a surface segment $\Delta^3\Sigma$ of a space-like three-surface in
Minkowski space is given by the relation
\be
\Delta N(x,p)=\int_{\Delta p}\int_{\Delta^3\Sigma}{\mathcal N}(x,p)
p^\mu \Delta^3\Sigma_\mu d^4p,
\ee
where
\be
{\mathcal N}(x,p)=\frac{1}{m_{eff}}\sum_{i=1}^{N}\int_{-\infty}^{\infty}
d\tau\delta^{(4)}\left[x-x_i(\tau)\right]
\delta^{(4)}\left[p-p_i(\tau)\right],
\label{Eq:function_N}
\ee
is a measure of the world line density.

It is easy to construct a new function ${\mathcal N}'(x,p)$ such that the
physical constraints, the mass-shell condition and positive energy are
factored out into the phases space measure
$dP=\frac{d^4p}{(2\pi)^3}\,2\theta(p_0))\delta(p^2-m_{eff}^2)$, in
such a way that $\int d^4p{\mathcal N}(x,p)=\int dP{\mathcal N}'(x,p)$ holds.
The construction goes as follows.

Owing to the fact that $p_i^0=\left(\p_i^2+m^2_{eff}\right)^{1/2}$ the delta
function containing $p^0$ can be factored out from (\ref{Eq:function_N}).
Then, the proper time variable $\tau$ is replaced with $\tau_i$ and
changing the integration variable from proper times $\tau_i$ to the time
variable $x_i^0$ with the help of
$d\tau_i=|dx_i^0/d\tau_i|^{-1}dx_i^0=(p_i^0/m)^{-1} dx_i^0$, we may integrate
with respect to these variables.

We obtain:
$
\displaystyle
{\mathcal N}'(x,p)=(2\pi)^3\sum_{i=1}^{N}d
\delta^{(3)}\left[\x-\x_i(t)\right]\delta^{(3)}\left[\p-\p_i(t)\right],
$
where the positions and momenta of the particles are written as functions of
time. Using the properties of the $\delta$ function one can show that both
functions satisfy the Boltzmann equation (\ref{Eq:colisionless}).

A statistical averaging over the particle positions and momenta
introduces the one particle distribution function:
\be
\avr{\mathcal N}=2\theta(p_0)\delta(p^2-m_{eff}^2)\avr{{\mathcal N}'}=:
(2\pi)^3\, 2\theta(p_0)\delta(p^2-m_{eff}^2)n(x,p).
\ee

\noindent
Next we present the evaluation of three integrals appearing in Eqs.
(\ref{effvev}), (\ref{Eq:jind2}).
\bea
\label{Eq:elso}
\hspace*{-0.75cm}
I_1&=&\int{d^4p\over (2\pi)^4} \Delta_0(p)=
\frac{1}{2\pi^2}\int_{-\infty}^\infty dp_0\int_0^\infty d|\p| |\p|^2  
\delta(p^2_0-\omega^2(\p))\left(\Theta(p_0)+\tilde n(|p_0|)\right),\\
\label{Eq:masodik}
\hspace*{-0.75cm}
I_2&=&\Im \int{d^4p\over (2\pi)^4}
\frac{k_0}{k\cdot p} 2\pi\delta(p^2-M^2)\frac{d\tilde n(|p_0|)}{dp_0},\\
\hspace*{-0.75cm}
I_3&=&\int {d^4p\over (2\pi)^4} \frac{\partial \Delta^{0}(p)}{\partial p^2}=
\frac{1}{2\pi^2}\int_\infty^\infty dp_0\int_0^\infty d|\p| |\p|^2
\frac{\partial^2}{\partial p^2}\delta(p^2_0-\omega^2(\p))
\left(\Theta(p_0)+\tilde n(|p_0|)\right).
\label{Eq:harmadik}
\eea
We use the property of the $\delta$ function:
\be
\delta(p^2_0-\omega^2(\p))=\frac{1}{2\omega(\p)}
\left(\delta(p_0-\omega(\p)+\delta(p_0+\omega(\p))\right),
\quad \textrm{with}\quad\omega(\p)=\left(\p^2+M^2\right)^{1/2}.
\label{Eq:deltaprop}
\ee

After the evaluation of the $p_0$ integral the two terms arising from the
use of (\ref{Eq:deltaprop}) turn out to be identical in the case of the
three integrals above, that can be easily seen using the change of variable
$\p\rightarrow -\p$. Here we also have assumed that the $\phi$-field
consists solely of a single mode, characterised by the momentum 4-vector
$(\omega ,{\bf k})$.

The temperature-dependent part of integral (\ref{Eq:elso}) is easy to
evaluate in the high temperature limit. In dimensionless variables it can be
rewritten as
\bea
I_1=\frac{1}{2\pi^2}\int_0^\infty d|\underline p|\frac{|\underline p|^2}
{\sqrt{|\underline p|^2+M^2}}\enspace\frac{1}{e^{\beta p_0}-1}
\biggl|_{p_0=\sqrt{|\underline p|^2+M^2}}
=\frac{M^2}{2\pi^2}\int\limits_{1}^{\infty} dy \frac{\sqrt {y^2-1}}
{e^{\beta M y}-1}.
\eea
Making use of a high temperature $(\beta M<<1)$ expansion  in the
above integral, namely
\be
F(u)=\int\limits_{1}^{\infty}dx\frac{\sqrt{x^2-1}}{e^{ux}-1}=\frac{2\pi^2}{u^2}
\left(\frac{1}{12}-\frac{u}{4\pi}+O(u^2\ln u)\right),
\ee
we obtain
\be
I_1(x)=\frac{1}{12}\frac{1}{\beta^2},
\ee
which reproduces the thermal mass expression for the low-$k$ modes
correctly.

For the integral (\ref{Eq:masodik}) one obtains:
\bea
I_2&=&\frac{1}{4\pi^2}{k_0\over|\k|}\Im\int_0^\infty d|\p|\int_{-1}^{1} dy\,
\frac{|p|^2}{\omega(\p)}\,\frac{1}{\omega(\p)\,
k_0/|\k|-|\p|y+i\varepsilon}\,\frac{d n_0(\omega(\p))}{d\omega(\p)}\nn
&=&
\frac{1}{4\pi^2}{k_0\over|\k|}\Im \int\limits_1^\infty\,dt \int_{-1}^{1} dy\,
\,\frac{\sqrt{t^2-1}}{k_0/|\k|\,t-\sqrt{t^2-1}\,y+i\varepsilon}
\,\frac{d\hat n_0(t)}{dt}\\\nonumber
&=&
-\frac{1}{4\pi}{k_0\over|\k|}\int\limits_{1}^{\infty}\,dt\int
\limits_{-1}^{1}dy\,\delta\left(y-\frac{k_0}{|\k|}\,
\frac{t}{\sqrt{t^2-1}}\right)\,\frac{d\hat n_0(t)}{dt}\,
=-\frac{1}{4\pi}{k_0\over|\k|}
\int\limits_{1\over{\sqrt{1-(k_0/|\k|)^2}}}^{\infty}\,dt
\frac{d\hat n_0(t)}{dt},
\label{Eq:imI1}
\eea     
where for the second line we used the substitution
$t=\omega(|p|)/M$ and have introduced 
$\hat n^{0}(t)=1/(e^{\beta t M}-1)\, \Theta(Mt-\lambda)$. For the third line 
we applied the Landau prescription 
\be
\frac{1}{z+i\epsilon}={\mathcal P}\frac{1}{z}-i\pi\delta(z)
\label{eq:pol}.
\ee

For the integral (\ref{Eq:harmadik}) we use the relation 
$\displaystyle\partial \delta(p^2)/\partial p^2=\frac{1}{2p_0}
\partial \delta(p^2)/\partial p_0$ and perform a partial integration in the
$p_0$ integral:
\bea
\nonumber
I_3&=&\frac{1}{4\pi^2}\int_0^\infty d|\p| |\p|^2 \int_{-\infty}^\infty dp_0 
\delta(p^2_0-\omega^2(\p))\left[
\frac{1}{p_0}\frac{d\tilde n(|p_0|)}{d |p_0|} \epsilon(p_0)-
\frac{\tilde n(|p_0|)}{p_0^2}-\frac{\Theta(p_0)}{p_0^2}
\right]\\
&=&\frac{1}{4\pi^2} \int_0^\infty d|\p|
\frac{|\p^2|}{\omega^2(\p)}
\left[
\frac{1}{\omega(\p)}\left(\frac{1}{2} +\tilde n(\omega(\p))\right)
- \frac{d\tilde n(\omega(\p))}{d\omega(\p)}\right].
\eea

\clearemptydoublepage

\chapter{\normalsize Equivalence of the one-loop perturbation theory and the
once iterated solution of the generalised Boltzmann-equations }
\label{app:pertth}

Instead of the generalised Boltzmann-equations described in
Subsection~\ref{sec:2pointfunction} we can use also perturbation theory
to evaluate the two-point correlation functions. Here we will
demonstrate the equivalence of the one-loop perturbation theory and the
iterative solution of the generalised Boltzmann-equations in the case
of $\Delta^{(1)}$.

In the perturbation theory we write
\be
\left<\phi_a(x)\phi_b(x)\right> =\left<\textrm{T}_c\,
  \phi^{(0)}_a(x)\phi^{(0)}_b(x)\,e^{-i{\cal S}_I}\right>_0,
\ee
where $\phi^{(0)}$ are the free fields, ${\cal S}_I$ is the
part of the action which describes the interaction between the different
field components in the presence of the background. $\textrm{T}_c$ stands
for the time ordering along a complex time path $C$ specified in
\cite{LeBellac}. At one loop, to linear order in $\Phi$ we need only
\be 
S_I=\frac{\lambda\bar\Phi}6\,\int_C dx_{0c}\int d^3x
\left[ \Phi_1\,(\phi_b)^2+ 2 (\Phi_b\phi_b) \phi_1 \right],
\ee
where the integration variable $x_{0c}$ represents the points on the
complex integration contour in the $t$ plane.  The field operator
contractions are performed with help of the matrix propagators
\begin{eqnarray}
iG_{ab}(x)=\left(\begin{array}[c]{cc}
        iG^C_{ab}(x)\quad & iG^<_{ab}(x) \cr
        iG^>_{ab}(x)\quad & iG^A_{ab}(x) \cr\end{array}\right) =
\left(\begin{array}[c]{cc}
   \langle\textrm{T}\phi_a(x)\phi_b(0)\rangle &
   \langle\phi_b(0)\phi_a(x)\rangle \cr
   \langle\phi_a(x)\phi_b(0)\rangle &
   \langle\textrm{T}^*\phi_a(x)\phi_b(0)\rangle \cr
\end{array}\right),
\end{eqnarray}
where T$^*$ denotes anti time ordering. The tree level propagators
are diagonal $G_{ab}(x)=\delta_{ab} G_a(x)$. Since the background
depends on the real (not the contour) time we find
\bea
\nonumber
  \left<\phi_a(x)\phi_b(x)\right>^{(1)} &\equiv&
  -i\left<\phi^{(0)}_a(x)\phi^{(0)}_b(x)\,{\cal S}_I\right> \\
 &=&
  \frac{\lambda\bar\Phi}3 \int\!d^4y\, [\Phi_1(y)\delta_{ab} +
  \Phi_a(y)\delta_{b1} + \Phi_b(y)\delta_{a1}] S_{ab}(x-y),
  \label{corr6eq}
\eea
where
\begin{equation}
  iS_{ab}(z) = G_a^C(z)G_b^C(z) - G_a^<(z) G_b^<(z).
\end{equation}
The propagators can be expressed with help of the spectral function
$\rho(p)=iG^>(p)-iG^<(p)$ as
\begin{equation}
G^<(p)=n(p_0)\rho(p),\qquad G^C(t,\p)= \Theta(t)\rho(t,\p) +
G^<(t,\p).
\end{equation}
Using the ($t,\p$) representation and finally performing time Fourier
transformation leads to
\begin{equation}
  S_{ab}(k) = \int\frac{d^3{\bf p}}{(2\pi)^3}\, \frac{dp_0}{2\pi}
  \,\frac{dp'_0}{2\pi} \,\frac{\rho_a(p_0,{\bf p}) \rho_b(p'_0,{\bf
  p}+{\bf k})}{k_0-p_0-p'_0+i\epsilon} \,(1+n(p_0) + n(p'_0)).
\label{kernelqu}
\end{equation}
Using free spectral functions $\rho_a(p)=(2\pi) \epsilon(p_0)
\delta(p^2-m_a^2)$ we arrive at the known expression
(see for example \cite{boyanovsky96b})
\begin{eqnarray}
  S_{ab}(k) =&& \int\frac{d^3{\bf p}}{(2\pi)^3}
  \,\frac1{4\omega_a\omega_b}\,\biggl[  \frac{1+n_a+n_b}
  {k_0-\omega_a-\omega_b + i\epsilon
} - \frac{1+n_a+n_b}
  {k_0+\omega_a+\omega_b + i\epsilon} +\nonumber\\
  &&\frac{n_a-n_b} {k_0+\omega_a-\omega_b + i\epsilon} + \frac{n_b-n_a}
  {k_0-\omega_a+\omega_b + i\epsilon} \biggr].
\end{eqnarray}

On the other hand introducing in (\ref{kernelqu})
$\Delta^{(0)}(p)=G^>(p)=(1+ n(p_0))\rho(p)= G^<(-p)$, next performing
one of the $p_0$ integrals, and shifting properly the {\bf p} integral
we obtain
\begin{equation}
  S_{ab}(k) =\int\frac{d^4 p}{(2\pi)^4}\left[\frac{\Delta_{aa}^{(0)}(p)}
  {k^2+p^2-2kp -m_b^2} + \frac{\Delta_{bb}^{(0)}(-p)}{k^2+p^2
  -2kp -m_a^2} \right].
\end{equation}
The $i\epsilon$ is assigned to $k_0$ by the Landau-prescription. Exploiting
that $\rho (p)\sim \delta(p^2-m^2)$
one can write
\begin{equation}
  S_{ab}(k) =\int\frac{d^4 p}{(2\pi)^4}\left[\frac{\Delta_{aa}^{(0)}(p)}{k^2
  -2kp -M^2} + \frac{\Delta_{bb}^{(0)}(p)}{k^2 + 2kp +M^2} \right],
\end{equation}
where $M^2=m_b^2-m_a^2$. Finally performing a $\pm k/2$ shift the
result is
\begin{equation}
  S_{ab}(k) =\int\frac{d^4 p}{(2\pi)^4}\, \frac{\Delta_{bb}^{(0)}(p-k/2) -
  \Delta_{aa}^{(0)}(p +k/2)}{2kp + M^2}.
\label{sab_exp}
\end{equation}

Introducing for the Fourier-transform of the left hand side
of Eq. (\ref{corr6eq}) the representation
\be
\langle\varphi_a\varphi_b\rangle=\int{d^4p\over (2\pi )^4}
\Delta^{(1)}(k,p),
\ee
we obtain from Eqs. (\ref{corr6eq}) and (\ref{sab_exp})
\begin{eqnarray}
  && 2kp \, \Delta_{aa}^{(1)}(k,p) = - \lambda_a \bar\Phi \Phi_1(k)
  \left[ \Delta_{aa}^{(0)}(p+k/2) - \Delta_{aa}^{(0)}(p-k/2)\right],
  \nonumber\\
  && (2kp +M_1^2) \, \Delta
_{1i}^{(1)}(k,p) = - \frac\lambda3 \bar\Phi
  \Phi_i(k) \left[ \Delta_{ii}^{(0)}(p+k/2) -
  \Delta_{11}^{(0)}(p-k/2)\right],
\end{eqnarray}
which exactly coincides with Eq. (\ref{corr1eq}).

\chapter{\normalsize The dynamics of the classical $O(N)$ model}
\label{calss}

An alternative approach for calculating characteristic quantities of
real-time correlation functions is based on real-time dimensional
reduction \cite{nauta,jako1} and solving the resulting classical
effective theory relevant for the modes with high
occupation numbers. On-shell as well as the off-shell damping rates were
successfully reproduced \cite{Aarts97,jako97} in the symmetric phase of
$\phi^4$ theory using this approach.
In this appendix we will discuss the application of
the classical approach to the $O(N)$ model and compare its results with the
exact one-loop dynamics.
The Lagrangian of the classical $O(N)$ theory has the following form:
\be
L_{cl}={1\over 2}{(\partial_{\mu}\tilde\Phi_a)}^2-{1\over 2}
m_{cl}^2(\Lambda)
{\tilde\Phi_a}^2-{\lambda_{cl}\over 24}{(\tilde\Phi_a^2)}^2-j_a\tilde\Phi_a.
\label{lclass}
\ee
The classical
equation of motion corresponding to this Lagrangian is:
\be
(\partial^2+m_{cl}^2(\Lambda))\tilde \Phi_a +
{\lambda_{cl}\over 6} \tilde \Phi_a (\tilde \Phi_b^2)+j_a=0.
\ee
The external currents $j_a(x)$ were introduced in order to prepare the
derivation of the classical response theory. They should not be confused
with the induced currents $J_a$ appearing in the effective quantum equations
of motion. The classical mass $m_{cl}(\Lambda)$ is different from the mass
parameter $m$ of the quantum theory (see Eq. (\ref{Lagr_dens})). The same is
true for the coupling $\lambda_{cl}$. The classical mass parameter depends
on the ultraviolet cutoff $\Lambda$. The results of the classical and
quantum calculation could be matched by a suitable choice of 
$m_{cl}(\Lambda)$ and $\lambda_{cl}$. In particular we shall see below 
(eq.(\ref{class_av})), that the ultraviolet divergences of the classical 
theory can be eliminated if the divergent part of $m_{cl}(\Lambda)$ is 
suitably chosen \cite{Aarts97,jako97,nauta,jako1}.
Therefore we will separate out the  divergent part from the classical
mass and write $m_{cl}^2(\Lambda)=m_T^2+\delta m^2(\Lambda)$
\footnote{The divergent part of the classical mass parameter will be treated
as interaction, similarly to quantum field theory.}. Note that $m_T$ is not
only the thermal mass, see Eq. ({\ref{Eq:osszematch}). 

In the broken symmetry phase one separates out the condensate
$\bar \Phi$,
\be
\tilde \Phi_a=\bP \delta_{a1}+\Phi_a
\ee
and the equations of motion read
\bea
\biggl(\partial^2+m_T^2+{\lambda_{cl}\over2} \bP^2\biggr) 
\Phi_1+{\lambda_{cl}\over 6}\Phi_1
(\Phi_a^2)+{\lambda_{cl}\over 2} \bP \Phi_1^2 +{\lambda_{cl}\over 3}\bP
\Phi_i^2+\delta m^2 \Phi_1+j_1=0,
\label{cleq1}\\
\biggl(\partial^2 + m_T^2 + {\lambda_{cl}\over 6} \bP^2\biggr)\Phi_i+
{\lambda_{cl}\over 6} \Phi_i (\Phi_a^2) +{\lambda_{cl}\over
3}\bP \Phi_1 \Phi_i+\delta m^2 \Phi_i+j_i=0.
\label{cleq2}
\eea
In addition the following initial conditions are imposed:
\be
\Phi_a(t=0,\x)=F_a(\x), \qquad \partial_t \Phi_a(t,\x)|_{t=0}=P_a(\x).
\ee
The expectation value of some quantity $O$ (e.g. some correlation
function ) is obtained by averaging over the initial conditions
with the Boltzmann factor determined by the classical Hamiltonian
$H_{cl}(P_a,F_a,\bP)$ corresponding to (\ref{lclass})
\bea
<O>&=&{1\over Z}\int DF_a DP_a O \exp(-\beta H_{cl}(P_a,F_a,\bP))\\
Z&=&\int DF_a DP_a \exp(-\beta H_{cl}(P_a,F_a,\bP)),
\eea
The explicit form of $H_{cl}(P_a,F_a,\bP)$ is
\bea
H_{cl}(P_a,F_a,\bP)=\int d^3 x \biggl[
{1\over 2} P_a^2+{1\over 2}{(\partial_i F_a)}^2+
{1\over 2} (m_T^2 +{\lambda_{cl}\over 2}\bP^2)F_1^2+{1\over 2}(m_T^2+
{\lambda_{cl}\over 6} \bP^2)F_i^2+\nonumber\\
{\lambda_{cl}\over 6} \bP F_1 {(F_b)}^2+
{\lambda_{cl}\over 24} {(F_a^2)}^2+(m_{cl}^2 \bP+{\lambda_{cl}\over
6}\bP^3) F_1+{1\over 2} m_{cl}^2 \bP^2+{\lambda_{cl}\over 24} \bP^4+
\delta m^2 F_a^2 \biggr],
\label{hclass}
\eea
where we have separated out the classical condensate $\bP$ since no
averaging over classical condensate is understood.
Since the classical condensate is separated both from the dynamical
fields and from the initial conditions the following equation holds:
\be
<\Phi_1(t=0,\x)>=<F_1(\x)>=0.
\ee
Using the explicit form of $H_{cl}$ (Eq. (\ref{hclass}))
this results at one-loop level in the following
equation for the classical condensate $\bP$:
\be
\biggl(m_{cl}^2(\Lambda)+{\lambda_{cl}\over 6} \bP^2\biggr)\bP+
{\lambda_{cl}\over 6}\bP \int {d^3 p\over {(2 \pi)}^3} \biggl[3{T\over
\p^2+m_T^2+{\lambda_{cl}\over 2}\bP^2}+(N-1){T\over
\p^2+m_T^2+{\lambda_{cl}\over 6}\bP^2}\biggr]=0.
\label{class_av}
\ee
Choosing $\delta m^2(\Lambda )$ to cancel the linearly divergent piece on
the left hand side this relation reduces in the high temperature limit
($m_T,\lambda_{cl} \bP\ll T$) to
\be
m_T^2+{\lambda_{cl}\over 6} \bP^2=0
\label{vac11}
\ee
and
\be
\delta m^2=-{\lambda_{cl}\over 6}(N+2){\Lambda T\over 2 \pi^2}.
\label{dm2}
\ee
Our procedure of solving the classical theory perturbatively closely follows
Ref. \cite{jako97}.
Equations
(\ref{cleq1}), (\ref{cleq2}) are rewritten in form of the following
integral equations:
\bea
&&
\Phi_1(x,j)=\Phi_1^0(x)+\int d^4 x' D_R^1(x-x')
\biggl({\lambda_{cl}\over 6} \Phi_1
(\Phi_a^2)+{\lambda_{cl}\over 2}\bP \Phi_1^2+{\lambda_{cl}\over 3}\bP
\Phi_i^2+\delta m^2 \Phi_1+j_1\biggr)\nonumber\\
&&
\Phi_i(x,j)=\Phi_i^0(x)+\int d^4 x' D_R^i(x-x')
\biggl({\lambda_{cl}\over 6} \Phi_i (\Phi_a^2) +{\lambda_{cl}\over 3}\bP 
\Phi_1 \Phi_i+\delta m^2 \Phi_i+j_i\biggr),
\label{inteq1}
\eea
where
\be
D_R^a(x-x')=-\theta(t) \int{d^3 q\over {(2 \pi)}^3}e^{i \q
\x}{\sin\omega_qt\over \omega_q}
\ee
is the classical retarded Green function \cite{jako97} with
$\omega_q=\sqrt{q^2+m_a^2}$, where $m_1^2=m_T^2+{\lambda_{cl}\over 2} \bP^2=
{\lambda_{cl}\over 3} \bP^2$ is the Higgs field mass and $m_i^2=m_T^2+
{\lambda_{cl}\over 6} \bP^2=0$ (we have used Eq. (\ref{vac11})).
Furthermore, $\Phi_a^0$ are the solutions of the free
equations of motion and $j$ in the argument of $\Phi_a$ refers to the
functional dependence on $j=(j_1,j_i)$.  Following Ref. \cite{jako97}
one introduces linear response functions
\be
H_{R}^{ab}(x-x')={\delta \Phi_a(x,j)\over\delta j_b(x')}.
\ee
Using the integral equation (\ref{inteq1}) one can derive a coupled set of
integral equations also
for the linear response functions $H_R^{ab}$ by functional differentiation
of eq. (\ref{inteq1})
(see Ref. \cite{jako97} for details). These integral equations can be
solved iteratively in the weak coupling limit. When solving the
equations, it is important to exploit the fact that only diagonal
components of $H_R^{ab}$ have terms ${\cal O}(\lambda^{0})$.  The
classical retarded response function is the ensemble average of
$H^{ab}_R$ with respect the initial conditions~:
\be
G_{ab}^{cl}(x-x')=<H^{ab}_R(x-x')>.
\ee
By eq.(\ref{inteq1}) one easily finds that it satisfies
a Dyson-Schwinger equation of the general form
\be
G_{ab}^{cl}(x-x')=D_R^a(x-x')\delta_{ab}+
\int d^4y d^4 y' D_R^{a}(x-y) \delta_{ad}
\Pi^{cl}_{dc}(y-y') G_{cb}^{cl}
(y'-x'),
\label{ds}
\ee
where $\Pi_{ab}^{cl}(y-y')$ is the classical self-energy.

In the perturbative expansion the average is done with
the free Hamiltonian and therefore all thermal $n$-point functions are
expressed as products of the two-point function of the free fields
(solutions of the free equation of motion).
The two point function of the free fields reads \cite{Aarts97, gpa} as
\be
\Delta_a^{cl,0}(x-x')=<\Phi^0(x)_a \Phi^0(x')_a>^0=
T \int {d^3 q\over {(2 \pi)}^3}e^{i \q
(\x-\x')} {1\over \omega_q^2} \cos(\omega_q (t-t'))
\ee
This classical two point function is analogous to the free two point
function of the quantum theory $\Delta_{aa}^{(0)}(x,x')$
Using eq. (\ref{inteq1}) one gets the following
self-energies for the Higgs and Goldstone fields at leading order in
the coupling constant $\lambda_{cl}$.
\bea
\Pi_{11}^{cl}(\omega,\k)&=&
\delta m^2+{\lambda_{cl}\over 6} \sum_a \Delta^{cl,0}_a(0,0)+
\nonumber\\
&&
{(\lambda_{cl} \bP)}^2 \int dt d^3y e^{i\omega t-i \k \y}\biggl(
\Delta_1^{cl,0}(y)D_R^1(y)+
{N-1\over 9} \Delta_i^{cl,0}(y) D_R^i(y)\biggr),
\nonumber\\
\Pi_{ii}^{cl}(\omega,\k)&=&\delta m^2+
{\lambda_{cl}\over 6} \sum_a \Delta^{cl,0}_a(0,0)+\nonumber\\
&&
{\biggl({\lambda_{cl} \bP\over 3}\biggr)}^2 \int dt d^3 y e^{i \omega t-i \k
\y} \biggl(\Delta^{cl,0}_1(y) D_R^i(y)+\Delta^{cl,0}_i(y) D_R^1(y) \biggr),
\label{piclass}
\eea
Using the explicit form of $\Delta^{cl,0}_a(0,0)$ and Eq. (\ref{dm2})
one can easily verify that
all divergencies present in the above expression cancel.
Furthermore, in the high temperature limit and at leading order
in the coupling constant only the last terms contribute in the expression of
$\Pi_{11}^{cl}(\omega,\k)$ and $\Pi_{ii}^{cl}(\omega,\k)$.
One has to evaluate integrals of the following intrinsic form
\begin{equation}
  \label{kernelcl}
\hspace*{-0.25cm}
  \int dt \int d^3 y e^{i \omega t -i \k \y} D_R^a(y) \Delta^{cl,0}_{b}(y)
  =\int_{\bf p}\int {dp_0\over 2\pi} {dp_0'\over 2 \pi}
  {\rho_a(p_0,\omega_a) \rho_b(p_0',\omega_b)\over \omega-p_0-p_0'+i \epsilon}
  \biggl(n^{cl}(p_0)+n^{cl}(p_0')\biggr),
\end{equation}
where $\omega_a=\sqrt{\p^2+m_a^2}$, $\omega_b=\sqrt{(\p+\k)^2+m_b^2}$ and
$\int_{\bf p}=\int{d^3p\over {(2 \pi)}^3}$.
The above expression coincides with the
result of the quantum calculation (\ref{kernelqu}), except the fact
the there is no $T=0$ contribution and the Bose-Einstein factors are
replaced by the classical distribution: $T/p_0$. Then it is easy to write
down
the explicit expression for the classical self-energies using the
formal analogy with the result of the one-loop quantum calculations.
For example for the
classical on-shell damping rate of the Goldstone modes one easily gets
the following expression:
\bea
\nonumber
\Gamma_i^{cl}(\k)&=&-{{\rm Im} \Pi^{cl}_{ii}(\omega=|\k|,\k)\over 2 |\k|}=
{\lambda_{cl}^2 \bP^2 \over 288 \pi^2 {| \k |}^2}
\int_{{m_{1}^2\over 4 |\k|}}^{{m_{1}^2\over 4 |\k|}+|\k|}
dp\, n^{cl}(p)\\
&=&
{\lambda_{cl}^2 \bP^2 T\over 288 \pi^2 {|\k|}^2}
\ln\left(1+{4 {|\k|}^2\over m_{1}^2}\right).
\label{giclass}
\eea

Now let us discuss the correspondence between classical and quantum
calculations. It was shown in Refs. \cite{Aarts97,jako97,nauta,jako1} that
the result of classical calculations can reproduce the high temperature
limit of the corresponding quantum results if the parameters
$m_{cl}^2(\Lambda)$ and $\lambda_{cl}$ are fixed to the values
determined
by static dimensional reduction. For our case this implies:
\be
m_{cl}^2(\Lambda)=m^2+{\lambda\over 6}(N+2)\biggl({T^2\over 12}-
{\Lambda T\over 2 \pi^2}\biggr),~~\lambda_{cl}=\lambda,
\label{Eq:osszematch}
\ee
where $m$ and $\lambda$ are the mass and the coupling constant of the
corresponding quantum theory. From the above equations it is easy to
see that the divergent part in $m_{cl}^2$ coincides with $\delta
m^2$ determined from the ultraviolet finiteness of the classical result
(cf. Eqs. (\ref{class_av}),(\ref{dm2}) and (\ref{piclass})) and
$m_1$ is the high temperature limit of $M_1$ (see Eqs. (\ref{vac11}) and
(\ref{avphi})).
For small values of $|\k|$ ($|\k| \ll m_{1}$) the logarithm in this
expression can be expanded and one obtains the result of
Eq. (\ref{goldstone_class}), which fails to reproduce the result of the
quantum calculation. The same is true for the whole imaginary part of
the Goldstone self-energy.  The high temperature limit of the
imaginary part of the Higgs self-energy (see Eq. (\ref{pis1})), on the
other hand, is well reproduced by the classical theory.

This result can be easily understood by looking at the effective
quantum equation of motion (\ref{effeqs}). In the equation for the
Goldstone fields the induced current $J_i$ has a
non-local contribution
from loop momenta $p\sim M_1^2/|\k|>>T$ (cf. (\ref{impis})).
However, no such term is present in the corresponding classical
equation of motion on one hand and the classical theory cannot describe
fluctuation with wave length much smaller than $T^{-1}$ on the
other hand. The induced current $J_1$ in the effective equation of
motion for the Higgs fields receives non-local contribution only from
the loop momenta around $p \sim M_1$ which can be described in the
framework of the classical theory.

\clearemptydoublepage

\chapter{\normalsize Evaluation of some integrals appearing in Subsection 
\ref{ss:non_eq_ld}}
\label{sec:appcalc}

In this Appendix we illustrate the steps of evaluation of the
integrals in Eq.~(\ref{pis}).

\paragraph{Imaginary parts.}

As example of the evaluation of a relevant integral we discuss in detail

\begin{equation}
  \textrm{Im} R_1(k,M)= \textrm{Im} \int \frac{d^4p}{(2\pi)^4}\,
  \frac{\Delta^{(0)}(p-k/2) - \Delta
^{(0)}(p+k/2)}{pk},
\end{equation}
where the Wigner-functions $\Delta^{(0)}$ can be either of type $11$ or
$ii$.  In order to implement the Landau-prescription we transform away any
$k$ dependence from $\Delta^{(0)}$  by shifting the $p$ integral by 
$\pm k/2$. Then we use
\begin{equation}
  \lim\limits_{\alpha\to 0}\textrm{Im}\frac1{x+i\alpha} = -\pi
  \epsilon(\alpha) \delta(x),
\end{equation}
and finally shift the integrals back. We write the 4D integration
measure as
\begin{equation}
  \int \frac{d^4p}{(2\pi)^4} = \frac1{8\pi^3}\int\limits_0^\infty\!
  dp\,p^2\!\int\limits_{-\infty}^\infty \!dp_0\!  \int\limits_{-1}^1
  \!dx,
\end{equation}
where $x=\hat{\bf p}\hat\k$ stands for the cosine of the angle between
the spatial momenta. The $x$ integration is trivial because it appears
in the Dirac-delta arising from the application of the principal value
theorem
\begin{equation}
  \int\limits_{-1}^1\!dx\,\delta(p_0k_0 - p|\k| x) = \frac1{p|\k|}\,
  \Theta(p|\k| - |p_0k_0|).
\end{equation}
Using the explicit form of the Wigner-functions $\Delta^{(0)}$
 (see
 Eq.~(\ref{propagator})) and the identity $\Theta(\omega)+n(|\omega|) =
\epsilon(\omega)(1+n(\omega))$ we find
\bea
\nonumber
  -\textrm{Im} R_1(k,M)&=&\frac1{4\pi|\k|}\int\limits_0^\infty\!dp\,p\!\!
  \int\limits_{-a}^a\!dp_0\, \epsilon(p_0\!-\!\frac{k_0}2)\,
  \epsilon(p_0\! +\! \frac{k_0}2)\times\\ 
&&\qquad\qquad\qquad\qquad\delta(p_0^2\! -\! S^2) \left[
  n(p_0\!-\!\frac{k_0}2) - n(p_0\!+\!\frac{k_0}2)\right],
\eea
where $a=p|\k|/|k_0|$ and $S^2=p^2+M^2-k^2/4$. The $p_0$ integration
over the Dirac-delta gives a constraint for the $p$ integration of the
form
\begin{equation}
  S<a \qquad\Rightarrow\qquad p^2\,\frac{k^2}{k_0^2} < \frac{k^2}4 -M^2.
\end{equation}
This can be fulfilled only for $k^2>4M^2$ (above the two-particle
threshold), or for $k^2<0$ (Landau damping). After elementary algebra
one can establish the value of the sign functions and one arrives at
the formula appearing in Eq.~(\ref{impis}).

A similar analysis can be performed for $\textrm{Im} R_i$, however, in
this case it proved to be more convenient to start from the equivalent
form
\begin{equation}
  \textrm{Im} R_i(k,M)= \textrm{Im} \int \frac{d^4p}{(2\pi)^4}\,
  \frac{\Delta_{11}^{(0)}(p-k) - \Delta_{ii}^{(0)}(p)}{2pk-k^2+M^2}.
\end{equation}
After implementing carefully the Landau prescription and performing
the $x$ integration as described before, we arrive at
\begin{equation}
  -\textrm{Im} R_i(k,M)=\frac1{8\pi|\k|}\int\limits_0^\infty\!dp\,p\!\!
  \int\limits_{b_-}^{b_+}\!dp_0\, \epsilon(p_0)\, \epsilon(p_0\!-\!k_0)\,
  \delta(p_0^2\! -\! p^2) \left[ n(p_0\!-\!k_0)-n(p_0)\right],
\label{gh_diff}
\end{equation}
where
\begin{equation}
  b_\pm = \frac{k^2-M^2}{2k_0} \pm p \frac{|\k|}{k_0}.
\end{equation}
The $p_0$ integration over the mass-shell delta-function again
restricts the domain of integration in the $p$ integral: $b_-<\pm
p<b_+$, which, however, does not restrict the possible values for
$k^2$. After the solution of these linear inequalities and the
analysis of the sign functions we arrive at the result appearing in
Eq.~(\ref{impis}).

\paragraph{Real parts.}

The relevant integrals in $\textrm{Re} R_1$ are
\begin{equation}
  I^{\pm} = \textrm{Re} \int\!\frac{d^4p}{(2\pi)^4}\,
  \frac{\Delta^{(0)}(p)}{pk \pm k^2/2},
\end{equation}
where the Wigner-functions $\Delta^{(0)}$ can be either of type $11$ or $ii$. 
With their help we find $\textrm{Re} R_1 =I^+ - I^-$. The real part comes from
the principal value integration. We decompose the integration measure
as in the calculation of the imaginary parts. When we use
Eq.~(\ref{propagator}) for $\Delta^{(0)}$, the value of $p_0$ is fixed
by the mass-shell delta function. The $x$ integration can be performed as
\begin{equation}
  \int\limits_{-1}^1\!dx\,{\cal P}\frac1{2p_0k_0 -2p|\k| x \pm k^2} =
  \frac1{p|\k|} \textrm{arth}(\frac{2p|\k|}{2p_0k_0 \pm k^2}),
\end{equation}
where $\textrm{arth}(x) =1/2 \ln|(1+x)/(1-x)|$. Using the
properties of the absolute value we find
\begin{equation}
  I^+(k_0,\k) = - I^{-}(-k_0,\k).
\end{equation}
Then we directly arrive at Eq.~(\ref{repis}).

We write $\textrm{Re} R_i$ in the form
\begin{equation}
  \textrm{Re} R_i = \textrm{Re} \int\!\frac{d^4p}{(2\pi)^4}\,
  \frac{\Delta_{11}^{(0)}(p)}{2pk + k^2 +M^2} - \textrm{Re}
  \int\!\frac{d^4p}{(2\pi)^4}\, \frac{\Delta_{ii}^{(0)}(p)}{2pk - k^2
  +M^2}.
\end{equation}
Then the previous scheme of calculation goes through directly.

The only problem still to be discussed is the zero temperature
contribution, which diverges logarithmically. The regularization and
renormalization prescriptions are better formalized in the language of
the propagators and couplings. Using the results of
Appendix~\ref{app:pertth} to go over to the perturbation theory, we
have to evaluate
\begin{equation}
  R=\int\frac{d^4p}{(2\pi)^4}\, \left[ G^C(p) G^C(p-k) -
    G^<(p) G^<(p-k) \right].
\label{req}
\end{equation}
The first term at $T=0$ is the usual time ordered product;
with finite four-dimensional cutoff one finds
\begin{equation}
  \int\frac{d^4p}{(2\pi)^4}\,G^C(p) G^C(p-k) =
  \frac{-1}{16\pi^2} \left[ 1 + \int\limits_0^1\!dx\, \ln\frac{|k^2
  x(1-x) -m_1^2 x -m_2^2 (1-x)|}{\Lambda^2} \right].
\end{equation}
The divergence is canceled by the coupling constant counterterm. In
modified minimal subtraction ($\overline{\textrm{MS}}$) scheme it reads
\begin{equation}
  \frac{-1}{16\pi^2}\int_0^1dx \ln\frac{|k^2 x(1-x) -m_1^2 x -m_2^2
  (1-x)|}{\mu^2}.
\end{equation}

In the second term of Eq. (\ref{req}) at $T=0$ we can use
$G^<(p)=\Theta(-p_0)(2\pi) \delta(p^2-m^2)$. Because of the delta
functions this piece yields finite result (as is expected by the
arguments of the renormalizability)
\begin{equation}
  \int\frac{d^4p}{(2\pi)^4}\,G^<(p) G^<(p-k) = \frac1{8\pi} \left\{
  \begin{array}[c]{ll}
    \sqrt{1-\frac{4m^2}{k^2}}\,\Theta(k^2-4m^2), & \textrm{if}\,
    ~m_1=m_2=m \cr (1-\frac{m^2}{k^2})\,\Theta(k^2-m^2), &
    \textrm{if}\, ~m_1=0,\,m_2=m. \cr
  \end{array}\right.
\end{equation}

%% file: results.tex
\section*{Summary of main results}

In my thesis I discuss the real time non-equilibrium dynamics of systems in
the broken symmetry phase. Beside reviewing the fields in which this kind of
investigation are relevant, I presented the equation  derived
to be used in the broken phase of scalar models and in the Abelian Higgs 
model. 

The main results of the analytical and numerical  investigations are as follows:

\noindent
{\bf 1.}
A classical kinetic theory was constructed for particles whose mass square
depends quadratically on the local amplitude of a real, low frequency scalar
wave. It was shown that the damping rate of the waves due to the scattering
of the particles agrees with the result of the quantum perturbation theory
in the broken phase of the one-component $\Phi$-theory.

\noindent
{\bf 2.}
The effective equation of motion of the order parameter of the real $\Phi^4$
theory extracted from numerical simulations follows the Maxwell construction
when a non-equilibrium first order phase transition occurs in the $2+1$
classical theory. The statistical features of the decay of the false vacuum
agrees with the results obtained by expanding around the critical bubble.

\noindent
{\bf 3.}
The damping rate of the quantum Goldstone modes was determined to leading
order in the $O(N)$ model with perturbation theory for low wave number $\bf k$
and in presence of small explicit symmetry breaking (h). It was shown that
the rate of relaxation vanishes non-analytically when $|{\bf k}|\rightarrow 0$
and $h\rightarrow 0$, as a direct consequence of Goldstone's theorem.
The discrepancy between the classical an quantum results was discussed.

\noindent
{\bf 4.}
The analytically determined large time asymptotic power law of the
relaxation of an arbitrary field configuration has been checked in a
numerical simulation of the $2+1$ dimensional $O(2)$ symmetric classical
field theory. In the linear regime of the relaxation the numerically
determined damping rate of the Goldstone mode was found to be in agreement
with the one calculated in the classical linear response theory. The real
time evolution of the symmetry breaking ground state into the symmetrical
one was checked. The suppression of the order expected on the basis of the
Mermin-Wagner theorem was checked to proceed through a diffusive mechanism.

\noindent
{\bf 5.}
The current density of the Maxwell equation has been determined to 
$\mathcal{O}(e^4)$ accuracy in scalar electrodynamics. This is induced by
the presence of a mean gauge field and also influenced by a symmetry
breaking homogeneous scalar background $\bar\Phi$. The result of quantum
perturbation theory displays a contribution proportional to $\bar\Phi^2$
in addition to the well-known ``hard thermal loop'' (HTL) expression.
The gauge-fixing parameter independent degrees of freedom were constructed.

%% file: eredmenyek.tex
\section*{A f\H obb eredm\'enyek \"osszefoglal\'asa}

A doktori \'ertekez\'esben a s\'ertett szimmetri\'aj\'u f\'azis
val\'osidej{\H u} nem-egyens\'ulyi di\-na\-mi\-k\'a\-j\'at vizsg\'alom. 
A val\'os idej\H u nem-egyens\'ulyi dinamika t\'emak\"or\'ehez 
kapcsol\'od\'o irodalom is\-mer\-te\-t\'ese  mellett bemutatom az \'altalam 
levezetett egyenleteket, melyek a  skal\'artereket illetve az abeli 
m\'ert\'ektereket tartalmaz\'o elm\'eletek s\'ertett szimmetri\'aj\'u 
f\'azis\'aban haszn\'alhat\'ok. Analitikus vizsg\'alatokat skal\'ar 
t\'erelm\'eletekben \'es az abeli Higgs modellben  v\'egeztem. Numerikus 
szi\-mu\-l\'aci\'okat a skal\'ar t\'erelm\'elet keretei k\"oz\"ott folytattam.

A f\H obb eredm\'enyeket a k\"ovetkez\H o pontokban foglalom \"ossze:

\noindent
{\bf 1.}
Megkonstru\'altam egy klasszikus kinetikus elm\'eletet az 
\"onk\"olcs\"onhat\'o egyens\'ulyi skal\'art\'er nagy\-frek\-ven\-ci\'as 
m\'odusait reprezent\'al\'o g\'azra. E g\'az relativisztikus skal\'ar 
r\'eszecsk\'einek t\"omege kva\-dra\-tikusan f\"ugg a hossz\'uhull\'am\'u 
h\'att\'er amplit\'ud\'oj\'at\'ol. Megmutattam, hogy a
hossz\'u\-hul\-l\'a\-m\'u m\'odusok csillap{\'\i}t\'asi r\'at\'aja, amely a
r\'eszecsk\'eken val\'o sz\'or\'od\'as k\"ovetkezm\'enye, a s\'ertett
f\'azis\-ban megegyezik a kvantum egy-hurok sz\'amol\'as eredm\'eny\'evel.

\noindent
{\bf 2.}
$2+1$ dimenzi\'os $\Phi^4$ skal\'ar elm\'elet nem-egyens\'ulyi
els{\H o}rend{\H u} f\'azis\'atalakul\'asa numerikus
szi\-mu\-l\'a\-ci\'oj\'a\-nak adataival egy effekt{\'\i}v mozg\'asegyenletet
konstru\'altam a rendparam\'eterre. Megmutattam, hogy a rendszer
dinamik\'aj\'at le{\'\i}r\'o effekt{\'\i}v mozg\'asegyenlet t\"ukr\"ozi a
Maxwell-\-konstruk\-ci\'ot: a rendparam\'eter a stabil \'ert\'ekek k\"oz\"ott
er{\H o}mentes mozg\'ast v\'egez. A metastabil $\longrightarrow$ stabil
\'atmenet id\H ostatisztik\'aj\'ab\'ol ad\'od\'o \'atmeneti r\'ata
\"osszhangban van a nukle\'aci\'os elm\'eletben sz\'amolt r\'at\'aval.

\noindent
{\bf 3.}
Az O(N) szimmetri\'aj\'u modell s\'ertett f\'azis\'aban kisz\'amoltam a
perturb\'aci\'osz\'am{\'\i}t\'as el\-s{\H o} rendj\'eben a
Goldstone-m\'odus csillap{\'\i}t\'asi r\'at\'aj\'at a $|\bf k|$ 
hull\'amsz\'am, illetve az explicit szim\-me\-tria\-s\'ert\'est jellemz{\H o} 
$h$ k\"uls{\H o} t\'er f\"uggv\'eny\'eben. Megmutattam, hogy a 
Goldstone-t\'etel \'ertel\-m\'e\-ben, a $|{\bf k}|\rightarrow 0$ m\'odusok  
csillap{\'\i}t\'asi r\'at\'aja  $h=0$-ra nem-analitikus m\'odon t{\H u}nik el. 
Kimutattam az elt\'er\'est a kvantumos \'es a klasszikus elm\'eletb{\H o}l 
sz\'amolt csillap{\'\i}t\'asi r\'ata  k\"oz\"ott.

\noindent
{\bf 4.}
$2+1$ dimenzi\'os numerikus szimul\'aci\'oval leellen{\H o}riztem az 
O(2) szimmetri\'aj\'u skal\'ar t\'er ana\-li\-ti\-kusan meghat\'arozott,
aszimptotikus id{\H o}kre vonatkoz\'o relax\'aci\'os t\"orv\'eny\'enek 
teljes\"ul\'es\'et. A relax\'aci\'o line\'aris szakasz\'aban a
Goldstone jelleg{\H u} szabads\'agi fok m\'ert csillap{\'\i}t\'asi
r\'at\'aj\'at \"osszhangban tal\'altam a klasszikus line\'aris
v\'alaszelm\'eletb{\H o}l sz\'amolt r\'at\'aval. Kimutattam a 
szim\-metrias\'ert{\H o} \'allapotnak a val\'osidej{\H u} 
\'atalakul\'as\'at a szimmetrikus alap\'allapotban. A Mer\-min--Wagner-t\'etel 
dinamikai megval\'osul\'as\'ara diff\'uzi\'os jellemz\'est adtam.

\noindent
{\bf 5.}
Skal\'ar elektrodinamik\'aban $\mathcal O(e^4)$ rend{\H u} pontoss\'aggal
meghat\'aroztam a hossz\'uhull\'am\'u m\'er\-t\'ek\-terek Maxwell egyenlet\'eben
szerepl{\H o} induk\'alt \'aramot. Az \'aramot a m\'ert\'ek \'atlagt\'er
induk\'alja \'es tartalmazza a szimmetrias\'ert\'est t\"ukr\"oz{\H o}
homog\'en skal\'ar h\'att\'er hat\'as\'at. Ennek $\Phi^2$ rend{\H u}
figyelembev\'etele t\'ulmutat az irodalomban szerepl{\H o}, un. ``hard thermal
loop'' (HTL) egyenleteken, hat\'asa fontos lehet a f\'azis\'atalakul\'as
le{\'\i}r\'asakor.  Megmutattam, hogy a fizikai szabads\'agi fokok
f\"uggetlenek a  m\'ert\'ekr\"ogz{\'\i}t{\H o} param\'etert{\H o}l.